\documentclass{article}

\usepackage{arxiv}

\usepackage[utf8]{inputenc} 
\usepackage[T1]{fontenc}    
\usepackage{hyperref}       
\usepackage{url}            
\usepackage{booktabs}       
\usepackage{amsfonts}       
\usepackage{nicefrac}       
\usepackage{microtype}      
\usepackage{lipsum}
\usepackage{graphicx,verbatim,float,subfigure,graphicx,color,siunitx}
\usepackage{amsmath}
\usepackage[dvipsnames]{xcolor}
\usepackage{makecell}
\graphicspath{ {./images/} }
\usepackage{amssymb}
\usepackage{threeparttable}
\usepackage{tablefootnote}
\usepackage[most]{tcolorbox}
\usepackage{svg}
\usepackage{multirow}  
\usepackage[export]{adjustbox}
\usepackage{soul}
\usepackage{textcomp}
\usepackage{multirow}
\usepackage{enumitem}
\usepackage{amsfonts}
\usepackage{bbm}
\usepackage{amsthm}
\newlist{Properties}{enumerate}{2}
\setlist[Properties]{label=Property \arabic*., font=\textbf, itemindent=*}

\usepackage{epstopdf}
\AppendGraphicsExtensions{.tif}

\title{\textit{Ar$\chi$i-Textile Composites}: Drapable Hybrid Woven Composites for Lightning Strike Protection}

\author{
  Hridyesh Tewani$^1$, Vincent Scheerer$^2$, Madison Owens$^2$, Emilio Cumbajin$^2$, Camila De Leon$^2$,
  \And
  MD Rashid Hussain$^3$, Pruthul Kokkada Ravindranath$^4$, Rachel Van Lear$^4$, David Jack$^4$, David Wallace$^3$ 
  \And
  Pavana Prabhakar$^{1,2}$  \\ \\
  $^1$ Dept. of Civil \& Env. Engineering, University of Wisconsin-Madison \\
  $^2$ Dept. of Mechanical Engineering, University of Wisconsin-Madison \\
  $^3$ Dept. of Electrical \& Computer Engineering, Mississippi State University, Starkville, MS 39759 \\
  $^4$ Dept. of Mechanical Engineering, Baylor University, Waco, TX 76704
}

\begin{document}
\maketitle

\begin{abstract}
Carbon fiber-reinforced polymers (CFRPs) have been extensively used in the aerospace and wind energy industries due to their superior specific mechanical properties and corrosion resistance. However, their higher electrical resistivity makes them susceptible to lightning strike damage, which necessitates the addition of a surface lightning strike protection (LSP) layer. Traditional LSP systems, such as copper mesh or expanded foil, reduce lightning strike damage, but are not easily drapable around complex geometries and may introduce delamination-prone regions within the composite. Here, \textbf{we propose a novel manufacturing strategy for architected hybrid composites as drapable LSP by weaving stainless steel yarns within the woven carbon fiber composites}. We varied the metal-to-carbon yarn ratio and stacking configuration to assess damage evolution under quasi-static arc exposures and simulated lightning strikes. Our results elucidate that incorporating hybrid layers into composites significantly reduced surface temperatures, through-thickness damage, and mass loss under both electric arc impacts. The composites with the proposed LSP layers also exhibited higher retention of flexural modulus and strength compared to the reference CFRP. Advanced air mobility (AAM) vehicles, which operate at lower altitudes, face significant safety challenges due to their high susceptibility to lightning strikes. \textit{Therefore, the proposed hybridized composites can be used as an efficient and drapable LSP around complex shapes in AAM vehicles, offering enhanced safety and protection}. 


\end{abstract}

\keywords{Hybrid composites \and Lightning strike protection \and Intralaminar hybridization \and Architected woven fabric }

\section{Introduction}\label{intro}
Carbon fiber-reinforced polymer (CFRP) composites have been increasingly used in industries seeking lightweight yet stronger solutions. Owing to their superior specific mechanical properties, fatigue resistance, and functional versatility, these composites are used as primary structural materials in the aerospace, marine, energy, and automotive industries. Despite their advantages over metal counterparts, CFRP composites are susceptible to catastrophic damage under lightning strikes due to their low and anisotropic electrical conductivity \cite{feraboli2009damage, feraboli2010damage, chakravarthi2011carbon}. Recently,  woven CFRP composites have gained attention for their excellent strength-to-weight ratio, tunable mechanical performance, and enhanced drapability \cite{bai2022physics, li2023impact}. The role of architectures in woven composites, like varying weave patterns or hybridizing with polymer fibers, on their mechanical response has been recently studied in depth \cite{feng2024physics, tewani2024archii}. However, the introduction of architectures by hybridizing carbon fiber fabrics with metallic fiber yarns for lightning strike protection remains unexplored. Understanding the role of incorporating metal fiber yarns into these woven architectures is critical for developing effective and drapable \textit{lightning strike protection (LSP)}. \textbf{Therefore, the primary objective of this study is to elucidate the effect of incorporating metallic (stainless steel) fiber yarns into carbon fiber fabrics on the electrical resistivity and performance of composites under electric arc impacts.}

Previously, extensive research has been performed on the effectiveness of metallic LSP on the dissipation of current during a lightning strike event \cite{kumar2020factors, das2021brief, wang2024challenges}. The state-of-the-art approach involves attaching metal-based linings or coatings to the surface of CFRP structures to provide a conductive path, enhancing structural safety under lightning strikes \cite{gagne2014lightning, rajesh2018damage}. Kawakami and Feraboli \cite{kawakami2011lightning} previously implemented a layer of woven copper mesh on the top layer as an effective LSP and examined the effect of scarf repairs on its residual performance. Their results showed that repairs with overlapping mesh aligned with the existing mesh performed comparably to pristine, non-repaired CFRP samples. However, the crossed wires in these meshes make it challenging to achieve a smooth surface in composites, and delamination can occur due to the weak interface between the metal and the composite. Guo et al. \cite{guo2019enhanced} conducted experimental and numerical studies on copper and aluminum expanded foils with varying densities as LSPs under 100 kA lightning current. Their findings demonstrated that adding these foils improved the conductivity of the composites and reduced lightning strike damage. The authors also suggested that the anisotropic conductivity of the expanded foil could be strategically oriented to optimize the effectiveness of LSP. Although metallic foils have shown efficacy against lightning strikes, they can create resin-rich areas within composite structures and introduce weak bonding between the composite and foil layers, leading to potential debonding \cite{gagne2014lightning, kumar2019interleaved}. Additionally, with the growing market of advanced air mobility (AAM), LSPs need to conform to complex and intricate structures of vehicles, an area where conventional metallic LSPs could face significant limitations \cite{bigand2025destructive}. 

To overcome these challenges associated with metallic LSPs, non-metallic materials with improved electrical conductivity have been proposed to enhance the lightning strike resistance of composites. Prior research includes the use of carbon nanotubes (CNTs) \cite{chakravarthi2011carbon, kumar2019interleaved, gou2010carbon,  zhang2019lightning, xia2020fabrication}, aligned carbon fibers \cite{kumar2023enhanced}, conductive resins \cite{hirano2016lightning}, and graphene \cite{wang2018fabrication, raimondo2018multifunctional}, to increase the conductivity of composite materials and limit the damage due to lightning. However, high manufacturing costs and comparatively low conductivity improvements have been previously reported \cite{das2022thickness,langot2023performance}. Hirano et al. \cite{hirano2016lightning} proposed using polyaniline (PANI)-based resin, which demonstrated reduced damage under lightning strikes (40 kA and 100 kA peak currents) and 90\% residual strength after the strike. Further, Das et al. \cite{das2022thickness} examined the threshold thickness required for PANI-based LSP. They determined that a minimum thickness of 345 $\mu$m of the LSP retained 83\% and 100\% flexural properties for glass fiber- and carbon fiber-reinforced polymers after lightning strike impacts. Although PANI-based composites displayed improved performance, they lacked mechanical properties. Moreover, incorporating a new resin system into the industrial manufacturing process poses significant challenges and requires further investigation. 

Recent advancements have proposed a strategic combination of carbon fibers with metallic reinforcements as effective LSP systems, such as non-woven nickel-coated carbon fiber (NCCF) veils \cite{langot2022comparative}. These veils are fabricated using chopped carbon fibers coated with nickel through chemical vapor deposition. As carbon fiber serves as the core material, adding a metal layer reduces resistivity while maintaining lower weight. Previous research has demonstrated the effectiveness of non-woven metal-coated carbon fiber veils under lightning strike conditions \cite{langot2023performance, zhao2018development}, showing that incorporating a layer of NCCF veils, particularly in the middle rather than the top of the composite, also improved both conductivity and interlaminar fracture toughness \cite{liu2021multifunctional}. Despite the effectiveness of non-woven metal-coated carbon fiber veils under lightning strikes, these materials pose significant challenges during manufacturing processes, high production costs, difficulty in bonding metals to carbon fibers, and the formation of resin-rich regions within composites. These issues could lead to a complicated repair process \cite{das2022thickness}. Furthermore, the uneven thickness of metal coatings on the carbon fibers may result in variable conductivities within the composite, compromising the overall reliability under lightning strikes. \textbf{Inspired by the combination of carbon and metal for LSPs, the current study focuses on incorporating metallic fibers with carbon fibers by interweaving them together}. This approach simplifies the integration of metal into carbon fiber composites using the existing weaving process, leveraging the flexibility of metallic yarns to enable a variety of weave architectures. As a result, the method yields more drapable and adaptable LSPs. Furthermore, this strategy is versatile and can be extended to other composites reinforced with glass, Kevlar, or natural fibers by choosing different metallic fiber yarns. The entire fabric can also be pre-impregnated with resin to create prepregs, streamlining its adoption within existing industrial manufacturing techniques. 

The primary goal of the current work is to enable a novel hybrid weave by integrating stainless steel yarns with carbon fiber yarns to mitigate lightning strike damage. Stainless steel is chosen as a model material, and the technique can be extended to other metallic yarns. Two types of hybrid weaves were developed, incorporating varying hybridization ratios of stainless steel to carbon fiber yarns. To assess their effectiveness as LSPs, one- and two-layer hybrid weave fabrics were added as top surface layers of the composites. The composites were subjected to two types of electric arc tests, quasi-static arc and simulated lightning strikes, to evaluate the performance of hybrid weaves. Post-testing, we visually inspected the extent of damage in the baseline composites and those with hybrid weaves. Additionally, residual strength tests were conducted to quantify the mechanical integrity of the composites after exposure to both electric arc damage, assessing the effectiveness of the proposed woven LSP. This approach provides a versatile, scalable, and effective LSP system compatible with existing composite manufacturing techniques.

\section{Motivation}\label{motiv}
\begin{figure}[ht]
\centering
\includegraphics[width=0.35\textwidth]{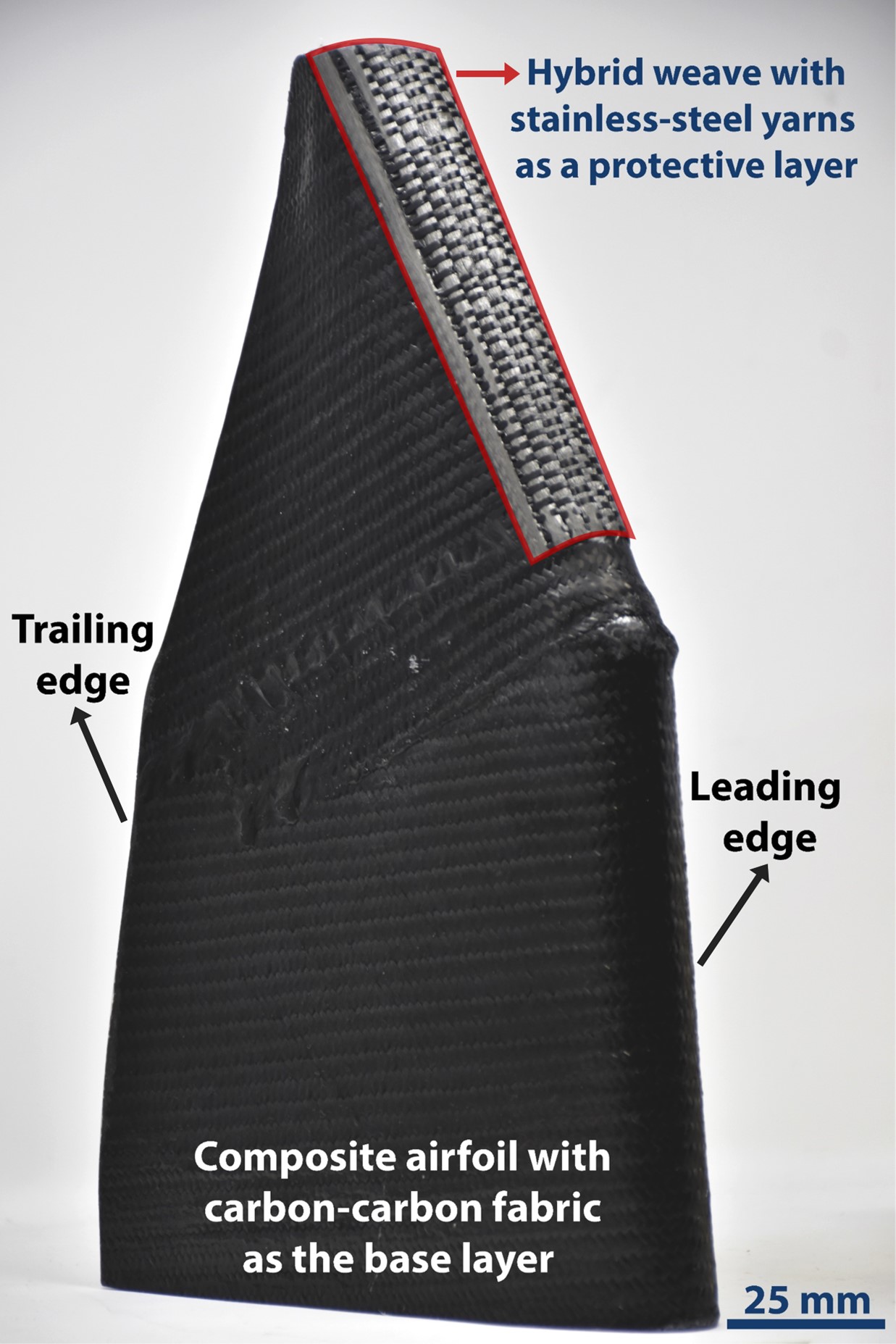}
\caption{A 3D-printed airfoil core wrapped with bi-directional carbon fabric composite, featuring a hybrid weave with stainless steel yarns, applied to the leading edge for enhanced lightning strike protection.}
\label{img:wing}
\end{figure}

In this study, we introduced novel hybrid fabrics composed of stainless steel and carbon fiber yarns to develop a new lightning strike protection (LSP) system. The application of the hybrid weaves as LSPs is demonstrated in \textbf{Figure} \ref{img:wing}.

Motivated by this, we aim to elucidate the impact of metallic fibers on the electrical resistivity of hybrid composites and the damage caused by arc impact. Although metallic reinforcements have been investigated in previous studies, the intralaminar incorporation of metal yarns in fabric to create drapable LSP has not been extensively investigated. We explored the influence of varying the hybridization ratio of stainless steel fibers to carbon fiber yarns in the developed fabrics. To assess the effectiveness of these hybrid fabrics, we subjected composite laminates made with hybrid weaves as sacrificial layers to two types of electric arc tests: (i) a low-amplitude current for a longer duration (quasi-static arc) and (ii) a high-amplitude current for a shorter duration (simulated lightning strike). 

The primary objectives of the current study are to address the following questions:

\begin{itemize}
    \item Can we hybridize carbon fiber fabric with metallic fiber yarns to create a drapable LSP?
    \item How does interweaving stainless steel yarns impact the electrical resistivity of composite materials?
    \item Can varying the hybridization ratio and fabric arrangement effectively optimize current dissipation and minimize damage caused by Joule heating?
\end{itemize}

\section{Methodology}\label{method}
This section first outlines the laminate design and the yarns used for weaving the hybrid fabrics in the current study. Next, we provide details on the weaving process, followed by a description of composite manufacturing. Further, we describe the electric arc and residual flexural tests conducted on these composites to assess the influence of weave hybridization. 

\subsection{Laminate Design}\label{lamdes}
Three categories of woven fabrics were chosen to design all laminates with 3K carbon fiber in the warp direction and:
\begin{itemize}
    \item all 3K carbon fiber in the weft direction \textit{(C/C)},
    \item all stainless steel fiber yarn in the weft direction \textit{(C/SS)}, and
    \item a combination of 24K carbon fiber and stainless steel fiber yarns in a 2:1 ratio in the weft direction \textit{(C/C-SS)}.
\end{itemize}

For tensile tests, four-layer laminates of each fabric type (C/C, C/SS, C/C-SS) were manufactured with a $[0/90]_s$ stacking, where 0 represents 3K carbon fibers in the warp direction. For the electric arc tests, three types of samples, each with six layers of fabric, were manufactured: (a) no hybrid weave (reference), (b) single layer of hybrid weave on the impact side, and (c) two layers of hybrid weave with $[0/90]$ orientation on the impact side. The samples with hybrid weaves were further categorized into two types corresponding to different hybrid weaves (C/SS - all stainless steel in weft direction and C/C-SS - alternating stainless steel and Carbon yarns in weft direction), shown in \textbf{Figure}~\ref{img:hybWeave}. These combinations resulted in five distinct sample configurations for arc tests, summarized in \textbf{Table} \ref{tab:weavesummLS}. 

\subsection{Materials}\label{met:mat}
\begin{figure}[ht]
\centering
\subfigure[]{
\includegraphics[width=0.5\textwidth]{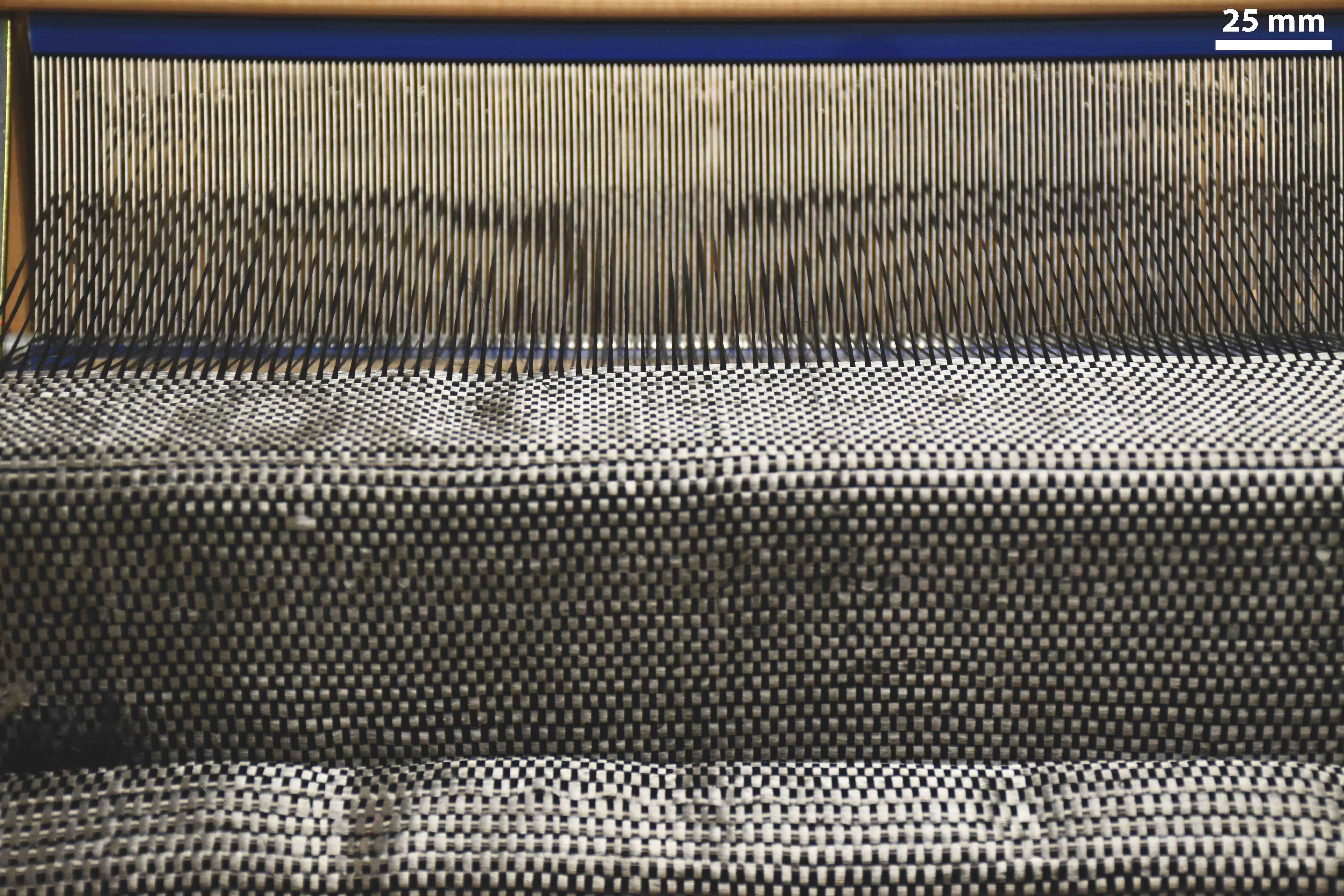}
\label{img:loom}
}
\subfigure[]{
\includegraphics[width=0.33\textwidth]{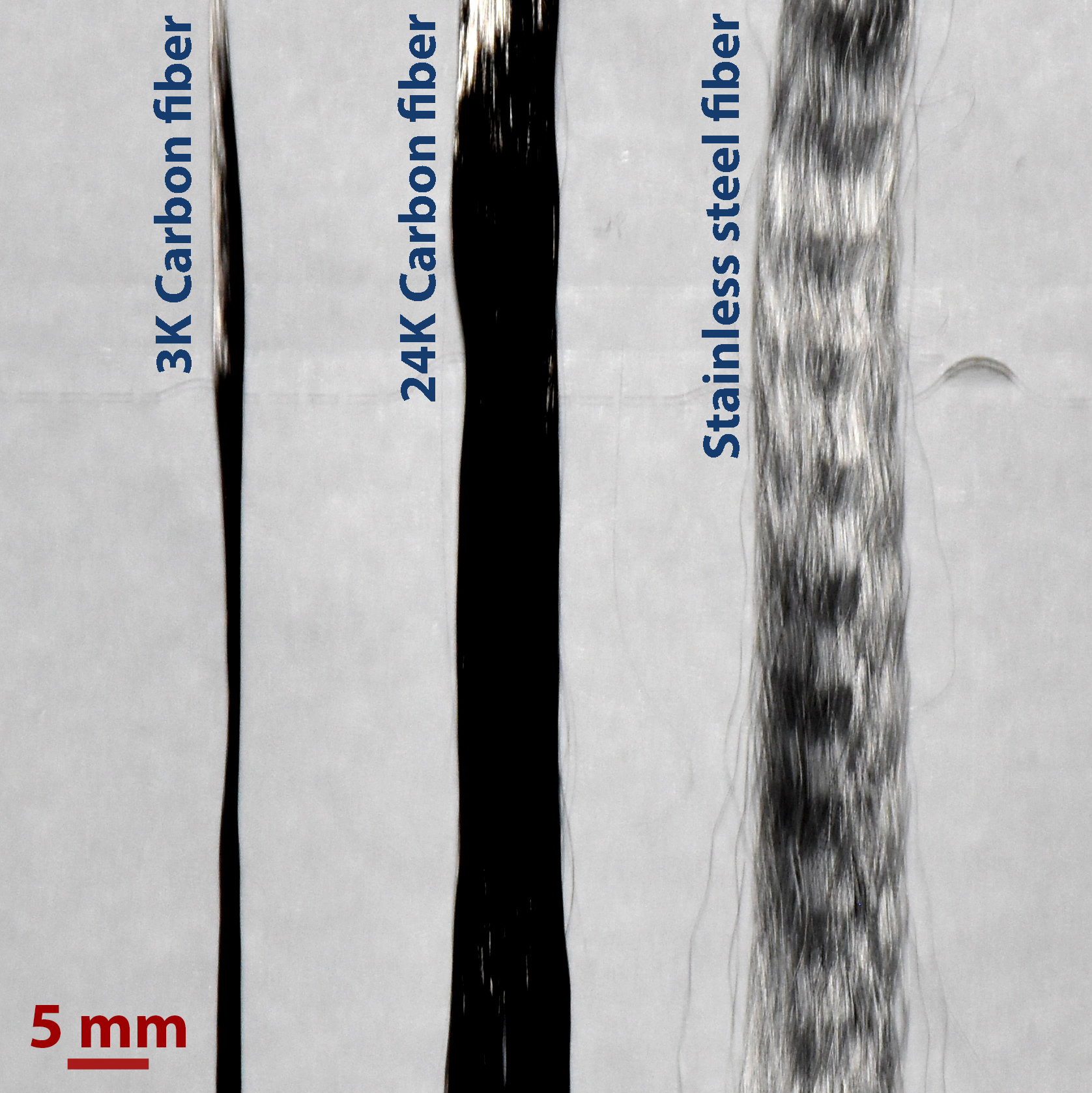}
\label{img:fibers}
}
\caption{(a) Loom setup with 3K carbon fibers in the warp direction (out-of-plane) with a density of 13 ends per inch (out-of-plane); and (b) 3K carbon fiber, 24K carbon fiber, and stainless steel fiber yarns (from left to right) used in the current study.}
\label{img:material}
\end{figure}

We sourced 3K carbon fiber yarn from ACP Composites (Toray T300B). This was used as the warp material on the loom with a density of 13 ends per inch, as shown in \textbf{Figure~\ref{img:loom}}. The conductive stainless steel yarns were purchased from Adafruit Industries, and the 24K carbon fiber yarns, used in C/C-SS fabric, were procured from FibreGlast Corp. Throughout this study, the fibers in the warp direction were the Toray T300B 3K carbon fiber yarn from ACP Composites. The materials used are shown in \textbf{Figure~\ref{img:fibers}}, including 3K and 24K carbon fiber yarns and stainless steel yarns. Additional details on the weaving process are provided in Section \ref{met:weave}. We obtained the reference 3K x 3K plain weave carbon fiber fabric from FibreGlast Corp for laminates manufactured for electric arc tests. 

\subsection{Manufacturing}\label{met:man}

\begin{figure}[ht]
\centering
\subfigure[]{
\includegraphics[width=0.4\textwidth]{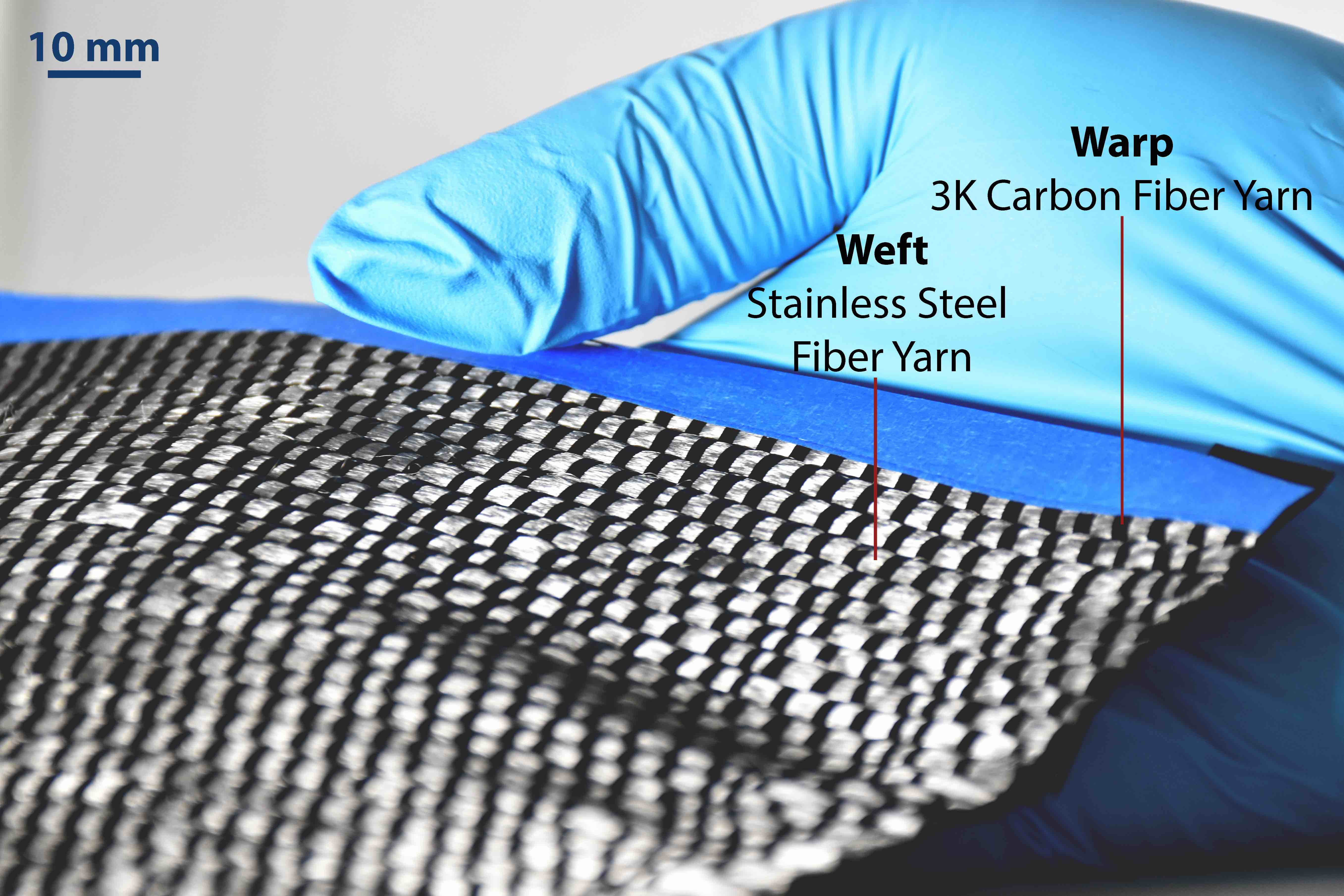}
\label{img:hybWeave1}
}
\subfigure[]{
\includegraphics[width=0.4\textwidth]{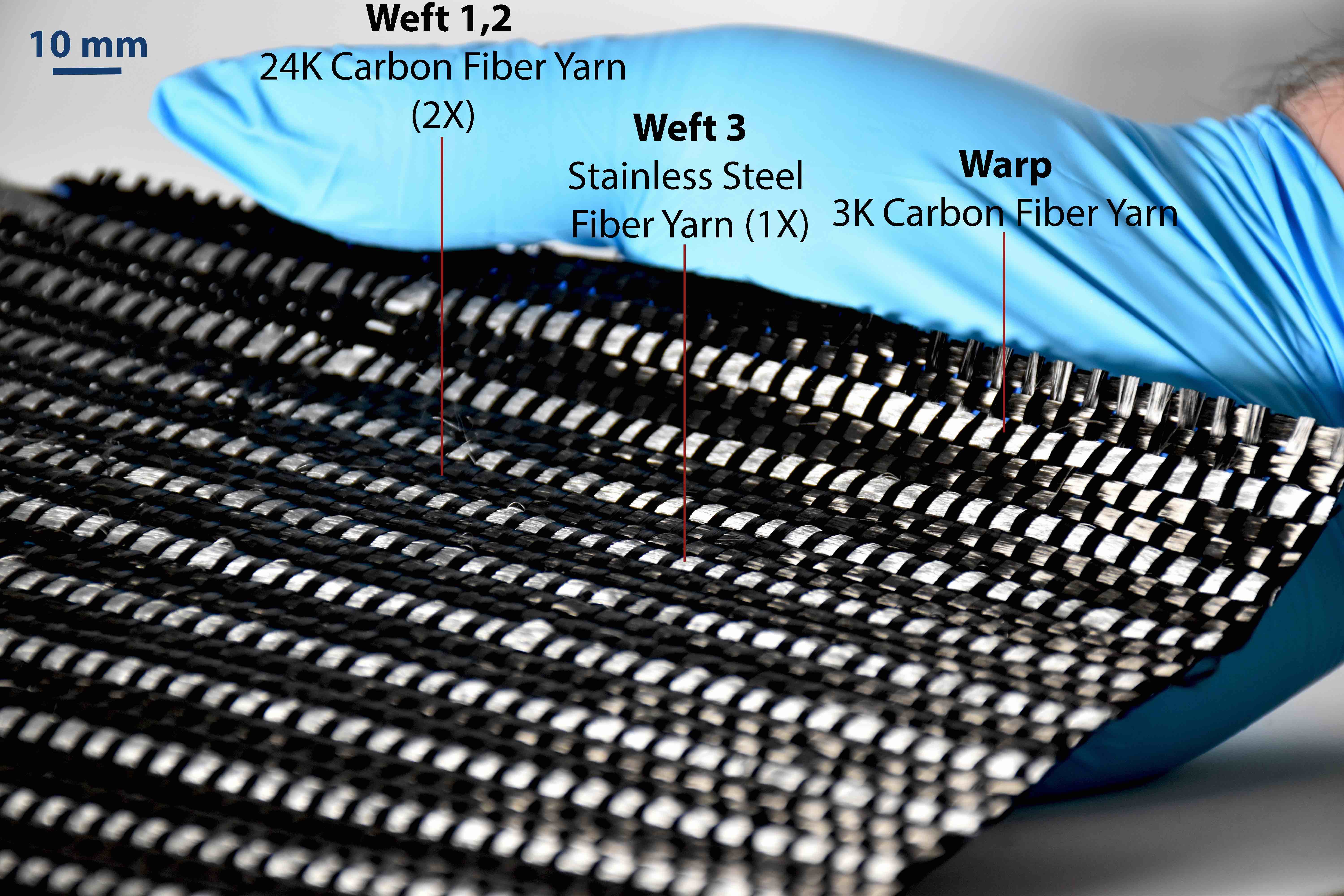}
\label{img:hybWeave2}
}
\caption{Hybrid weaves with 3K carbon fiber in the warp directions and (a) all stainless steel fiber yarn in the weft (C/SS); and (b) a combination of 24K carbon fiber and stainless steel fiber yarns in a 2:1 ratio in the weft (C/C-SS).}
\label{img:hybWeave}
\end{figure}

\paragraph{Weaving and drapability}\label{met:weave}
In this work, we chose the plain weave pattern and incorporated architectures by strategically hybridizing the carbon fiber fabric with stainless steel yarns. A hand loom with 3K carbon fiber in warp, as shown in \textbf{Figure~\ref{img:loom}}, was used to weave the fabric. Three fabric types were woven using different weft yarns, following the designs provided in Section \ref{lamdes}. The two hybridized weaves, with varying ratios of stainless steel yarns, C/SS and C/C-SS, are illustrated in \textbf{Figure \ref{img:hybWeave}}. In this study, we define \textit{hybrid weaves} as fabrics with stainless steel yarns interwoven with carbon fiber yarns.

To evaluate the drapability of the woven fabrics, one end of a single fabric layer (150 mm × 50 mm) was clamped between two rigid blocks with the other end extending freely to form a cantilever \cite{de2010experimental}. The fabric was allowed to bend under its own weight. The experimental setup is illustrated in \textbf{Figure}~\ref{img:resistmeas} (a). The draping angle ($\phi$) was measured as the angle formed between the settled weave and a horizontal reference line in ImageJ \cite{rueden2017imagej2}. The draping angle was also normalized by the fabric's density. 

\paragraph{Composite preparation}\label{met:compman}
To manufacture woven composites, we used West System 105 epoxy resin with West System 209 extra-slow hardener in a 3:1 volume ratio. The composites were fabricated using a hand layup technique to impregnate the fabric, followed by vacuum bagging to apply pressure on the entire assembly. We then let the composite cure at room temperature for 48 hours. 

    \begin{table}[h!]
    \centering
    \renewcommand{\arraystretch}{1.2}
    \caption{Summary of samples for the arc testing in the current study. \\ *0 direction corresponds to the direction of 3K carbon fiber in the weave.} 
    \resizebox{\columnwidth}{!}{
    \begin{tabular}{lp{2cm}p{2cm}p{2cm}p{2cm}p{2cm}}
    \hline
        \textbf{Material} & \textbf{Number of C/C fabric} & \textbf{Number of hybrid fabric}  & \textbf{Thickness $(mm)$} & \textbf{Density $(g/cm^3)$} & \textbf{Nomenclature}   \\ \hline
        
        \textit{Carbon/Carbon (Reference)} & 6 & 0 & 1.60 & 1.27 & C/C   \\ 
        
        \textit{Carbon/stainless steel} & 5 & 1 & 2.203 & 1.64 & C/SS1L   \\ 
    
        \textit{Carbon/stainless steel} & 4 & 2 $[0/90]^*$ & 2.75 & 1.83 & C/SS2L   \\ 
        
        \textit{Carbon/stainless steel (reduced)} & 5 & 1 & 2.267 & 1.41 & C/C-SS1L   \\ 
    
        \textit{Carbon/stainless steel (reduced)} & 4 & 2 $[0/90]^*$ & 2.843 & 1.47 & C/C-SS2L  \\  \hline
    \end{tabular}
    }
    \label{tab:weavesummLS}
    \end{table}

\subsection{Electrical resistivity measurement}\label{met:resist} 
\begin{figure}[h!]
\centering
\subfigure[]{
\includegraphics[width=0.387\textwidth]{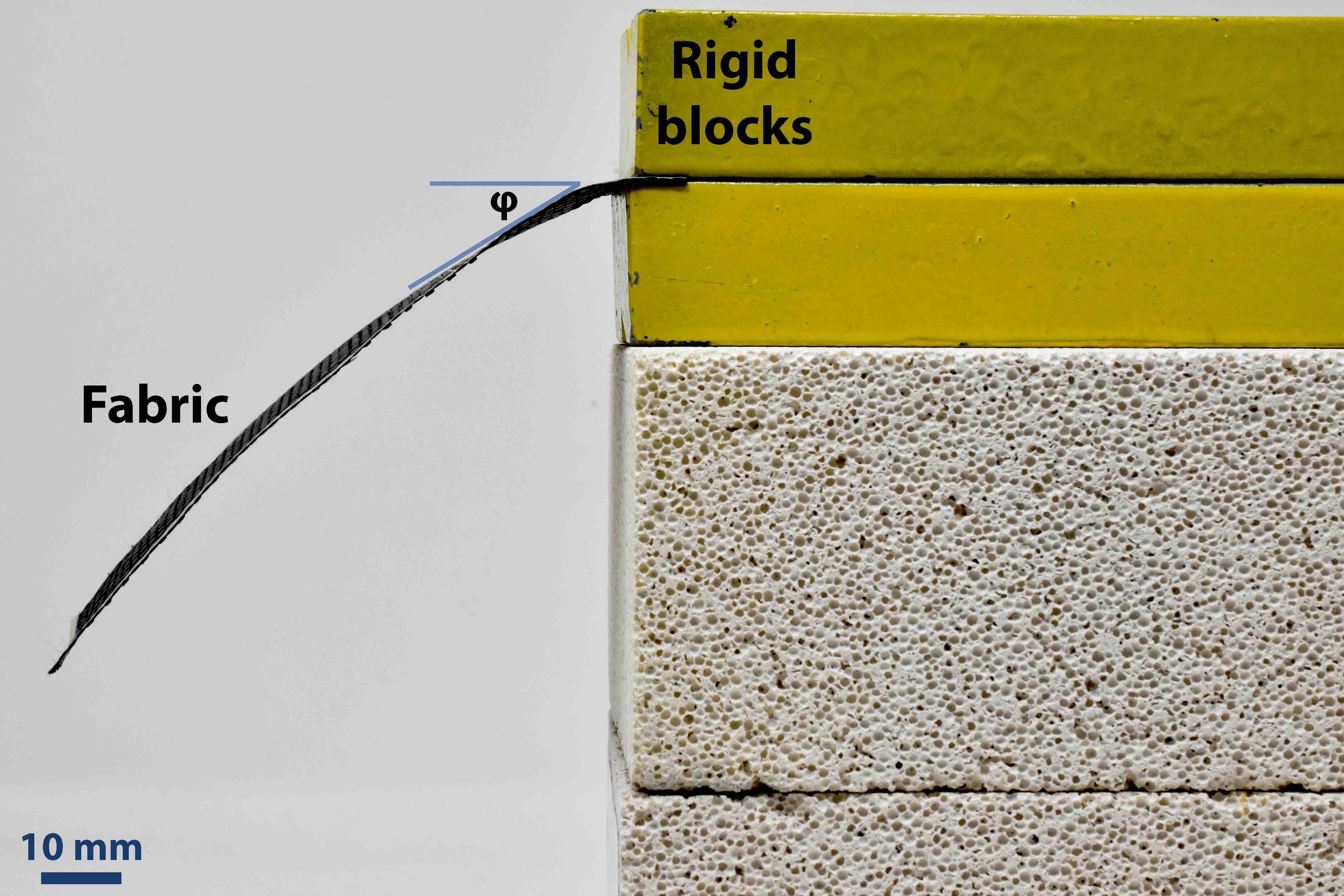}
}
\subfigure[]{
\includegraphics[width=0.46\textwidth]{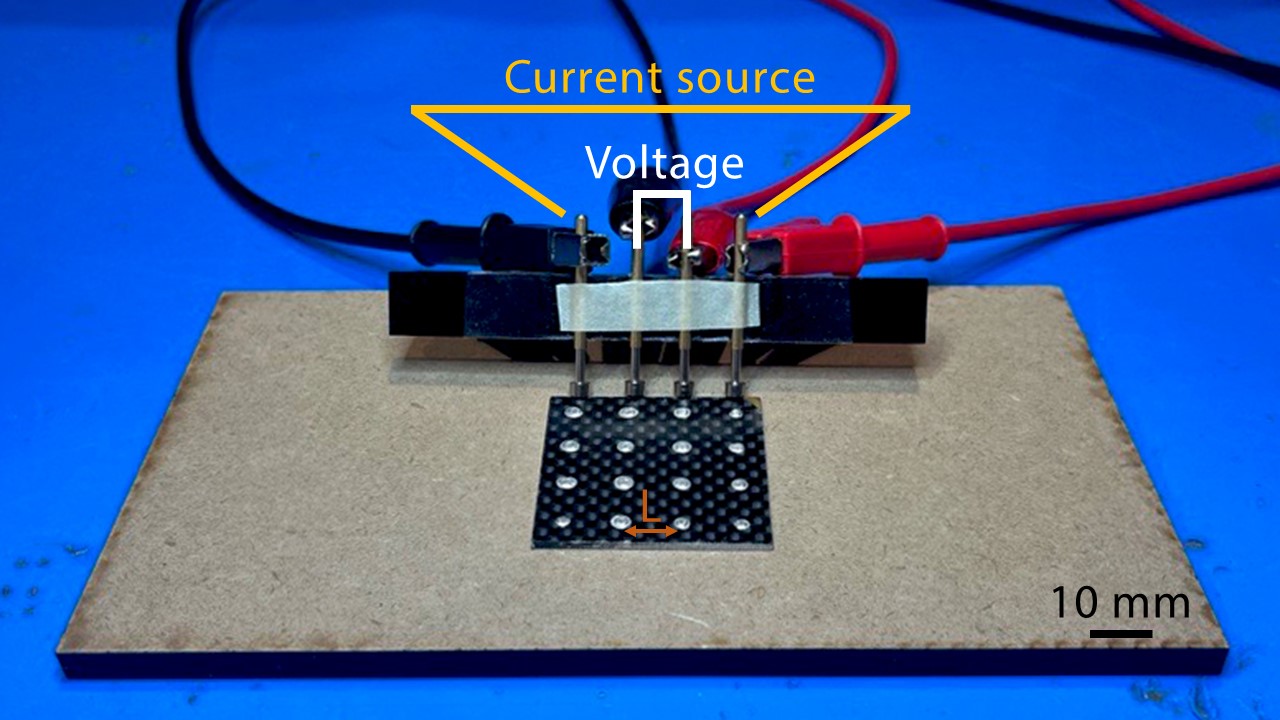}
}
\caption{Test setups for evaluating (a) the draping angle ($\phi$) of a fabric and (b) the electrical resistivity of a composite using the 4-probe method.}
\label{img:resistmeas}
\end{figure}

We measured the in-plane electrical resistivity of composites using the 4-probe method as shown in \textbf{Figure}~\ref{img:resistmeas} (b) \cite{kumar2019interleaved, lampkin2019epoxy}. Three types of composite samples were prepared with (i) 4 layers of C/C fabric (reference), (ii) a single layer of C/SS hybrid fabric, and (iii) a single layer of C/C-SS hybrid fabric. The reference C/C laminate was fabricated with four plies to achieve a similar thickness to the hybrid laminates. Each sample was cut to a size of 40 mm × 40 mm, and the surface was polished and cleaned before applying silver epoxy to minimize contact resistance. A grid of 4x4 measurement points was used for each sample, ensuring consistent placement of 4 probes. A Keysight E3631A power supply was used to apply a direct current (DC) of 100 $\mu$A through the outer two probes, while a Keysight 34450A Digital Multimeter recorded the voltage ($V$) between the inner two probes. The electrical resistance ($R$) was calculated using the $R = V/I$ relationship. We determined the resistivity ($\kappa$) using the formula $\kappa = \frac{RA}{L}$, where $R (\Omega)$ is the measured resistance, $A (cm^2)$ is the cross-sectional area of the sample, and $L$ is the length between the inner probes (~0.75 cm).

\subsection{Arc tests}\label{met:arc} 

\begin{figure}[ht]
\centering
\subfigure[]{
\includegraphics[width=0.35\textwidth]{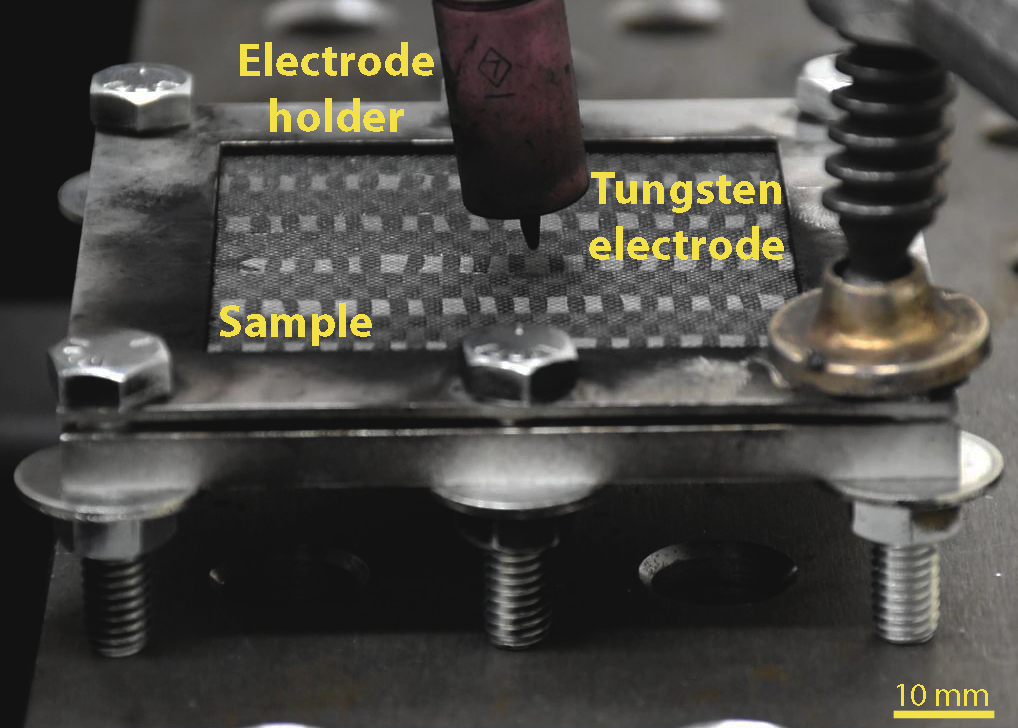}
}
\subfigure[]{
\includegraphics[width=0.35\textwidth]{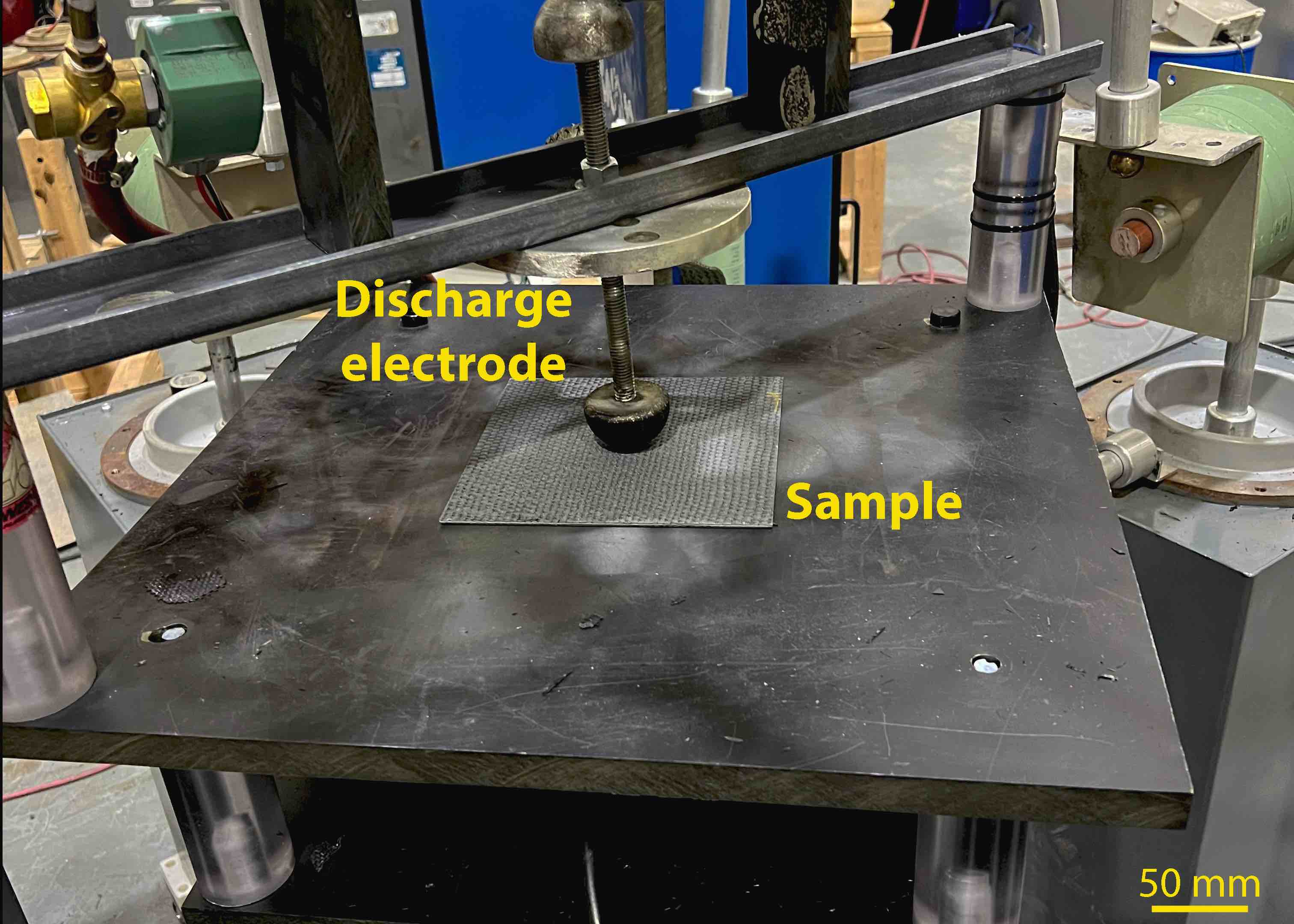}
}
\caption{Experimental setups for (a) quasi-static arc tests using a TIG welder; and (b) simulated lightning strikes conducted at Mississippi State University}
\label{img:arctestsetup}
\end{figure}

We subjected the samples listed in \textbf{Table} \ref{tab:weavesummLS} to two types of electrical arc impacts: (i) low current for a longer duration (we call it quasi-static) and (ii) high current for a shorter duration (simulated lightning strike). The experimental setups for both types of arc tests are presented in \textbf{Figures} \ref{img:arctestsetup}(a,b). These tests aimed to evaluate the performance of the hybrid weave layers under electrical impacts and assess the effects of hybridization. 

\paragraph{Quasi-static arc}\label{met:weld}
We used the Miller Electric Syncrowave 250 DX TIG welder to perform the quasi-static arc tests. The samples, with in-plane dimensions of 76.2mm x 76.2mm (3in x 3in), were tested using a direct current electrode negative (DCEN) with a linearly increasing current ranging from 0 to 30 A over 2 seconds. Argon gas supply was switched on throughout the arc tests for all samples. We mounted the samples in the fixture, ensuring a 5 mm gap between the sample and the welding arc nozzle for all quasi-static tests. A FLIR E76 thermal camera with a maximum temperature range of 150 °C was used to monitor the temperature on the impacted surface of the composites.  

\paragraph{Simulated lightning strike}\label{met:light}
We performed simulated lightning strikes on the composite samples listed in \textbf{Table} \ref{tab:weavesummLS}, each with in-plane dimensions of 177.8mm x 177.8mm (7in x 7in), using the high-current impulse generator at the High-Voltage Laboratory, Mississippi State University \cite{yousefpour2021design}. The generator can produce components A and D of the Society of Automotive Engineering (SAE) standard lightning waveform \cite{feraboli2009damage}. In the current study, we generated component A of the lightning strike waveform using a current discharge magnitude of 40 kA \cite{kumar2018effect, jang2025protection}. A gap of 30 mm was maintained between the top and the bottom electrodes, with 2-3 mm between the composite sample and the top electrode. The final impact current for all samples in \textbf{Table} \ref{tab:weavesummLS} ranged between 39.47 ± 0.75 kA. After 1 minute of impact, we used a Fluke TiS20 infrared camera to capture images of the surface and observe the Joule heating effect on the composite samples caused by the lightning strikes. The grounding during the test was attached only to the back side of all samples, ensuring the current propagated through the material to maximize damage and simulate the worst-case scenario.

\subsection{Four-point bend tests}\label{met:flex}
We conducted 4-point bend (4PB) tests on the composite samples after the quasi-static arc impacts and simulated lightning strikes in compliance with ASTM D6272 \cite{ASTM6272}. 4PB tests were chosen to eliminate shear between the loading rollers and to avoid direct loading in the central damaged area. After each arc test, the impacted samples were sectioned into 1-inch wide strips, using the impact location as the origin. For quasi-static arc impact samples, a span-to-depth ratio of 16:1 was chosen, whereas a higher ratio of 64:1 (larger composite sample size) was selected for the samples subjected to simulated lightning strikes. The samples were carefully placed on the fixture with the impact side as the bottom surface. All 4PB tests were conducted on an MTS Criterion 43 test frame, equipped with a 50 kN load cell, at a displacement rate of 1 mm/min. 

\section{Results and Discussion}\label{resdis}
In this section, we examine the effect of hybridization on the drapability, electrical resistivity, and current dissipation in composites, comparing those with and without hybrid weaves. First, we discuss the influence of hybrid weaves on the bulk resistivity of these composites. Then, we analyze the results from the quasi-static arc tests, followed by assessments from the simulated lightning strikes. We inspected each sample to evaluate the extent of damage, followed by 4-point bend tests to assess the residual mechanical properties after the electric arc tests.

\subsection{Fabric Drapability}\label{res:draping}
\begin{figure}[ht]
\centering
\subfigure[]{
\includegraphics[width=0.31\textwidth]{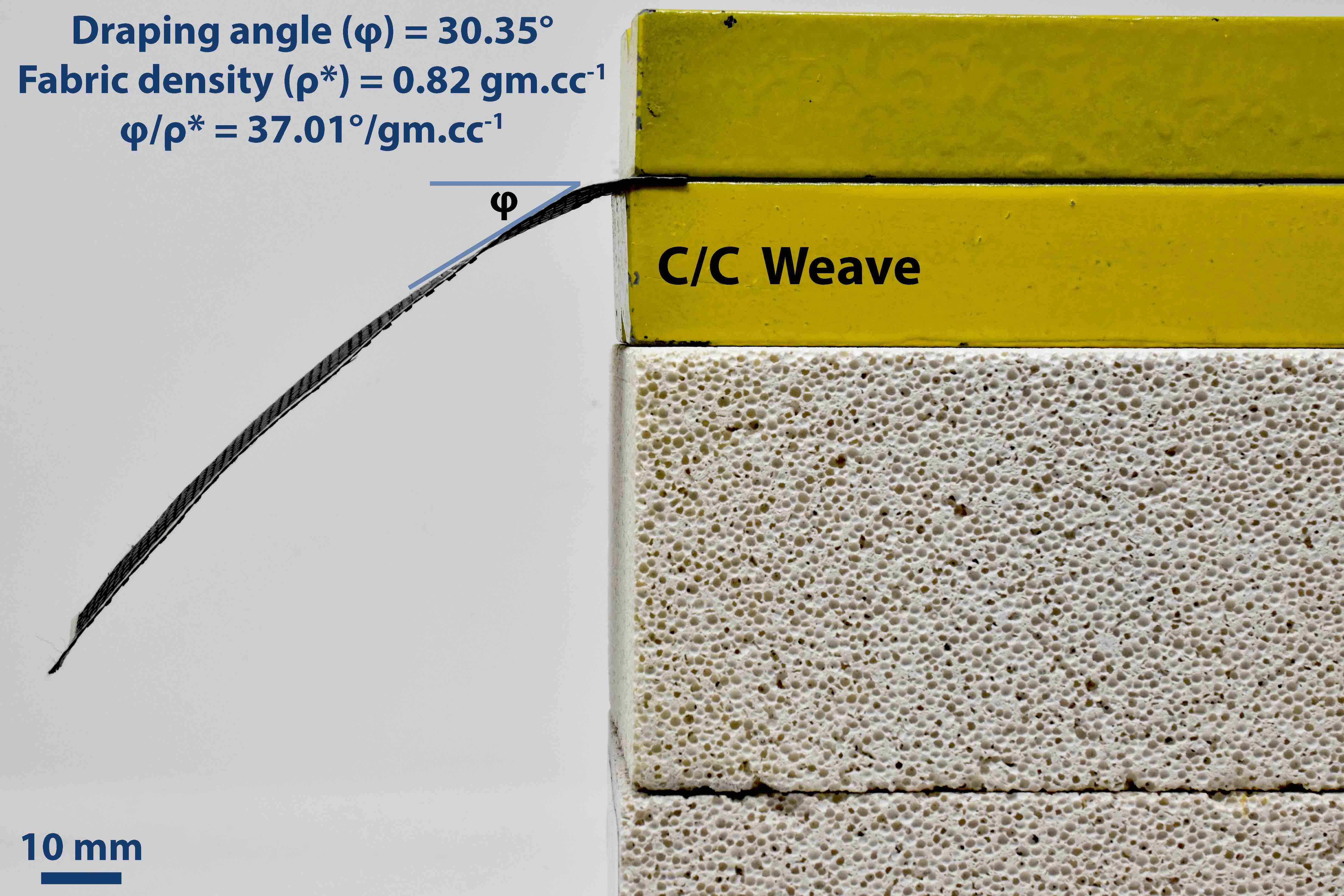}
}
\subfigure[]{
\includegraphics[width=0.31\textwidth]{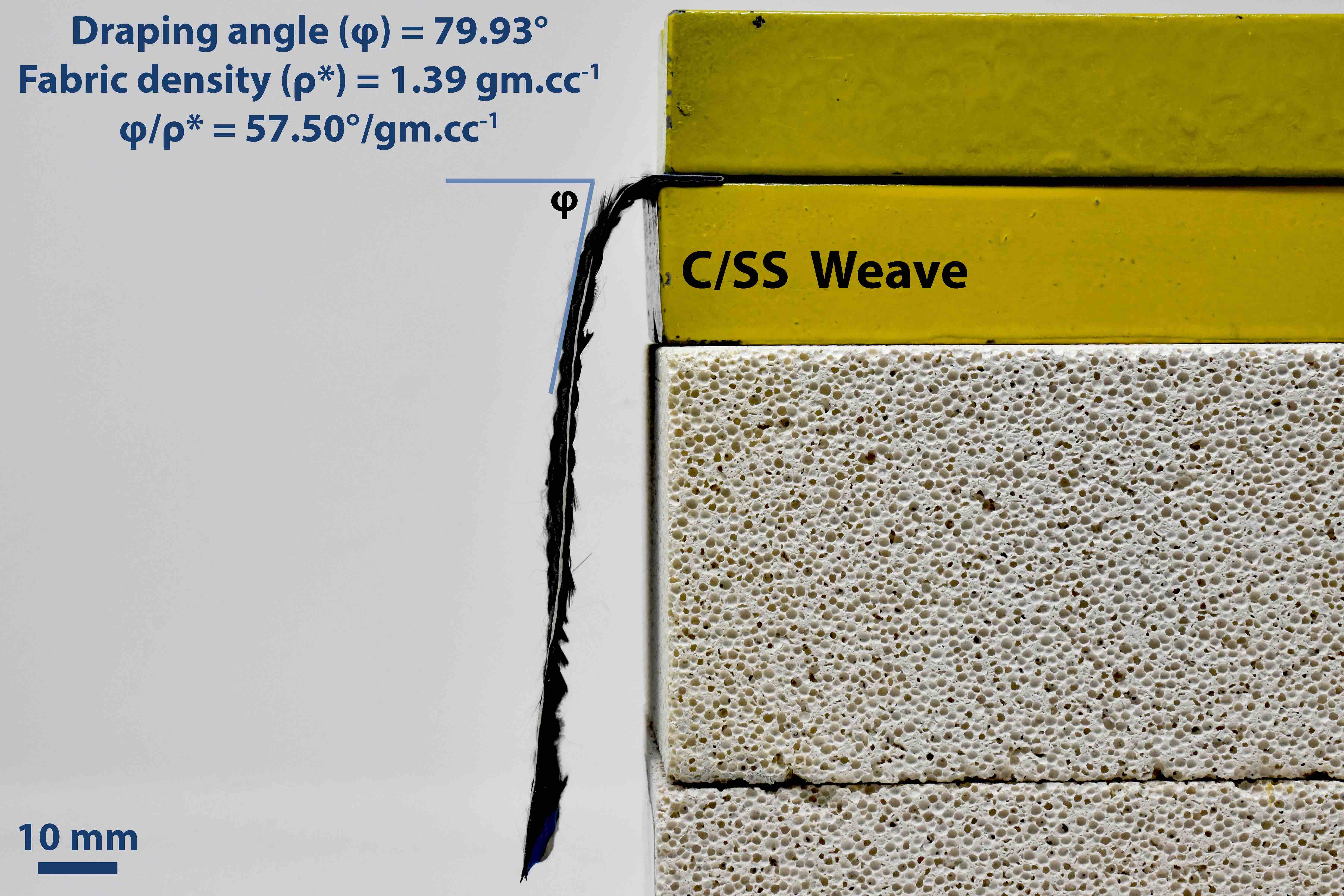}
}
\subfigure[]{
\includegraphics[width=0.31\textwidth]{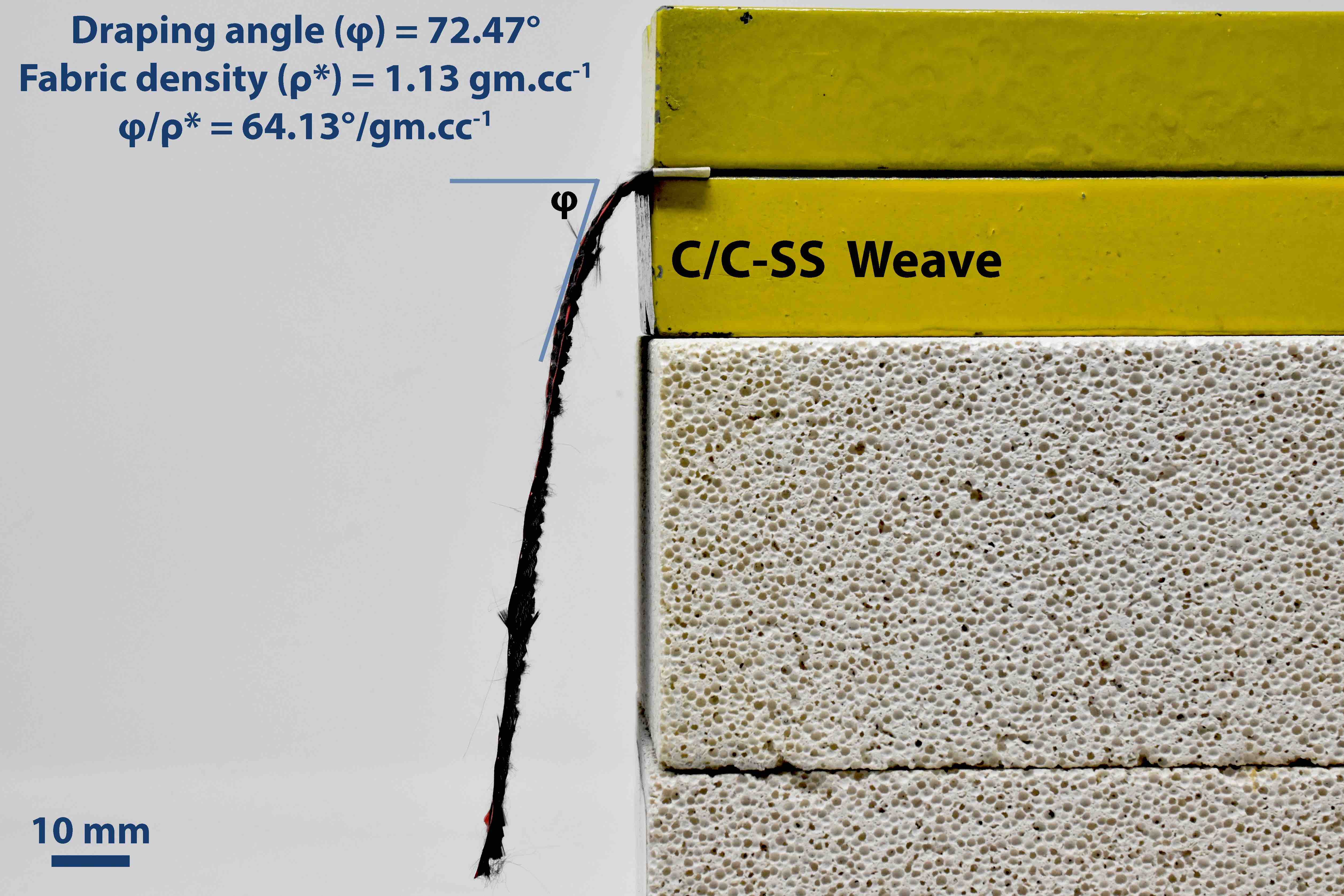}
}
\caption{Drapability of (a) C/C; (b) C/SS; and (c) C/C-SS weaves. The draping angle ($\phi$), defined as the angle between the relaxed fabric and the horizontal reference line, was normalized by the fabric density ($\rho^*$) to determine the normalized draping angle ($\phi/\rho^*$).}
\label{img:Draping}
\end{figure}

We measured the drapability of all fabrics, and as shown in \textbf{Figure}~\ref{img:Draping}, the C/SS weave exhibited the highest draping angle of 79.93$^{\circ}$, followed by the C/C-SS weave at 72.47$^{\circ}$, and the C/C weave at 37.01$^\circ$. After normalizing the draping angle by fabric density, both hybrid weaves continued to demonstrate superior drapability compared to the C/C weave. However, the C/C-SS weave showed the highest normalized draping angle of 64.13$^\circ / gm.cc^{-1}$, while the C/SS weave exhibited 57.50$^\circ / gm.cc^{-1}$, and the C/C weave remained at 37.01$^\circ / gm.cc^{-1}$. This reversal between the two hybrid weaves is attributed to the lower density of the C/C-SS weave compared to the C/SS weave. Although we could not find explicit drapability data in the literature for conventional metallic LSPs, the results presented demonstrate that both hybrid weaves exhibited higher drapability than the conventional C/C plain weave. This superior drapability indicates their suitability for applications involving complex geometries.

\subsection{Electrical resistivity}\label{res:resist}
We measured the in-plane electrical resistivity of the reference C/C composites and the hybrid weave composites to evaluate the influence of incorporating stainless steel yarns with carbon fiber yarns. The composites with a single layer of C/SS and C/C-SS fabrics exhibited resistivities of 0.15 $\Omega.cm$ and 0.18 $\Omega.cm$, respectively. In comparison, the reference C/C composite showed a significantly higher resistivity of 0.56 $\Omega.cm$, which is 3.73 and 3.11 times greater than that of the C/SS and C/C-SS hybrid composites, respectively. This substantial reduction in resistivity with the inclusion of metallic yarns is expected to decrease Joule heating at the arc impact location and enhance current dissipation. Hence, resulting in less damage to a composite structure. The damage tolerance and related assessments are discussed next in detail in \textbf{Section} \ref{res:visdamage}. Among the two hybrid weaves, the C/SS composite demonstrated a lower resistivity compared to the C/C-SS composite, attributed to the presence of all stainless steel yarns in the weft direction. A summary of the average electrical resistivity values for all composites is presented in \textbf{Table} \ref{tab:resistsumm}. 

    \begin{table}[ht]
    \centering
    \renewcommand{\arraystretch}{1}
    \caption{Summary of in-plane electrical resistivities for the reference C/C composite and hybrid weave composite laminates.} 
    \resizebox{\columnwidth}{!}{
    \begin{tabular}{lccc}
    \hline
        \textbf{Material} & \textbf{Fabric} & \textbf{Composite thickness $(mm)$} & \textbf{Electrical resistivity ($\Omega.cm$)}    \\ \hline
        
        \textit{Carbon/Carbon (Reference)} & C/C & 1.40 & 0.56 $\pm$ 0.08 \\ 
        
        \textit{Carbon/stainless steel} & C/SS & 1.16 & 0.15 $\pm$ 0.02 \\ 
    
        \textit{Carbon/stainless steel (reduced)} & C/C-SS & 1.19 & 0.18 $\pm$ 0.05 \\  \hline
    \end{tabular}
    }
    \label{tab:resistsumm}
    \end{table}

\subsection{Visual damage analysis}\label{res:visdamage}
We visually inspected the damage caused by both types of electric arc tests using the digital pictures captured under UV lighting. The UV lighting highlighted the epoxy damage on both the impacted and the back faces of the composites. First, we will compare the performance of composites listed in \textbf{Table} \ref{tab:weavesummLS} subjected to quasi-static arc impact, followed by simulated lightning strike tests. 

\paragraph{Quasi-static arc:}\label{res:TIGimp}

\begin{figure}[h!]
\resizebox{1\textwidth}{!}
{   \begin{tabular}{ccccc}
    \toprule
        \textbf{Material}  & \multicolumn{2}{c|}{\textbf{Quasi-static Arc}} & \multicolumn{2}{|c}{\textbf{Lightning Strike}} \\ 
          & \textbf{Impact side} & \textbf{Back side} & \textbf{Impact side} & \textbf{Back side} \\  \midrule
        \textit{C/C} & 
        \includegraphics[width=0.150\textwidth]{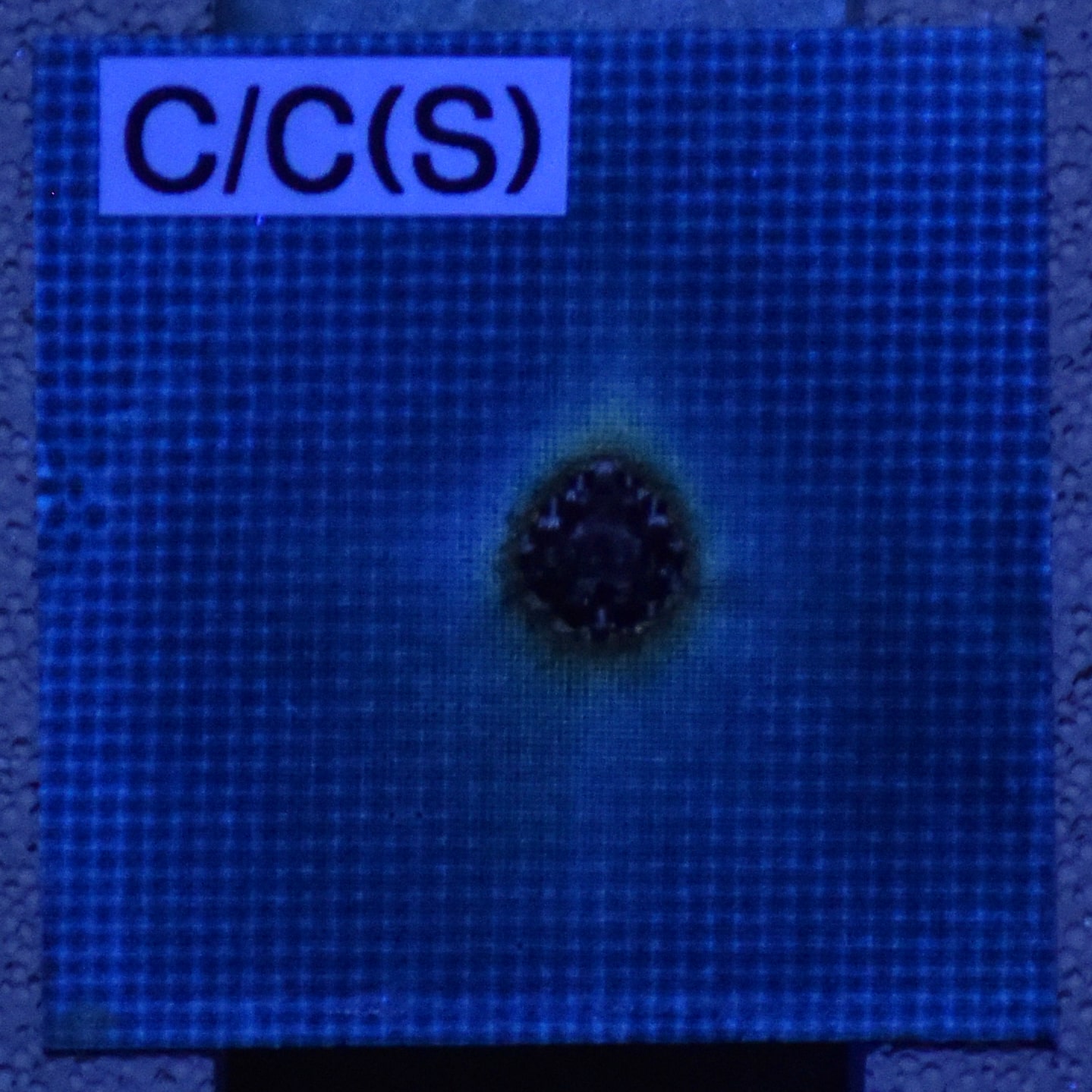} (a1) &  
        \includegraphics[width=0.150\textwidth]{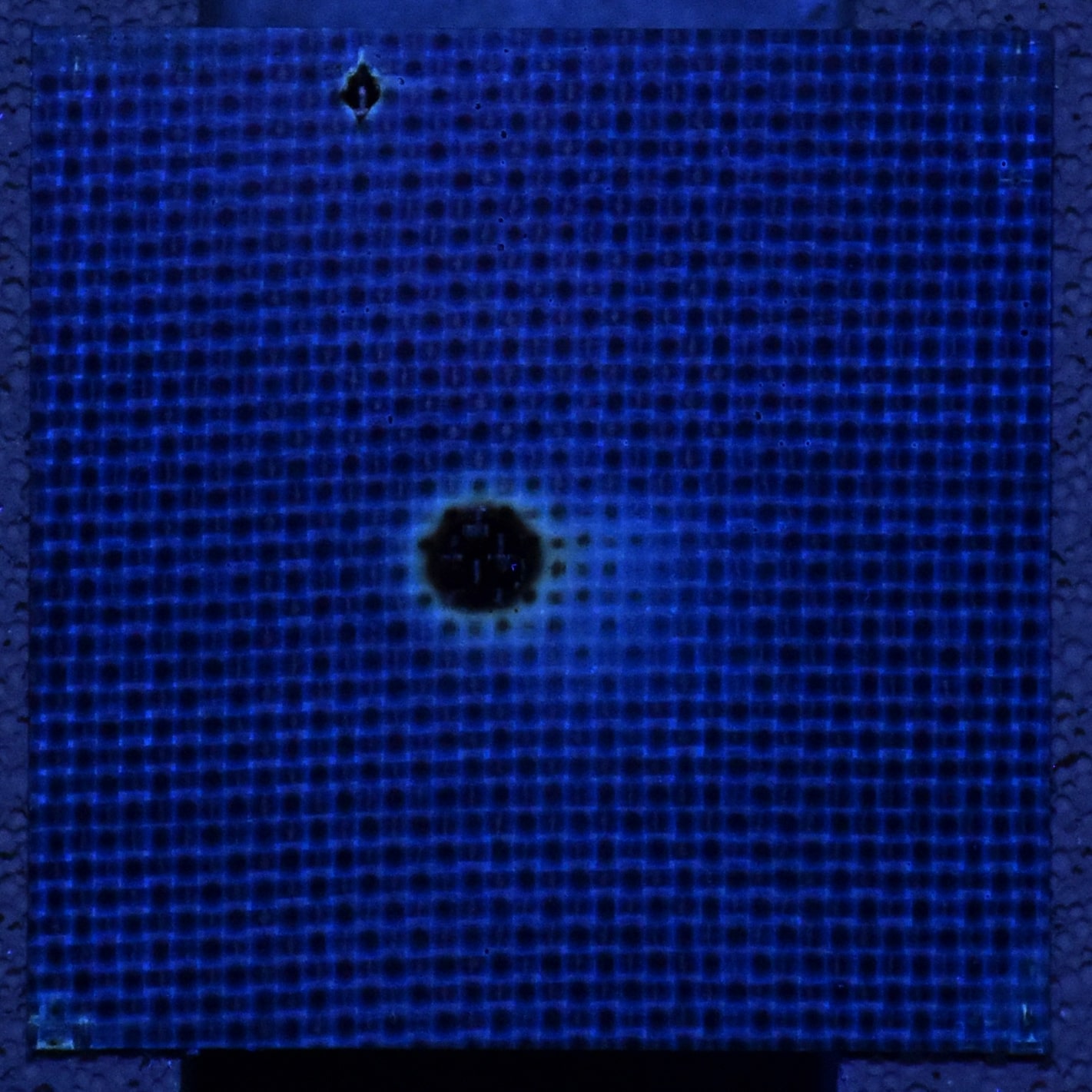} (b1) & 
        \includegraphics[width=0.150\textwidth]{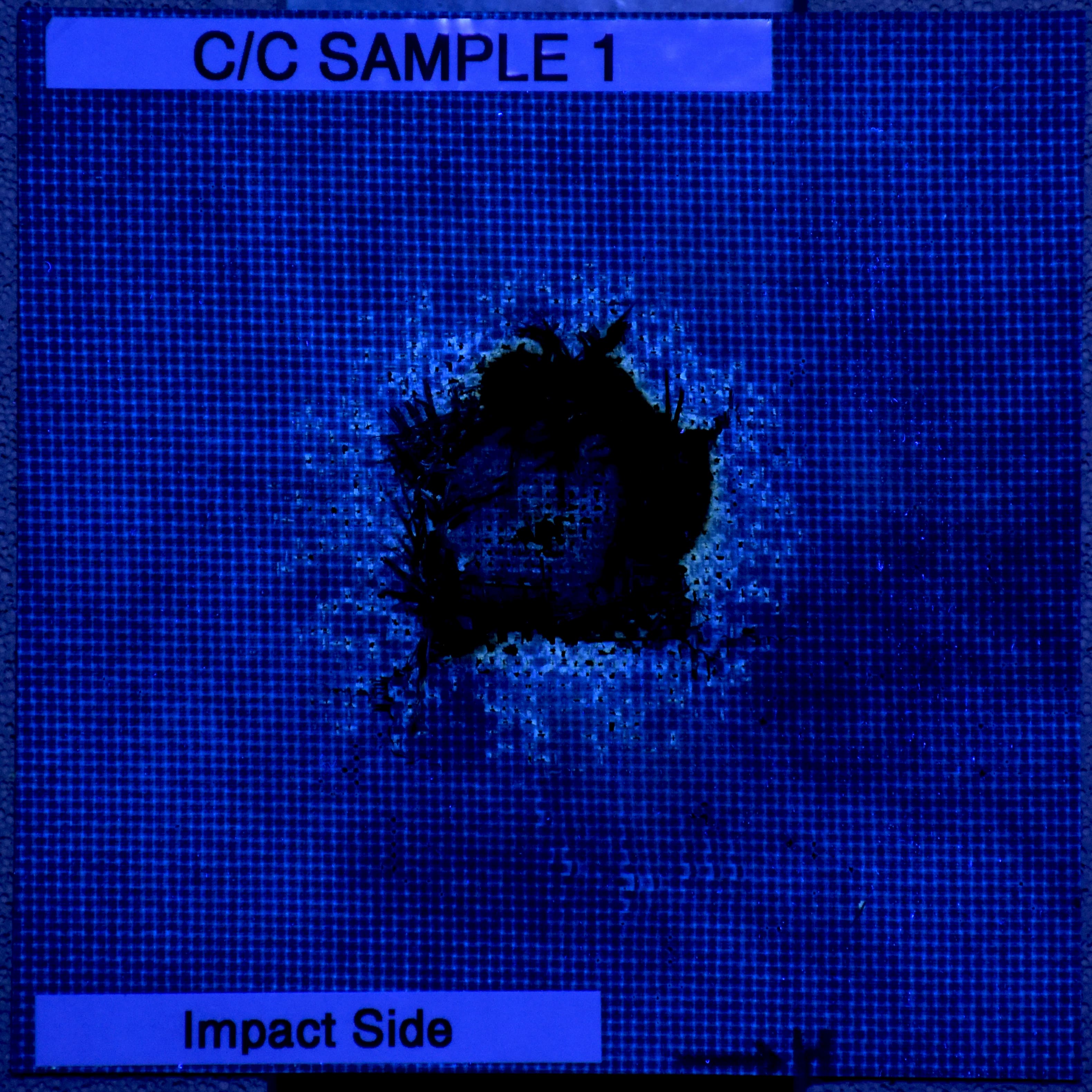} (a2) & 
        \includegraphics[width=0.150\textwidth]{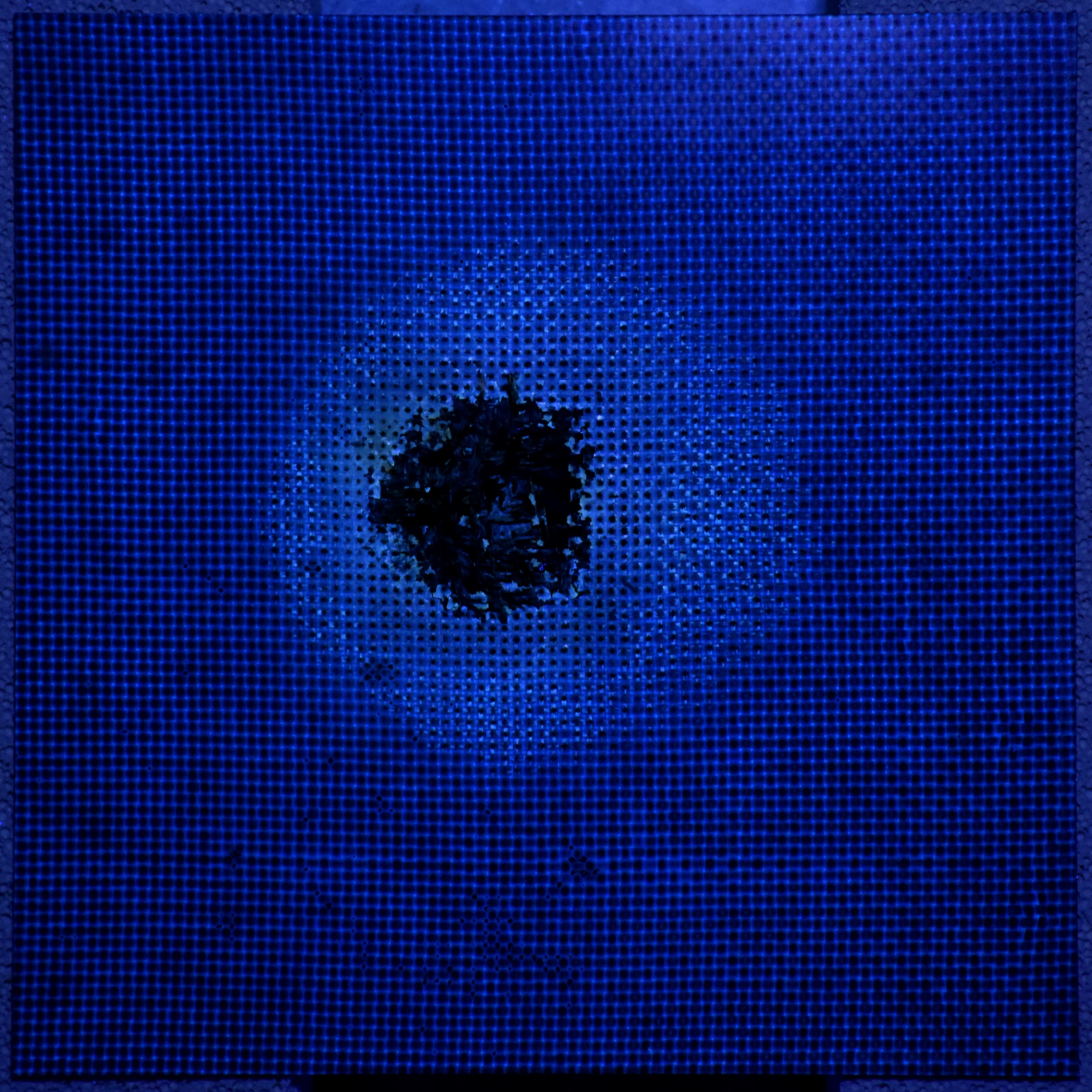} (b2) \\ 
        \textit{C/SS1L} & 
        \includegraphics[width=0.150\textwidth]{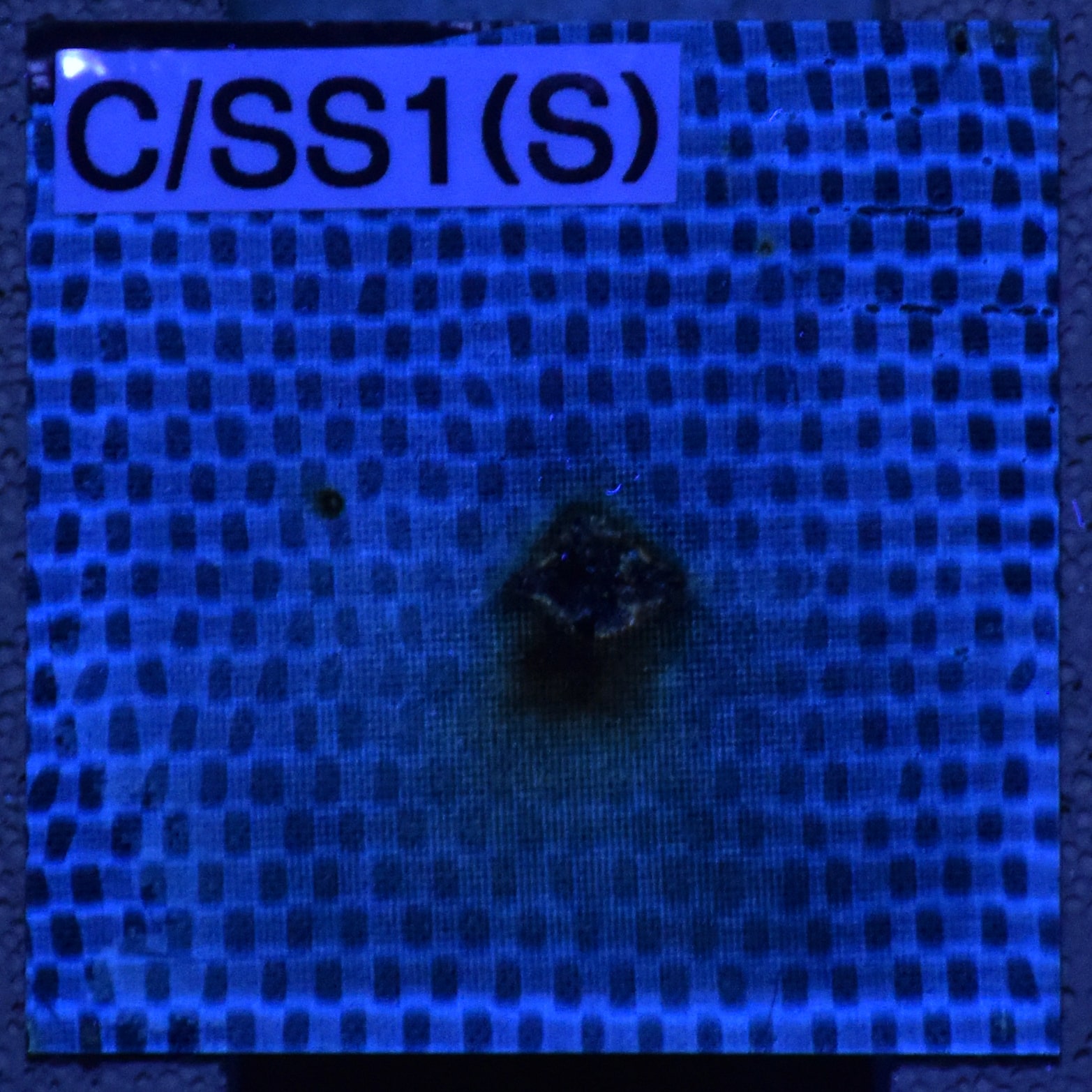} (c1) &  
        \includegraphics[width=0.150\textwidth]{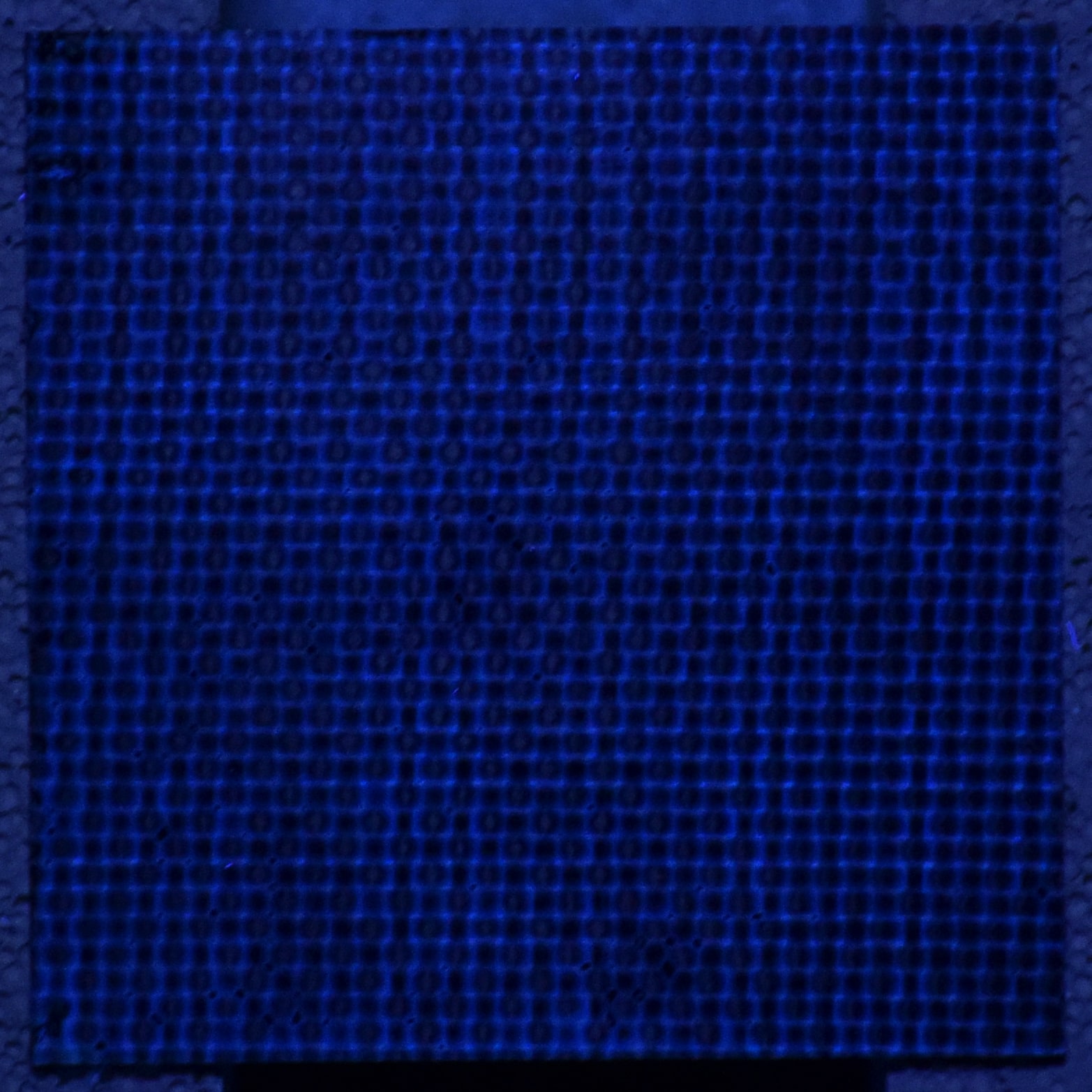} (d1) & 
        \includegraphics[width=0.150\textwidth]{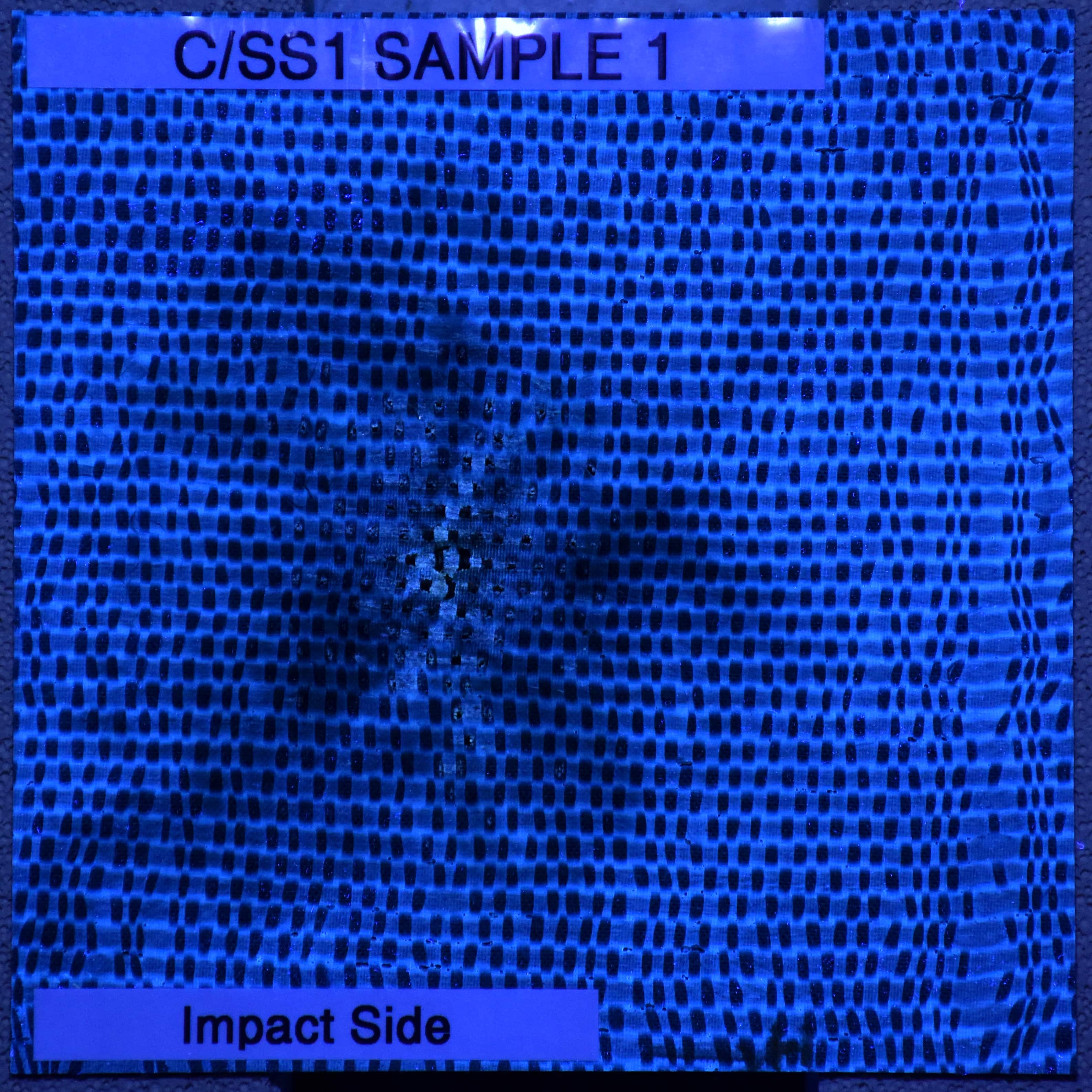} (c2) & 
        \includegraphics[width=0.150\textwidth]{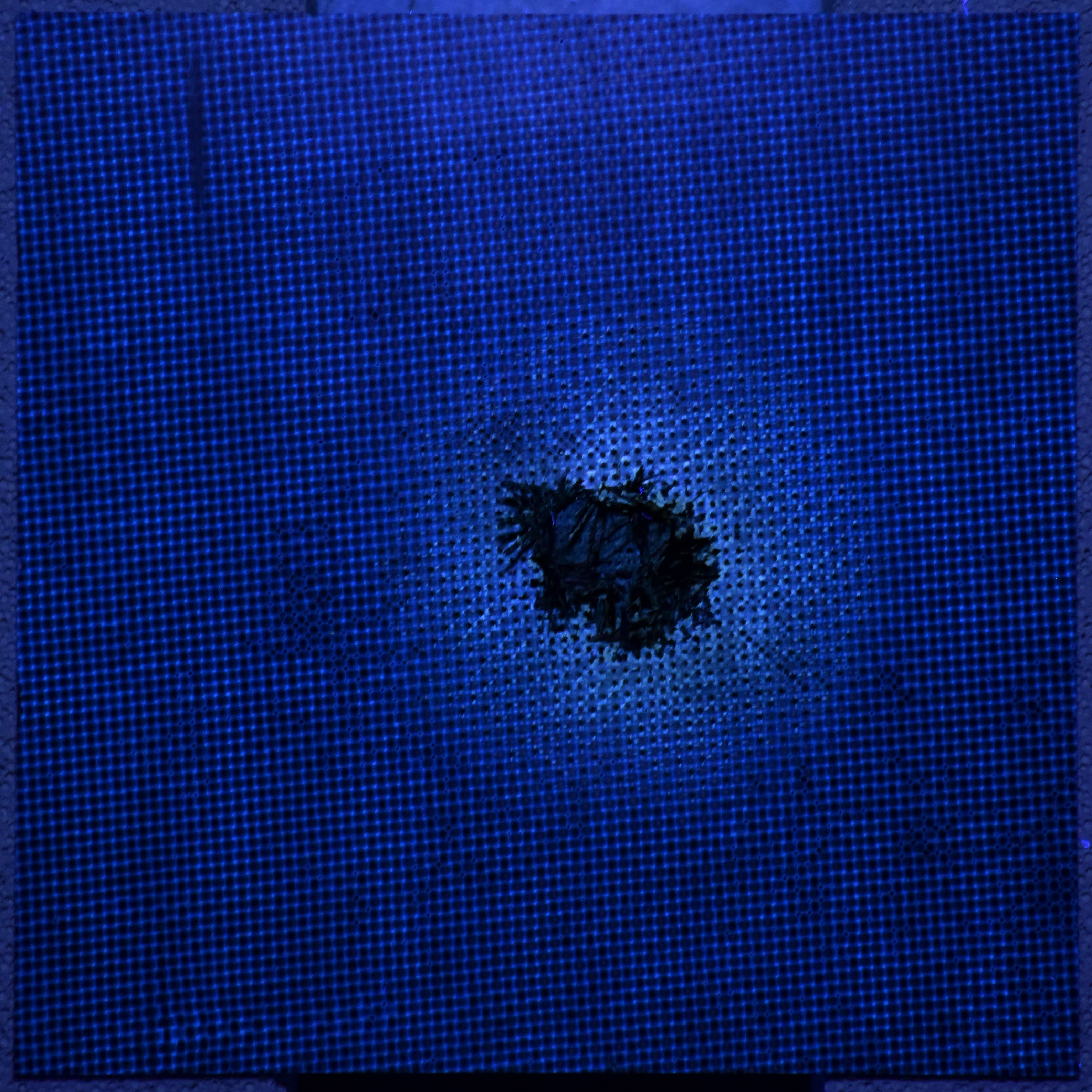} (d2)  \\   
        \textit{C/SS2L} & 
        \includegraphics[width=0.150\textwidth]{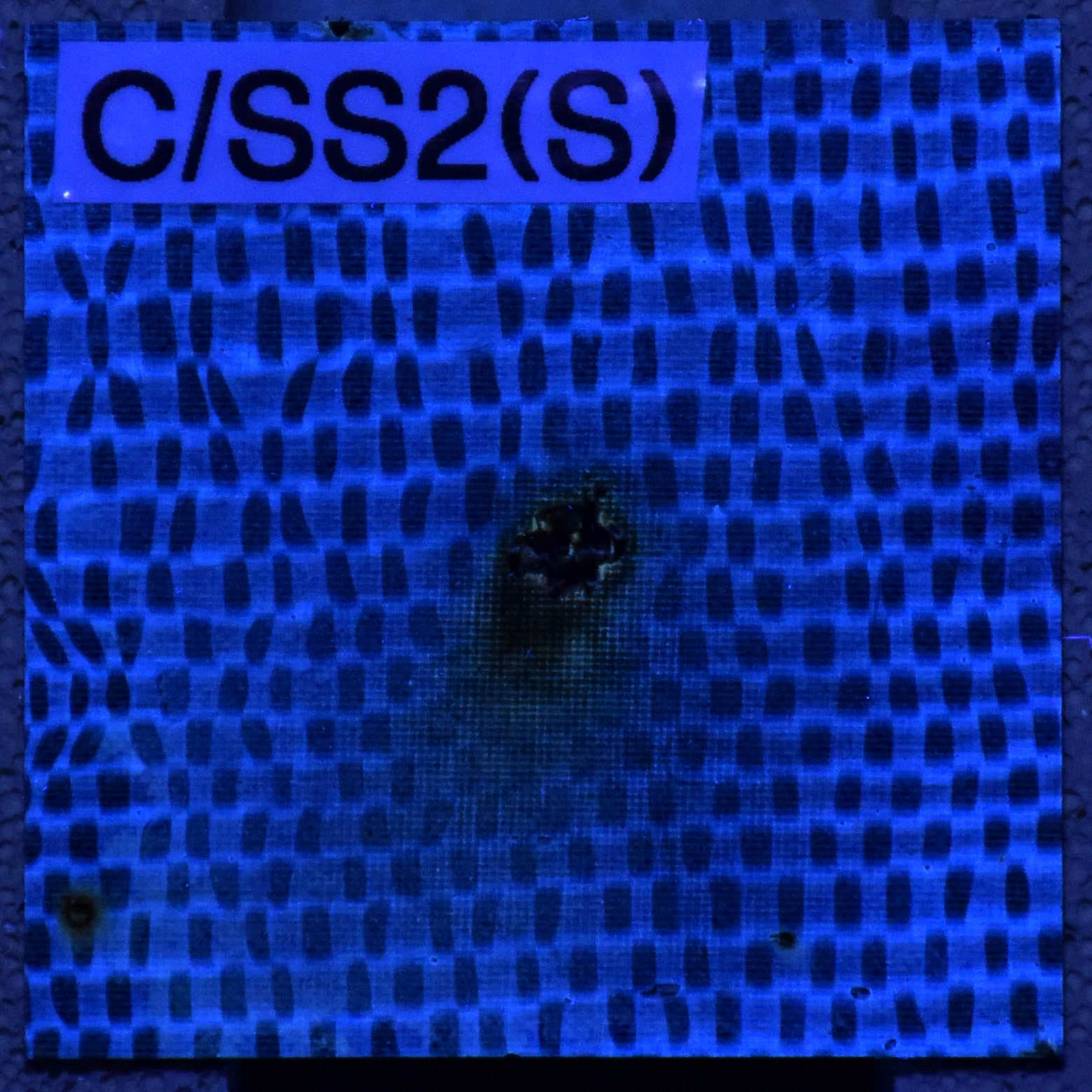} (e1) &  
        \includegraphics[width=0.150\textwidth]{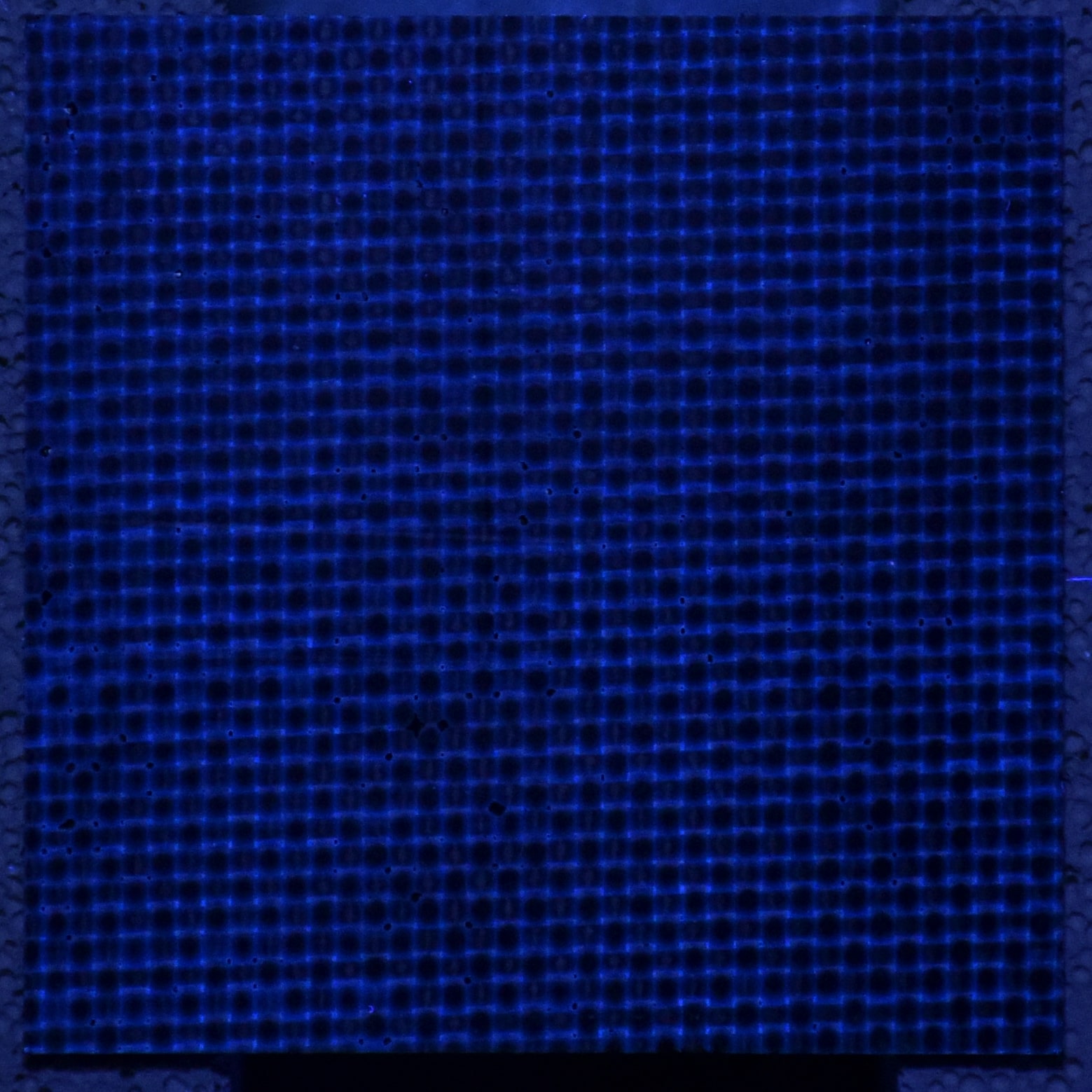} (f1) & 
        \includegraphics[width=0.150\textwidth]{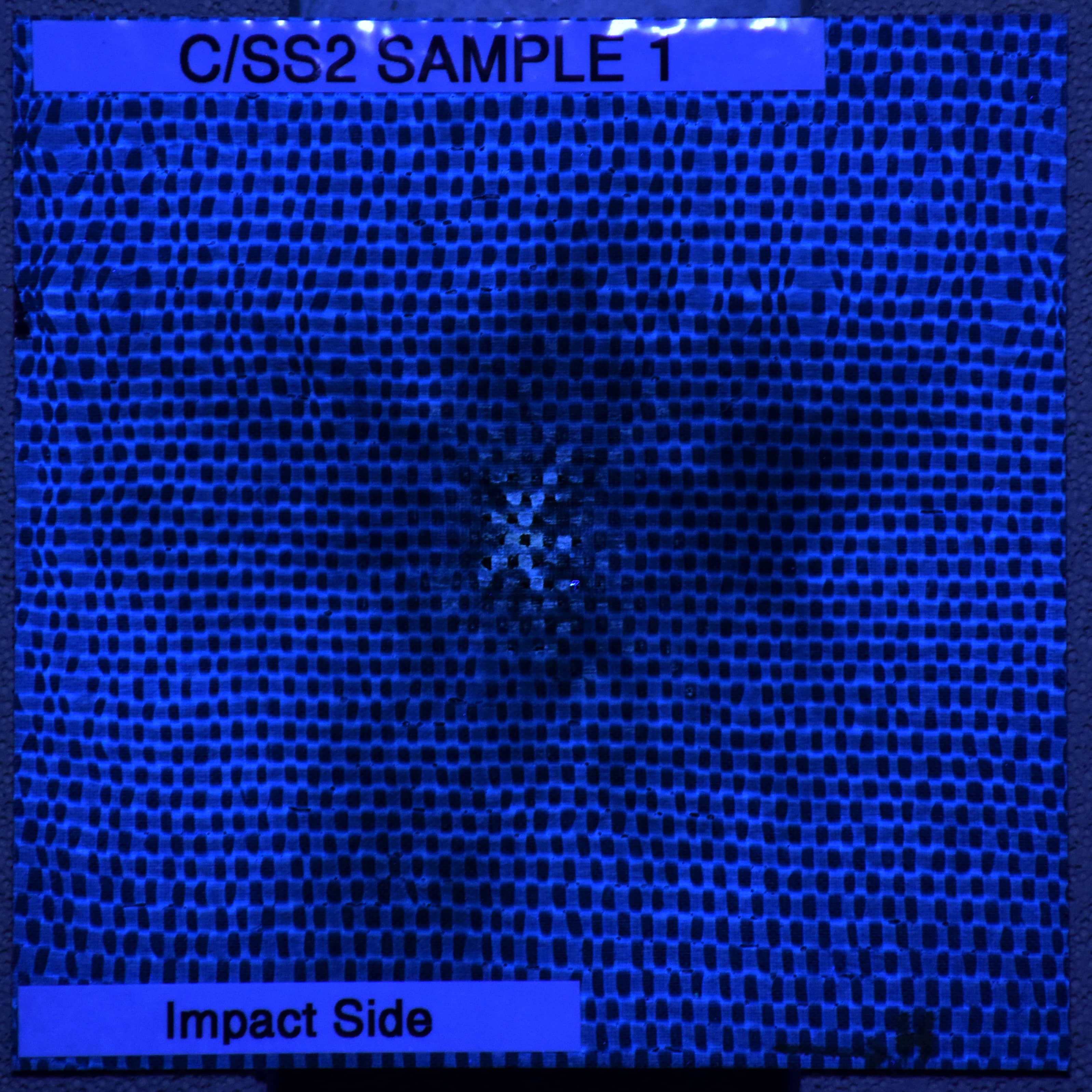} (e2) & 
        \includegraphics[width=0.150\textwidth]{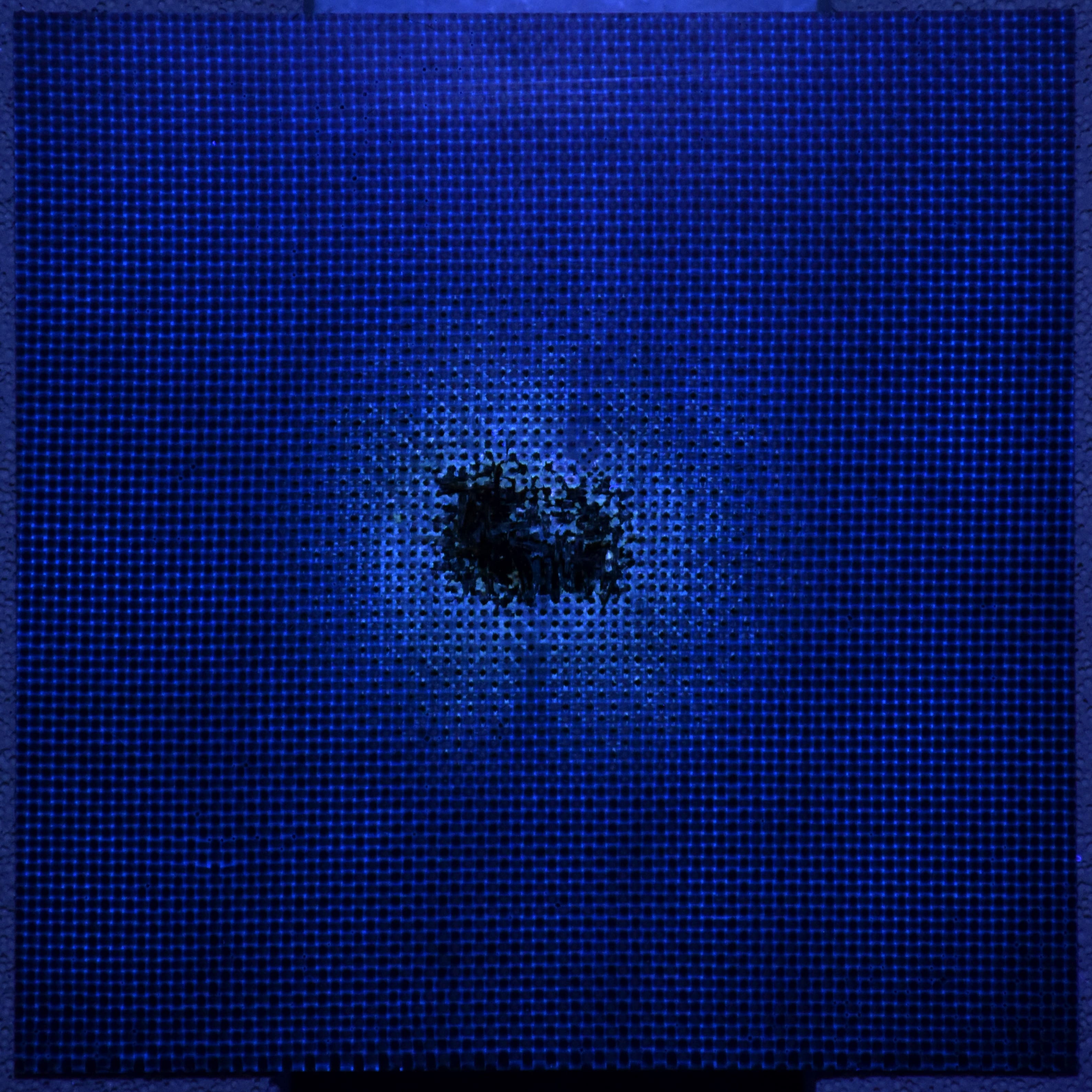} (f2) \\ 
        \textit{C/C-SS1L} & 
        \includegraphics[width=0.150\textwidth]{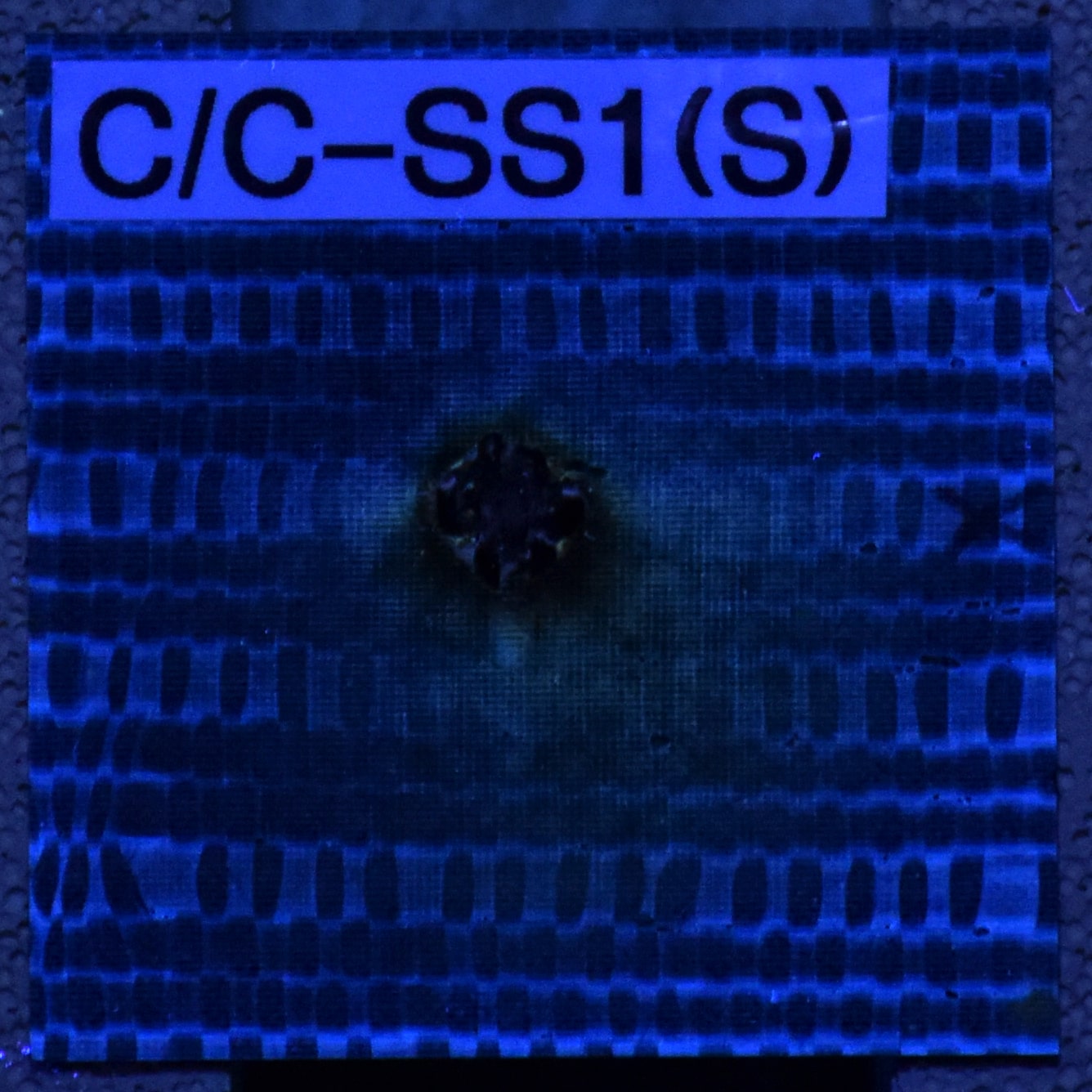} (g1) &  
        \includegraphics[width=0.150\textwidth]{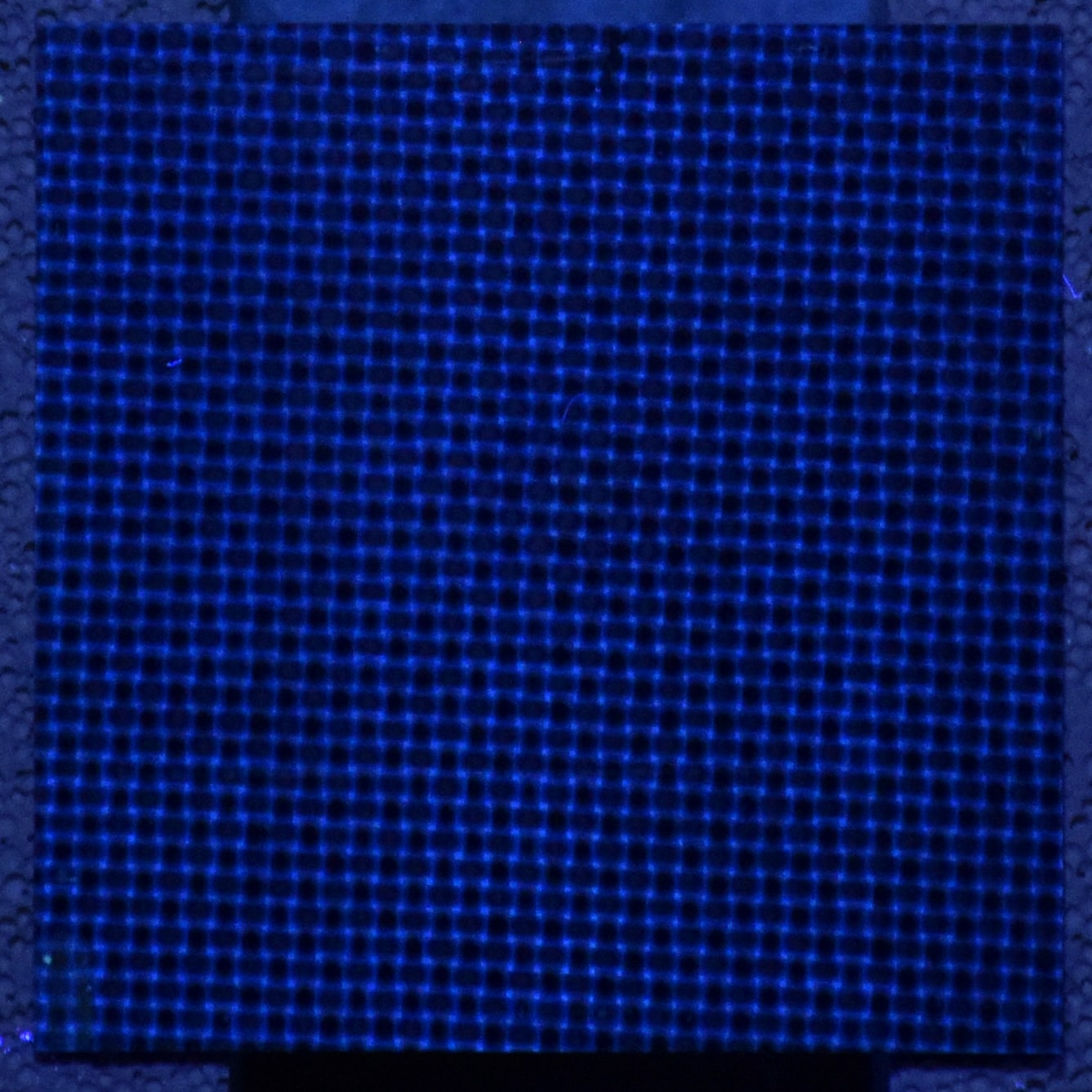} (h1) & 
        \includegraphics[width=0.150\textwidth]{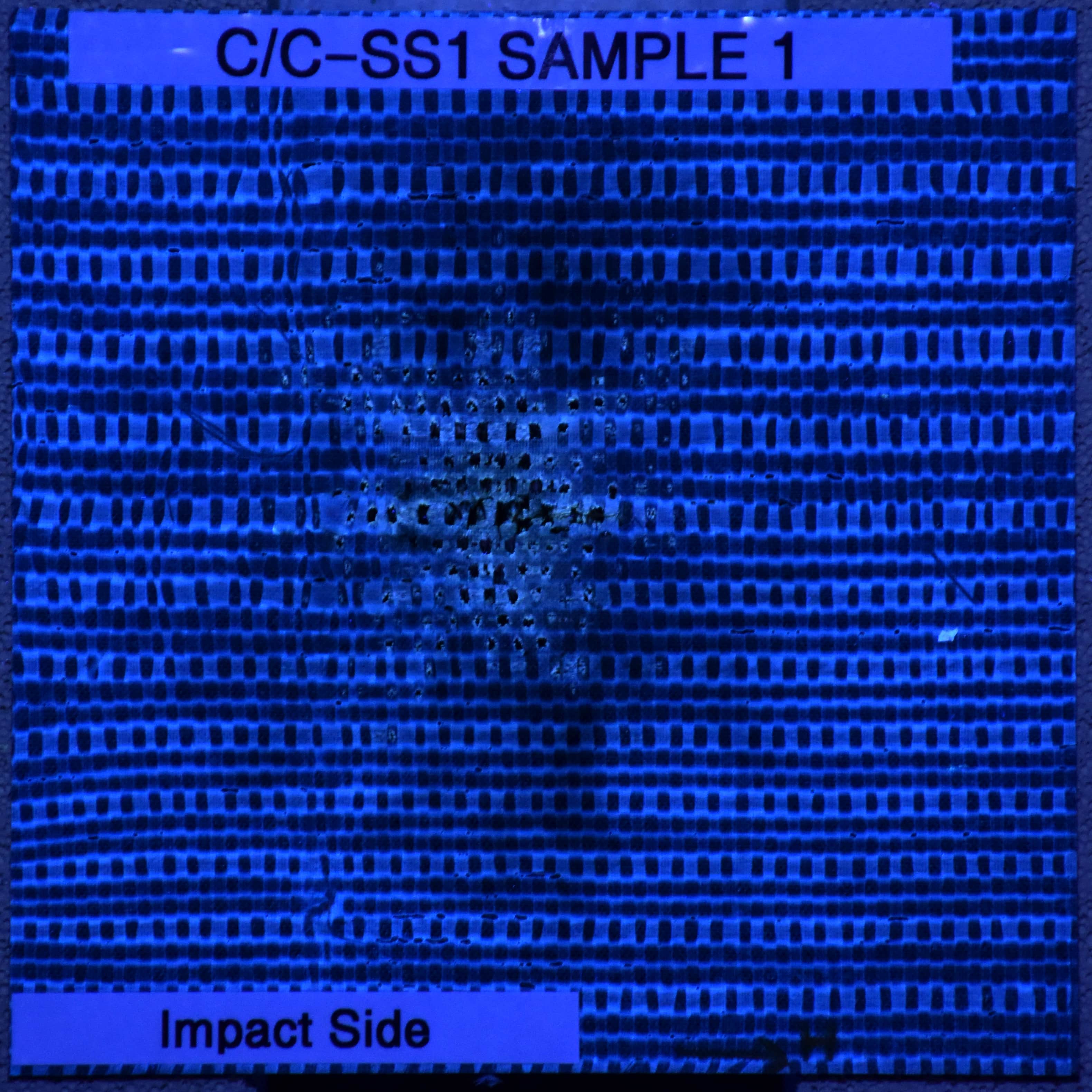} (g2) &  
        \includegraphics[width=0.150\textwidth]{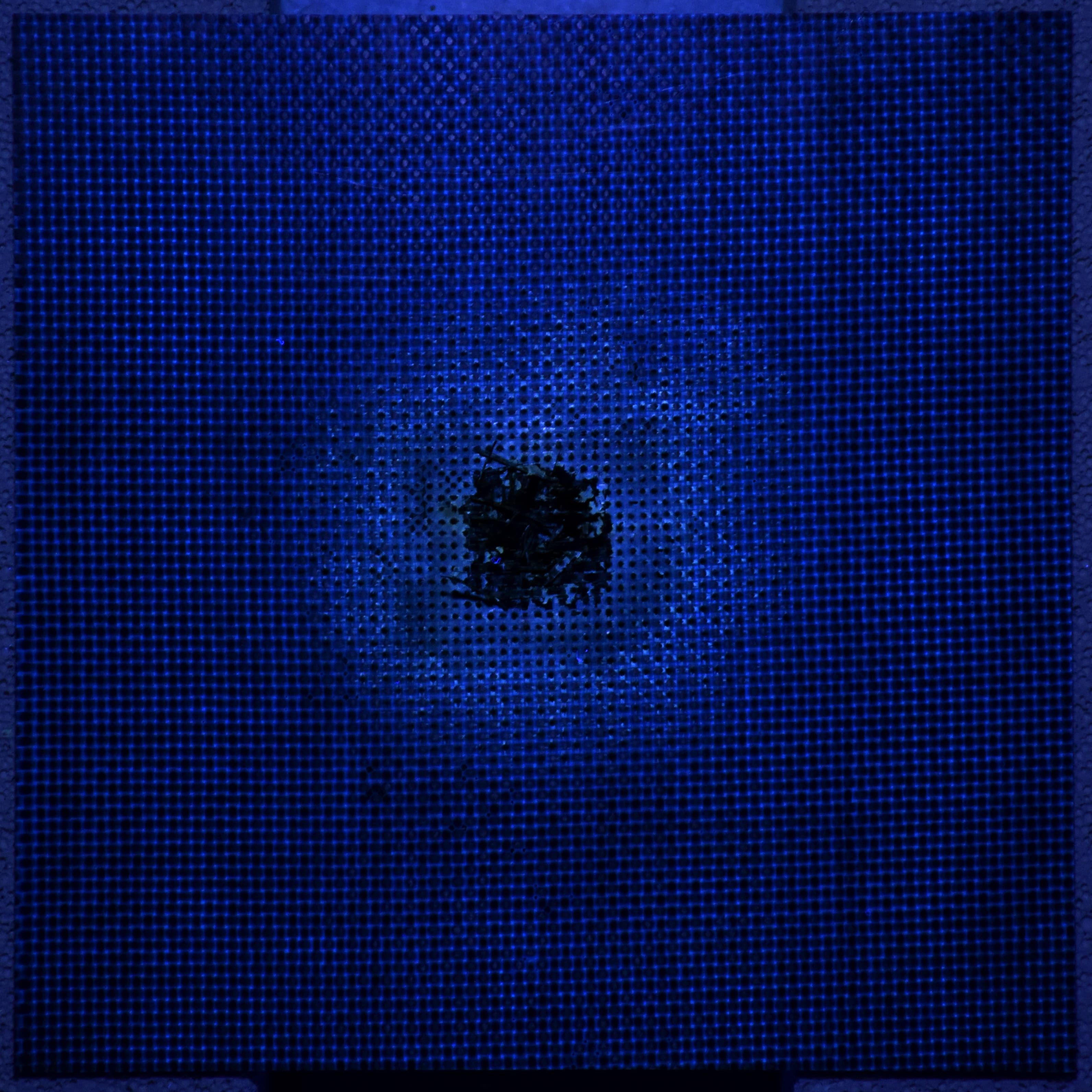} (h2)  \\
        \textit{C/C-SS2L} & 
        \includegraphics[width=0.150\textwidth]{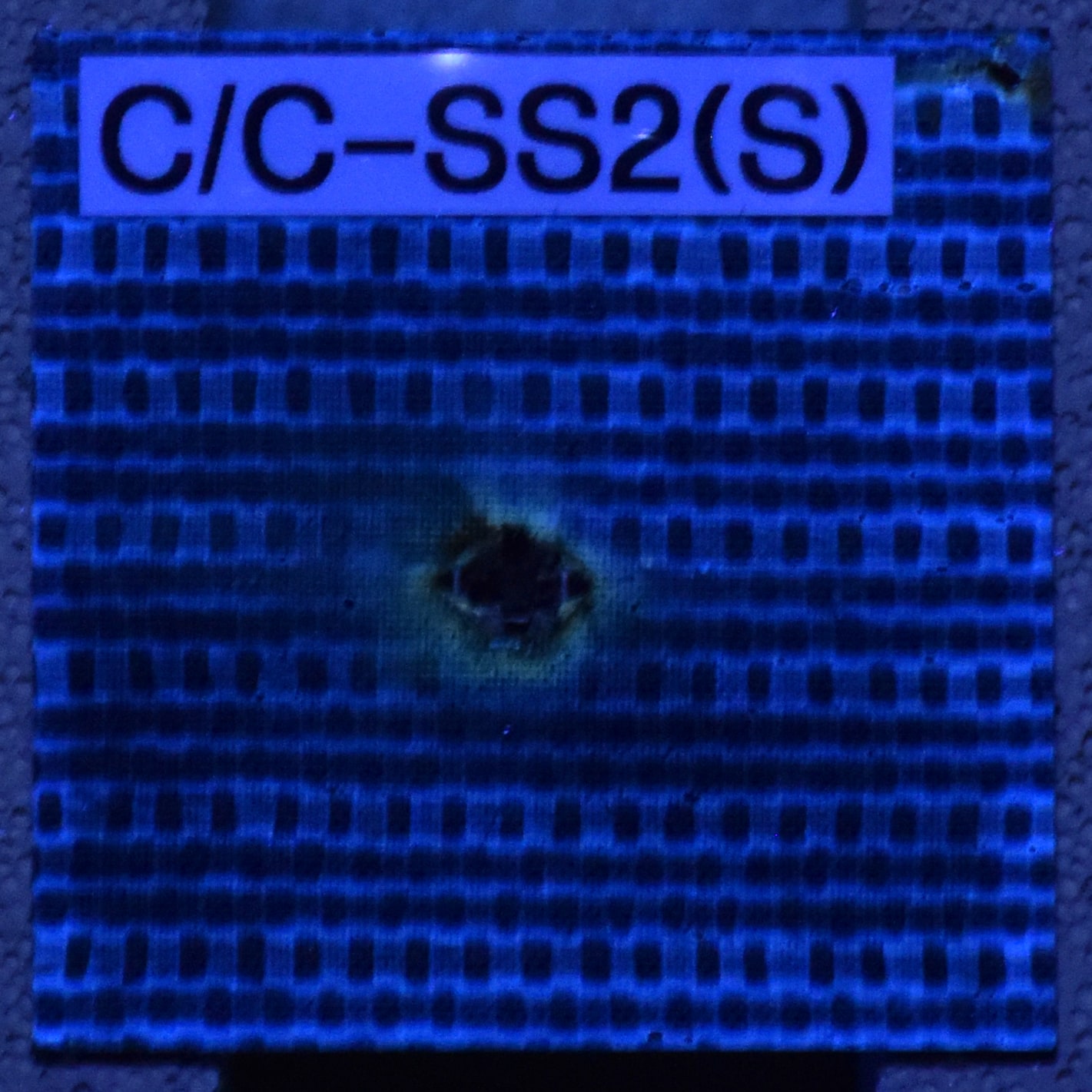} (i1) &  
        \includegraphics[width=0.150\textwidth]{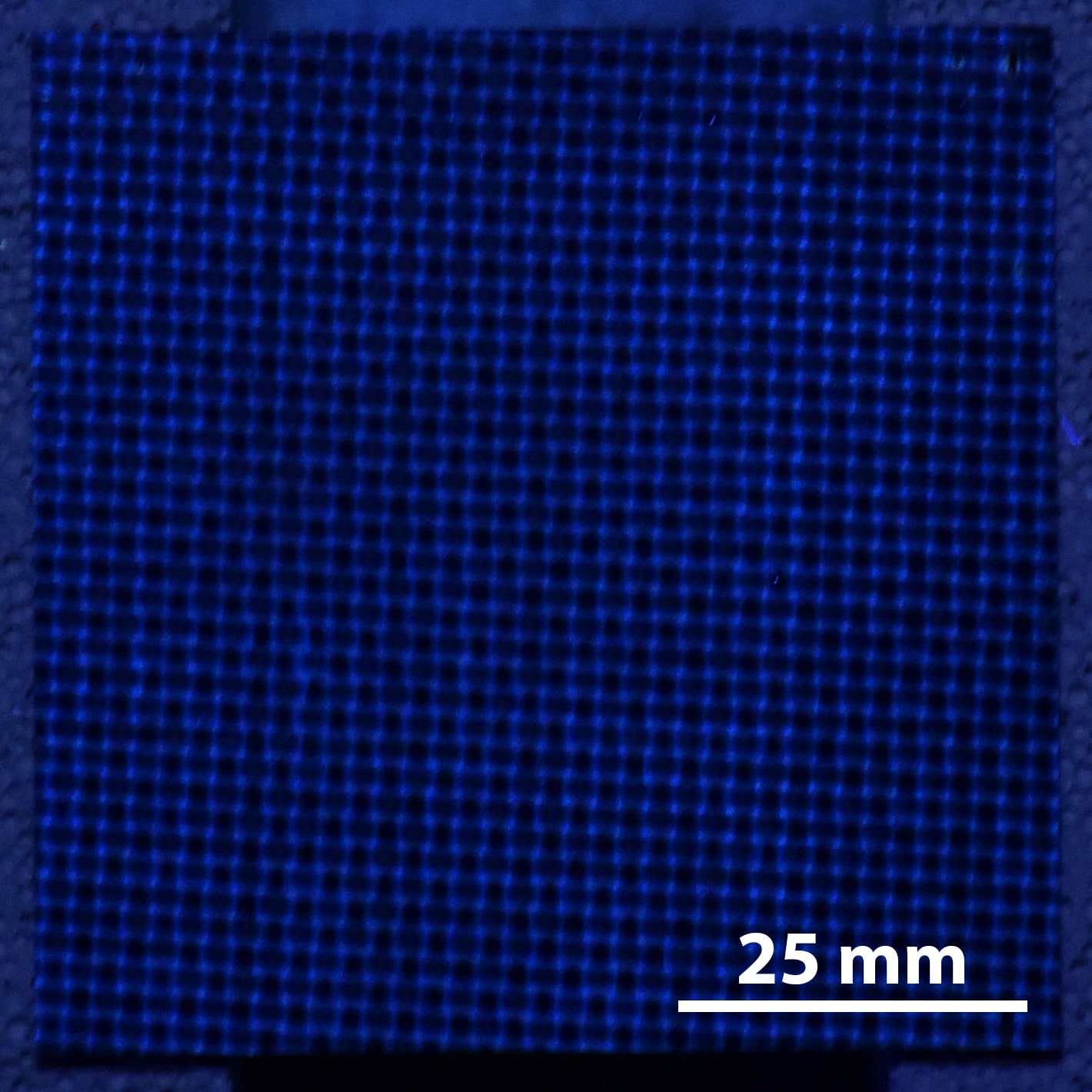} (j1) & 
        \includegraphics[width=0.150\textwidth]{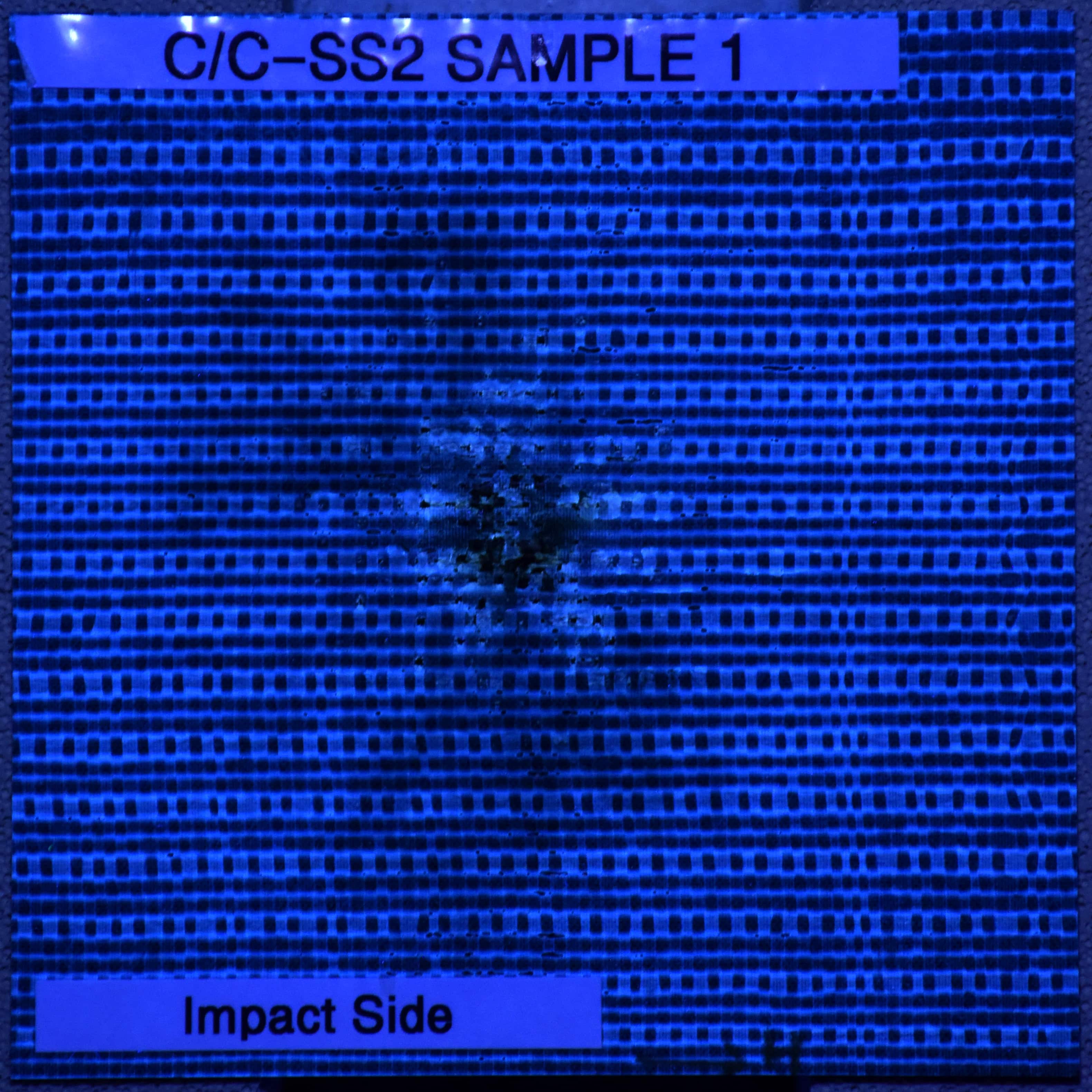} (i2) &  
        \includegraphics[width=0.150\textwidth]{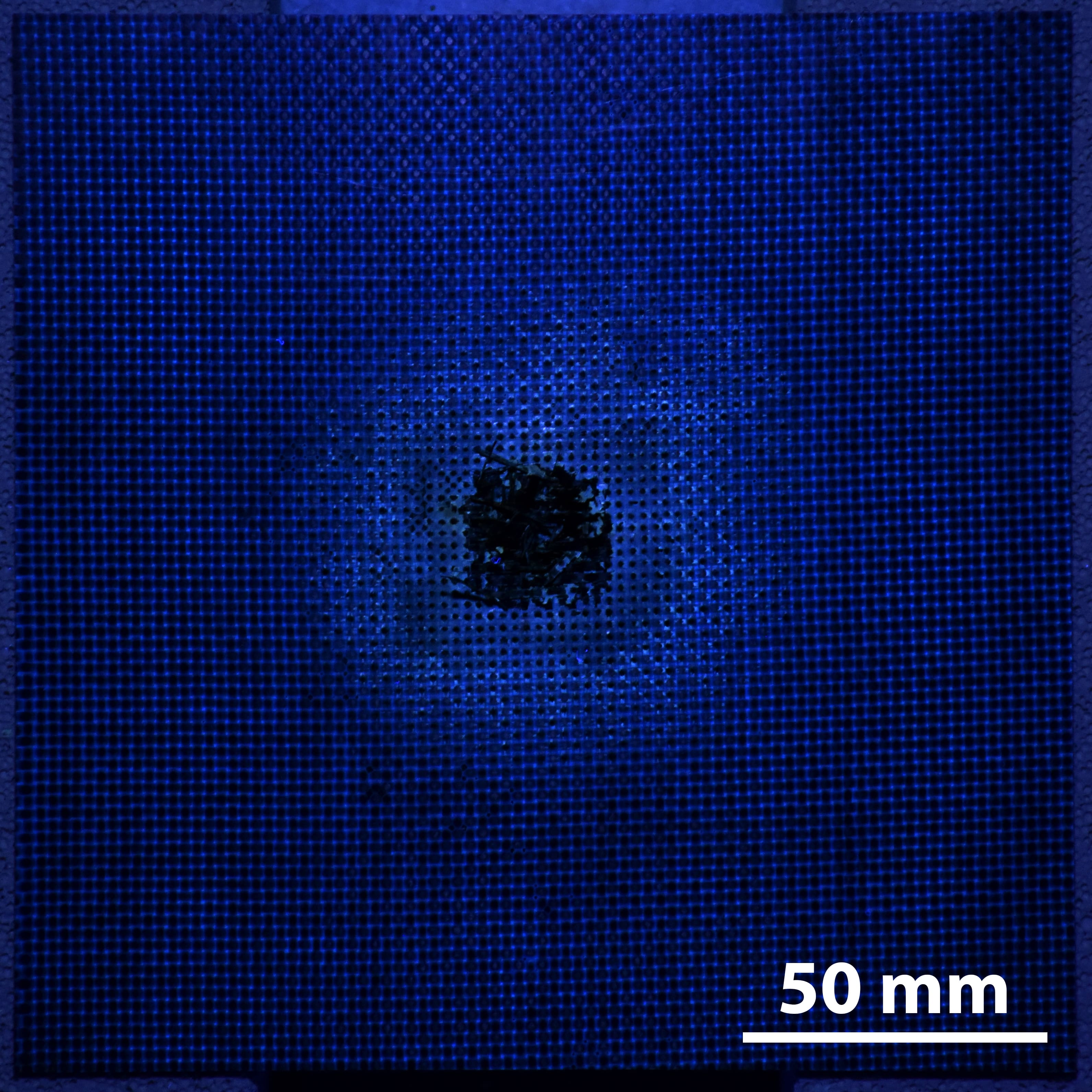} (j2)  \\
        \bottomrule  
    \end{tabular}
}
\caption{Impacted side and back side of the composite surfaces highlight damage in all samples from Table \ref{tab:weavesummLS} from quasi-static arc impact (a1-j1) and simulated lightning strike (a2-j2). The sample sizes from quasi-static arc impact are shown for 75mm x 75mm samples, and the dimensions were 177.8mm x 177.8mm for simulated lightning strikes.}
\label{img:ImpTIGUV}
\end{figure}

From \textbf{Figures}~\ref{img:ImpTIGUV}(a1-j1), we observed that the damage was more localized (both in-plane and through-thickness) in the samples with hybrid weaves compared to the C/C samples under quasi-static arc impact. The damage areas for the composites after quasi-static arc impacts are summarized in \textbf{Table} \ref{tab:damagesummLS}. We used ImageJ to mark the damage area to measure the damage areas on both the impact side and the back side of the laminate \cite{rueden2017imagej2}. The C/C sample manifested a circular damaged region due to the equal current dissipation in the warp and weft directions \cite{kumar2023enhanced}. We also observed that only C/C samples displayed full through-thickness damage (\textbf{Figure} \ref{img:ImpTIGUV}(b)), attributed to the high resistivity of the composites. This high resistivity prevented effective current dissipation, resulting in a higher temperature buildup and Joule heating in the C/C samples. Using the FLIR camera, we captured the surface temperature following the arc impact, which showed a temperature buildup of 150 $^{\circ}$C after 3 seconds of impact, as highlighted in \textbf{App. Figure} \ref{siimg:thermalTIG}(a). Our FLIR camera has an upper limit of 150 $^{\circ}$C, so the actual temperature buildup could be higher. 

In samples with a single layer of hybrid weaves (C/SS1L and C/C-SS1L), we observed that the damage on the impact side was more elongated along the direction of stainless steel fibers. Furthermore, there was a larger surface damage area on C/C-SS1L due to the fewer steel fibers dissipating the current compared to C/SS1L. Recall that C/C-SS1L has a single layer of hybrid fabric at the impact surface, with all carbon yarns in the warp direction, while the weft has stainless steel yarns after every two carbon fiber yarns. Whereas C/SS1L has all stainless steel along the weft direction in the 1-layer hybrid fabric. From the FLIR camera, we also observed that the peak recorded temperature dropped to 97.6 $^{\circ}$C and 110 $^{\circ}$C for C/SS1L and C/C-SS1L, respectively. A strong horizontal line of the heat path was visible on the surface, corresponding to stainless steel yarns in the weave. This heat signal along a line is shown in \textbf{App. Figures} \ref{siimg:thermalTIG}(b,c) corresponding to C/SS1L and C/C-SS1L, respectively.

\begin{table}[h!]
    \centering
    \renewcommand{\arraystretch}{1.4}
    \caption{Summary of damage quantification for composites subjected to quasi-static arc and simulated lightning strike impacts.} 
    \resizebox{\columnwidth}{!}{
    \begin{tabular}{lcccccc}
    \hline
         \textbf{Material} &  \multicolumn{3}{c|}{\textbf{Quasi-static Arc}} & \multicolumn{3}{|c}{\textbf{Lightning Strike}}  \\ \hline
         
         & \textbf{Surface Temperature ($^{\circ} C$)} & \multicolumn{2}{c}{\textbf{Damage Area ($mm^2$)}} & \textbf{Surface Temperature ($^{\circ} C$)} & \multicolumn{2}{c}{\textbf{Damage Area ($mm^2$)}}    \\ 
         
         & (max. after 3 seconds) &  Impact side & Back side & (max. after 60 seconds) & Impact side & Back side    \\ \hline
        
        \textit{C/C} & 150+ & 496.27 & 126.08 &  88.2 & 6207.33 & 6162.48   \\ 
        
        \textit{C/SS1L} & 97.6 & 117.08 & 0 & 63.2  & 2490.66 &  3441.93   \\ 
    
        \textit{C/SS2L} & 93.5 & 101.23 & 0 & 57.8  & 1013.18 &  2884.96   \\ 
        
        \textit{C/C-SS1L} & 110 & 202.64 & 0 & 71.7 & 3095.79 & 3957.91     \\ 
    
        \textit{C/C-SS2L} & 105 & 145.33 & 0 & 60.0 & 2064.87 & 3608.44   \\  \hline
    
    \end{tabular}
    }
    \label{tab:damagesummLS}
    \end{table}
    
For composites with two layers of hybrid weaves (C/SS2L and C/C-SS2L), the surface damage area was smaller than the single-layer hybrid weave samples, which is attributed to stainless steel yarns in both primary orthogonal directions. This distribution again resulted in a circular damage shape similar to the C/C composite samples. The temperature buildup on the surface was further reduced for these samples. However, the drop was less significant than the reduction observed between C/C samples and those with one hybrid layer, as summarized in \textbf{Table} \ref{tab:damagesummLS}. As observed previously, a strong heat signal is observed along the direction of stainless steel yarns in the first hybrid layer on the impact side (\textbf{App. Figures} \ref{siimg:thermalTIG}(d,e)). 

\paragraph{Simulated lightning strike:}\label{res:LSimp}

\textbf{Figures} \ref{img:ImpTIGUV}(a2-j2) illustrate the damage profiles in composites with hybrid weaves compared to pristine C/C samples subjected to simulated lightning strikes. Since the top layer was not grounded, the simulated lightning strikes caused through-thickness damage in all samples. However, as discussed in Section \ref{res:TIGimp}, no through-thickness damage occurred in composites with hybrid weaves when their top layers were grounded. The damage areas observed on the impacted and back surfaces in all composites subjected to the simulated lightning strikes are summarized in \textbf{Table} \ref{tab:damagesummLS}. Upon inspecting the impacted side, we found that C/C composites exhibited fiber breakage, ply-lift, fiber sublimation, and resin burn-off. We measured the mass loss after the simulated strike and observed a reduction of 2.98 \%. Additionally, a peak surface temperature of 88.2 $^{\circ}$C was recorded 1-minute post-impact using a Fluke camera.

In samples containing a single layer of hybrid weave (C/SS1L and C/C-SS1L), surface damage on the impact side decreased by 59.88 \% and 50.13 \%, respectively, relative to the C/C composites. Similarly, damage area reductions were observed on the back side of these hybrid composites, as shown in \textbf{Table} \ref{tab:damagesummLS}. Temperature imaging with a Fluke camera showed lower peak surface temperatures of 63.2 $^{\circ}$C and 71.7 $^{\circ}$C for C/SS1L and C/C-SS1L, respectively, measured 1 minute after the strike. Further, the analysis indicated a minimal mass loss of 0.07 \% and 0.114 \% for C/SS1L and C/C-SS1L, respectively. We attribute this decrease in temperature rise and mass loss to the lower electrical resistivity of hybrid weaves containing stainless steel yarns. The relatively higher surface temperature and mass loss observed in the C/C-SS1L composite compared to the C/SS1L composite are due to a lower amount of stainless steel yarn in the C/C-SS1L hybrid weave. 

In composites with two layers of hybrid weaves (C/SS2L and C/C-SS2L), we found that the surface damage area on the impact side further reduced by 83.67 \% and 68.74 \%, respectively, compared to C/C samples. The reductions were also significant compared to their corresponding single-layer hybrid weave versions, with damage areas being 59.32\% lower in C/SS2L compared to C/SS1L, and 33.30\% lower in C/C-SS2L compared to C/C-SS1L. Consequently, we observed less damage on the back sides of these samples relative to those with single-layer hybrid weave, C/SS1L and C/C-SS1L. The surface temperatures on the impacted side dropped to 57.8 $^{\circ}$C and 60.0 $^{\circ}$C for C/SS2L and C/C-SS2L, respectively. These temperatures are lower compared to the samples with single-layer hybrid weaves. The Fluke thermal images for all samples, captured 1 minute after the strikes and showing the maximum recorded temperatures, are presented in \textbf{App. Figure} \ref{siimg:thermalLS}. Finally, mass loss measurements showed reductions of 0.038 \% and 0.061 \% for C/SS2L and C/C-SS2L, respectively, compared to C/C samples. These values are roughly half the values observed in single-layer hybrid weave composites. 

\begin{figure}[ht]
\centering
\subfigure[]{
\tcbox[colframe=black, boxrule=1pt, colback=white, boxsep=0pt, arc=0mm]{\includegraphics[width=0.4\textwidth]{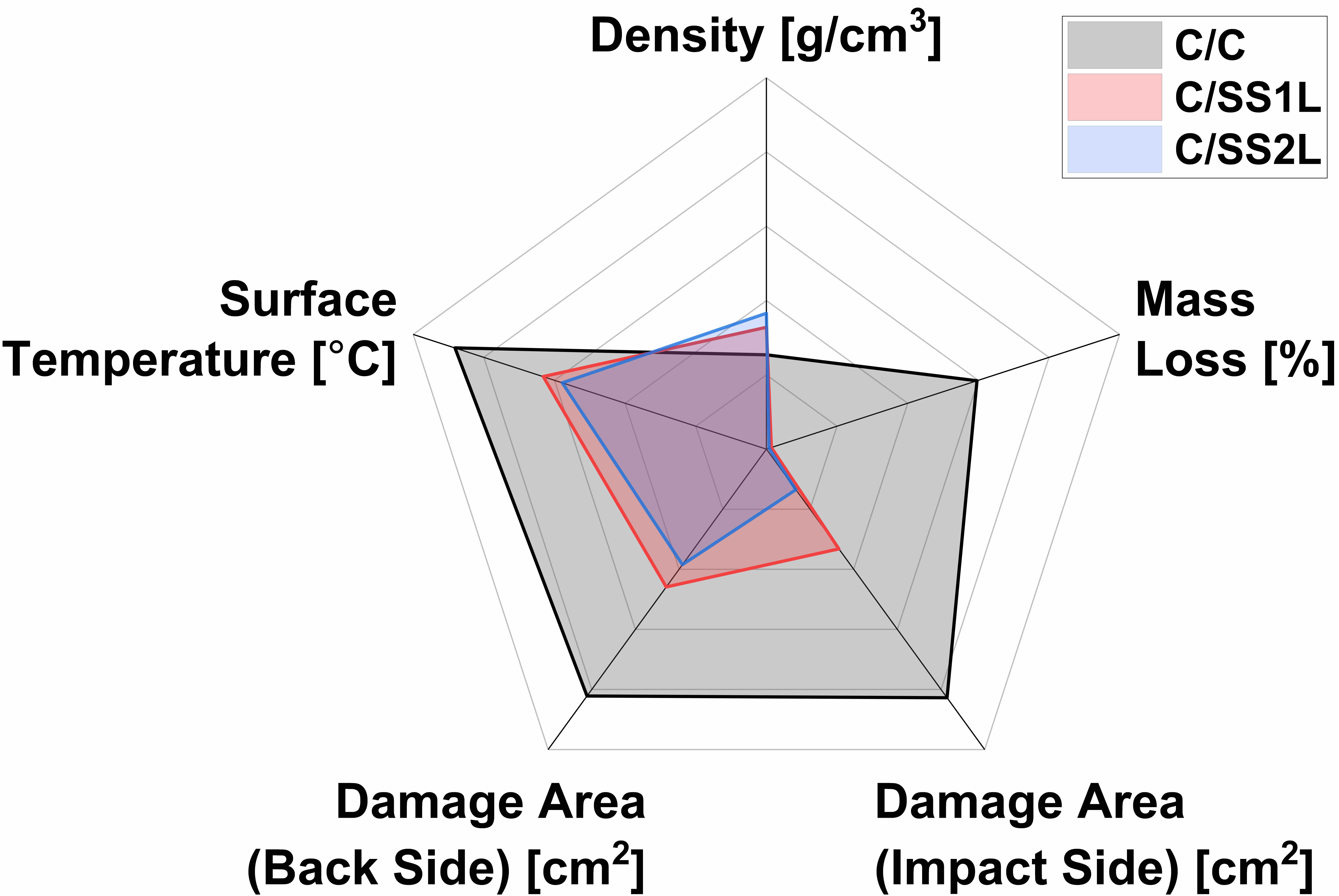}}
}
\subfigure[]{
\tcbox[colframe=black, boxrule=1pt, colback=white, boxsep=0pt, arc=0mm]{\includegraphics[width=0.4\textwidth]{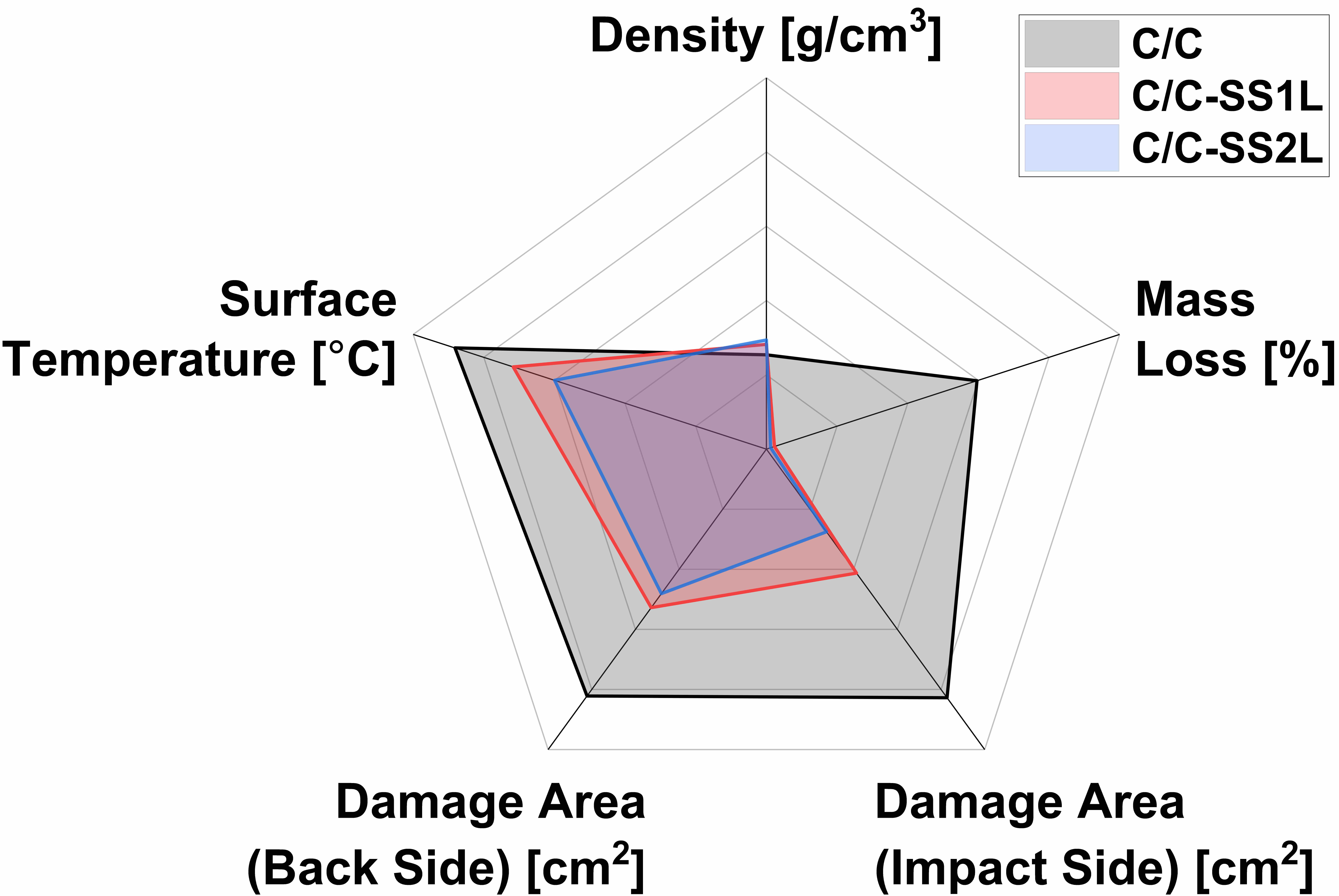}}
}
\caption{Summary plots comparing density and damage quantification for C/C composites with (a) C/SS1L and C/SS2L; and (b) C/C-SS1L and C/C-SS2L composites. Axes ranges are as follows: density ($kg/m^3$) and mass loss ($\%$) from 0 to 5. damage area on both impact and back sides ($cm^2$) from 0 to 7500, and surface temperature ($\circ$) from 0 to 100.}
\label{img:SpiderPlots}
\end{figure}

We attribute the enhanced lightning strike protection in the two-layer hybrid weave composites compared to single-layer hybrid composites to the presence of stainless steel yarns in both orthogonal directions, unlike in single-layer hybrid composites, which are present only in one direction. These trends are visually summarized in comparative plots shown in \textbf{Figure} \ref{img:SpiderPlots}, which highlight substantial improvements in mass loss, surface temperature, and damage area, despite a moderate increase in the density for the composites with hybrid weaves relative to the baseline C/C laminate.

\subsection{Internal damage analysis}\label{res:ctimaging}

\begin{figure}[ht]
\centering
\includegraphics[width=1\textwidth]{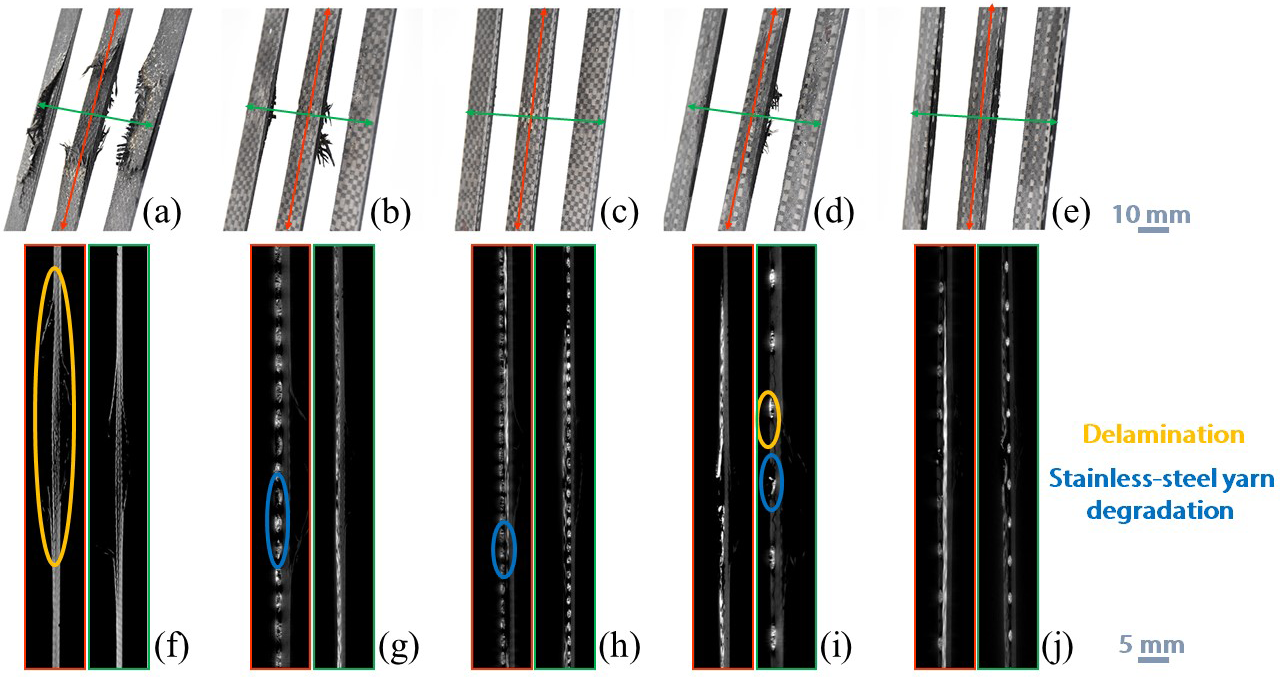}
\caption{Cross-sectional images of the central region of damaged composite samples subjected to 40 kA simulated lightning strikes: (a) C/C; (b) C/SS1L; (c) C/SS2L; (d) C/C-SS1L; and (e) C/C-SS2L. Corresponding sagittal (red) and axial (green) micro-CT scans at the impact sites, illustrating the extent of damage for (f) C/C; (g) C/SS1L; (h) C/SS2L; (i) C/C-SS1L; and (j) C/C-SS2L.}
\label{img:uCTLS}
\end{figure}

\begin{figure}[ht]
\centering
\includegraphics[width=0.7\textwidth]{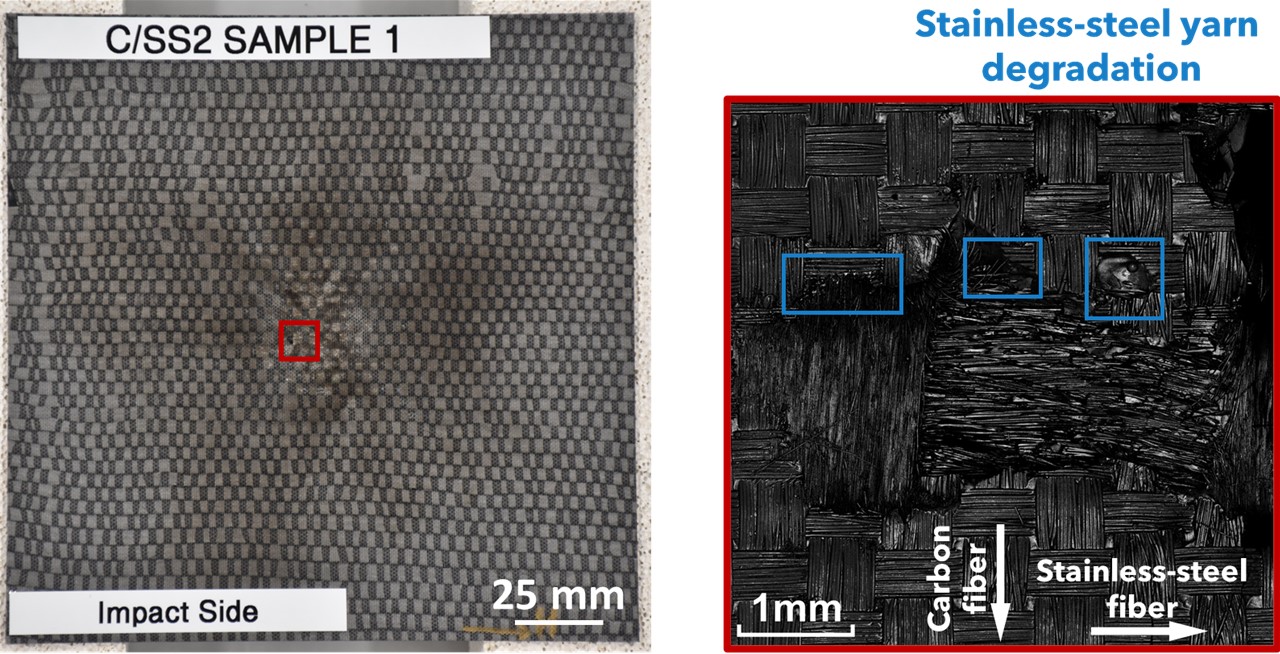}
\caption{Impact side of the C/SS2L sample subjected to a 40 kA simulated lightning strike (left), along with a microscopic view (right) highlighting stainless steel yarn degradation (blue rectangles) as observed in \textbf{Figure}~\ref{img:uCTLS}.}
\label{img:steeldamage}
\end{figure}

To further evaluate the internal (through-thickness) damage caused by the simulated lightning strikes, we performed micro-computed tomography (micro-CT) on the damaged samples using the North Star Imaging X3000 Micro-focus CT System at Baylor University. The scanning parameters were set to a voltage of 200 kV and a current of 500 $\mu$A, with a focal spot size of 100 $\mu$m. Cross-sectional images of the damaged composites and their corresponding micro-CT scans are shown in \textbf{Figure}~\ref{img:uCTLS}. We can observe that the extent of damage in the reference C/C composite is noticeably higher than in the hybrid weave samples. Specifically, the C/C composite exhibited extensive pyrolysis, resulting in delamination across all layers. In contrast, the hybrid weave composites exhibited intact central layers, with delamination primarily confined to the last layer. This was attributed to the absence of grounding on the top surface to maximize lightning strike damage in all composites. Localized degradation of stainless steel fibers is also observed at the strike location in composites with the hybrid weaves. To further investigate this degradation, we performed surface microscopy using the LEXT OLS5100 laser microscope. From \textbf{Figure}~\ref{img:steeldamage}, we observed localized damage in the metallic fibers at the strike location. Specifically, we identified both localized melting and brittle fracture in the exposed stainless steel fibers. The melting is attributed to the extreme thermal spike during the simulated lightning strike, which can momentarily exceed the melting point of stainless steel. Additionally, the mechanical impact of the lightning strike, occurring after matrix pyrolysis around the fibers, could have contributed to the brittle fracture observed in the stainless steel fibers. 

In summary, according to Table \ref{tab:damagesummLS}, C/C composites exhibited the highest surface temperature and sustained through-thickness damage under quasi-static arc impacts. On the other hand, composites with hybrid weaves exhibited significantly lower surface temperatures and resisted the through-thickness damage due to the presence of conductive stainless steel yarns. After simulated lightning strike tests, all composites experienced higher overall damage compared to the quasi-static arc tests as anticipated. However, the composites with hybrid weaves demonstrated better performance, as evidenced by reduced damage areas on both impact and back surfaces, as well as lower surface temperatures, compared to the reference C/C composite.

\subsection{Residual flexural test}\label{res:TIGFRes}
\begin{figure}[ht]
\centering
\includegraphics[width=0.7\textwidth]{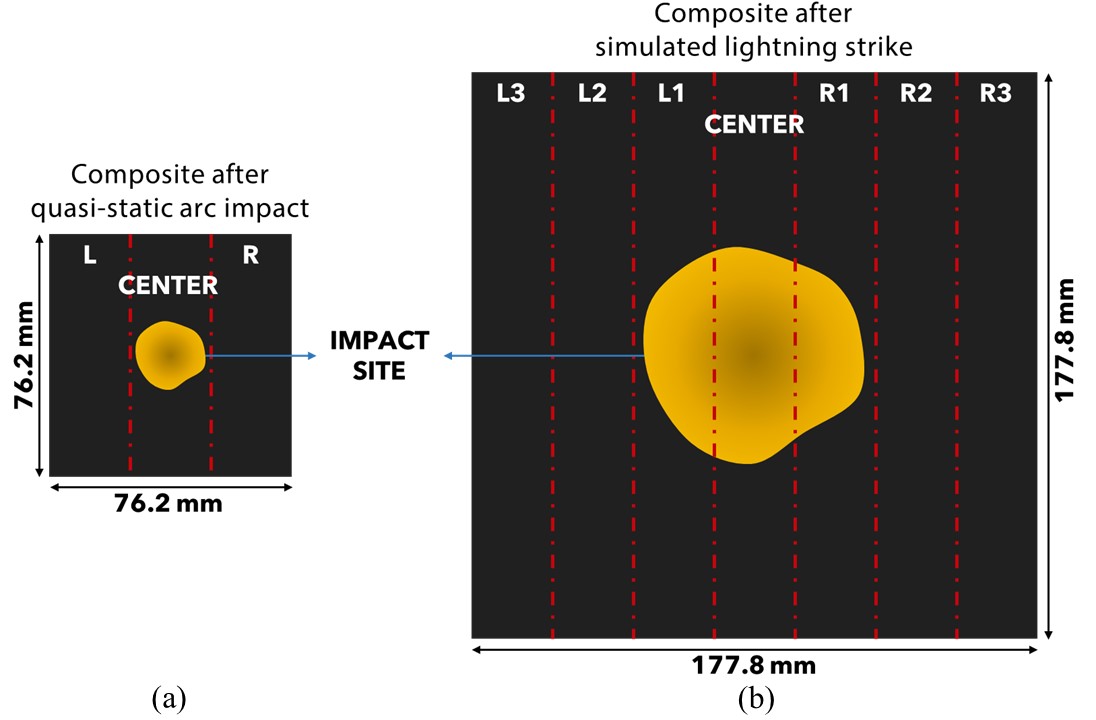}
\caption{Illustration of samples extracted from damaged composites for residual flexural tests after (a) quasi-static electric arc impact (3 samples) and (b) simulated lightning strike (7 samples).}
\label{img:ressampleillus}
\end{figure}

We performed four-point bend (4PB) tests to assess the effect of the quasi-static arc impact and simulated lightning strikes on the residual mechanical performance of all damaged composites. Samples were extracted from the damaged composites as shown in \textbf{Figure}~\ref{img:ressampleillus}. The two farthest samples from either side of the impact site were used to calculate the average pristine flexural properties of the composite, establishing the baseline. Each 4PB test sample maintained a width of 25.4 mm and a span-to-depth ratio of 16:1 for those obtained from the composites impacted with quasi-static arc and 64:1 for those obtained from composites impacted with simulated lightning strike. We labeled the samples from the left and right as shown in \textbf{Figure}~\ref{img:ressampleillus}, with the center sample aligned with the strike site.

\paragraph{Quasi-static arc:}\label{res:resTIGimp}

\begin{figure}[h!]
\centering
\subfigure[]{
\includegraphics[width=0.45\textwidth]{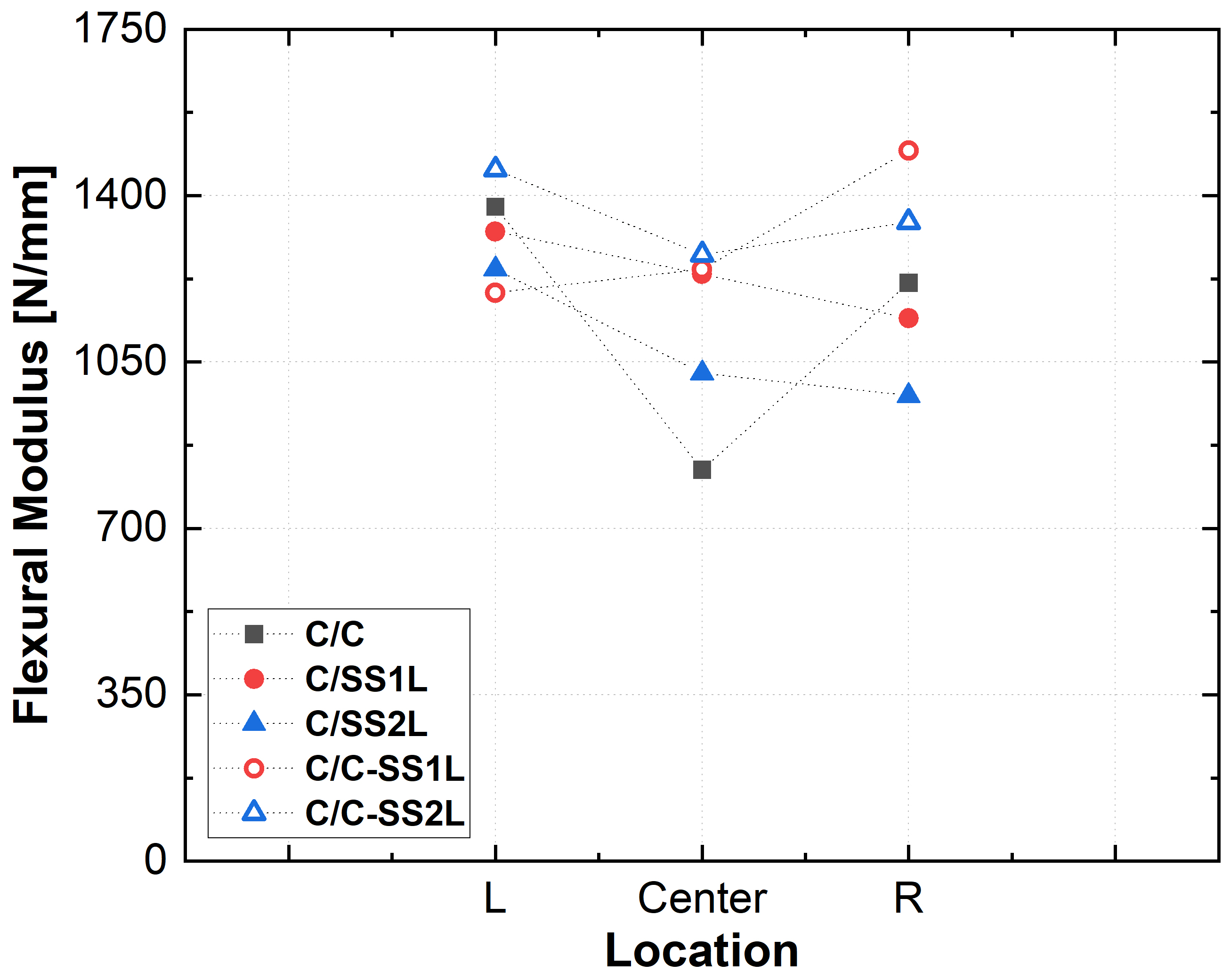}
\label{img:ResTIG1}
}
\subfigure[]{
\includegraphics[width=0.45\textwidth]{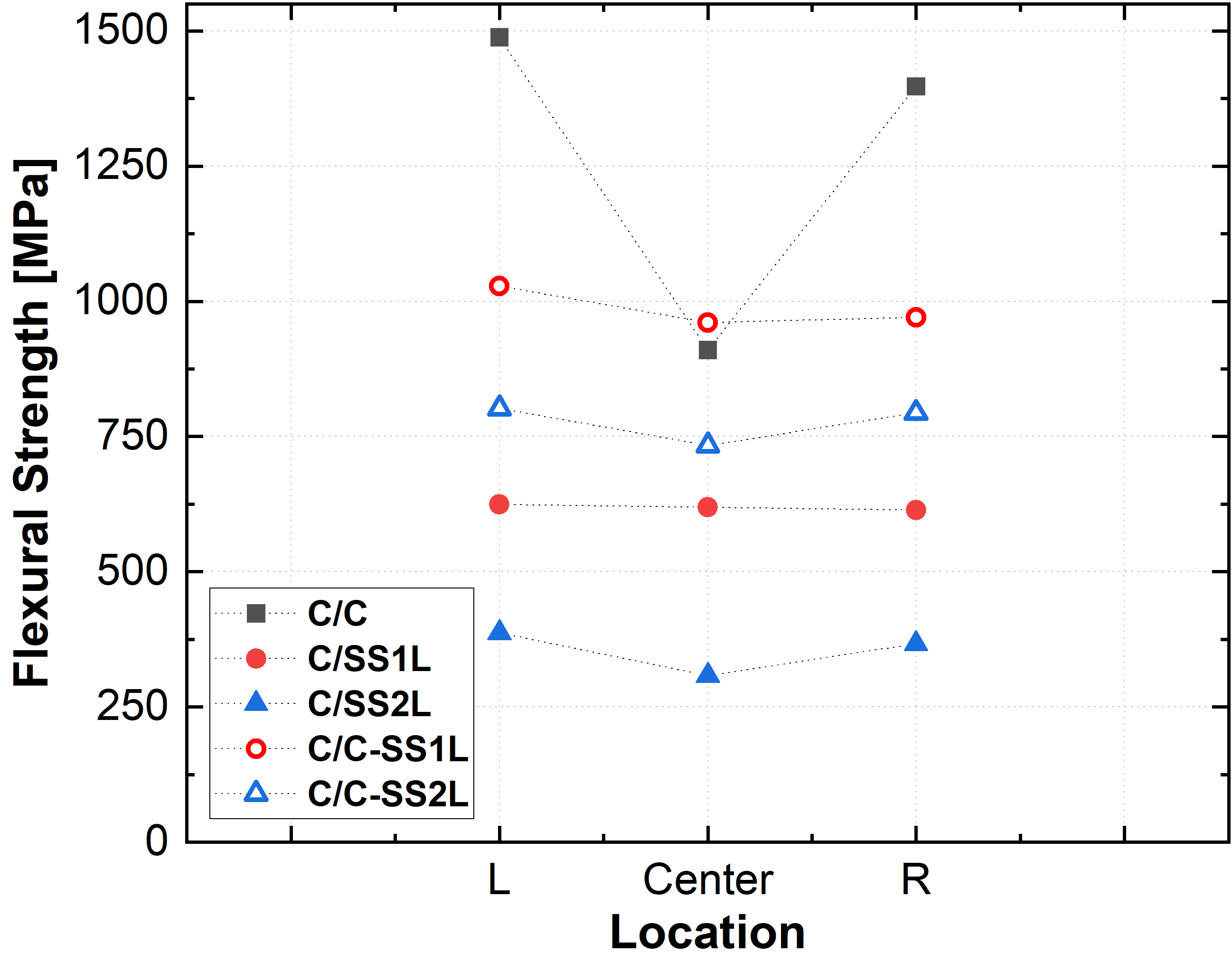}
\label{img:ResTIG2}
}  
\caption{Comparison of (a) flexural modulus and (b) flexural strength of three samples extracted from composites exposed to quasi-static arc impacts.}
\label{img:ResTIG}
\end{figure}

\begin{figure}[ht]
\centering
\subfigure[]{
\includegraphics[width=0.4\textwidth]{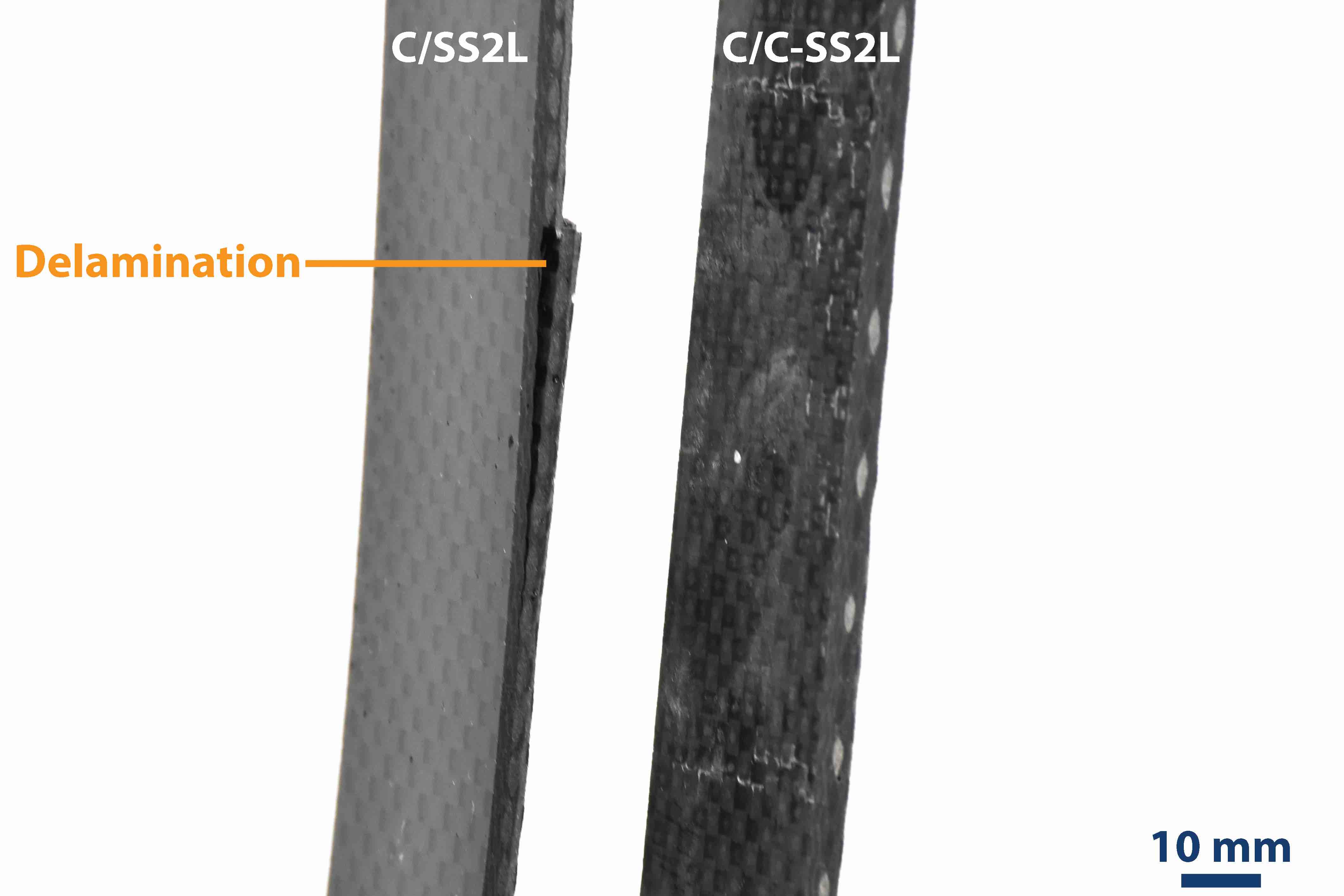}
}
\subfigure[]{
\includegraphics[width=0.4\textwidth]{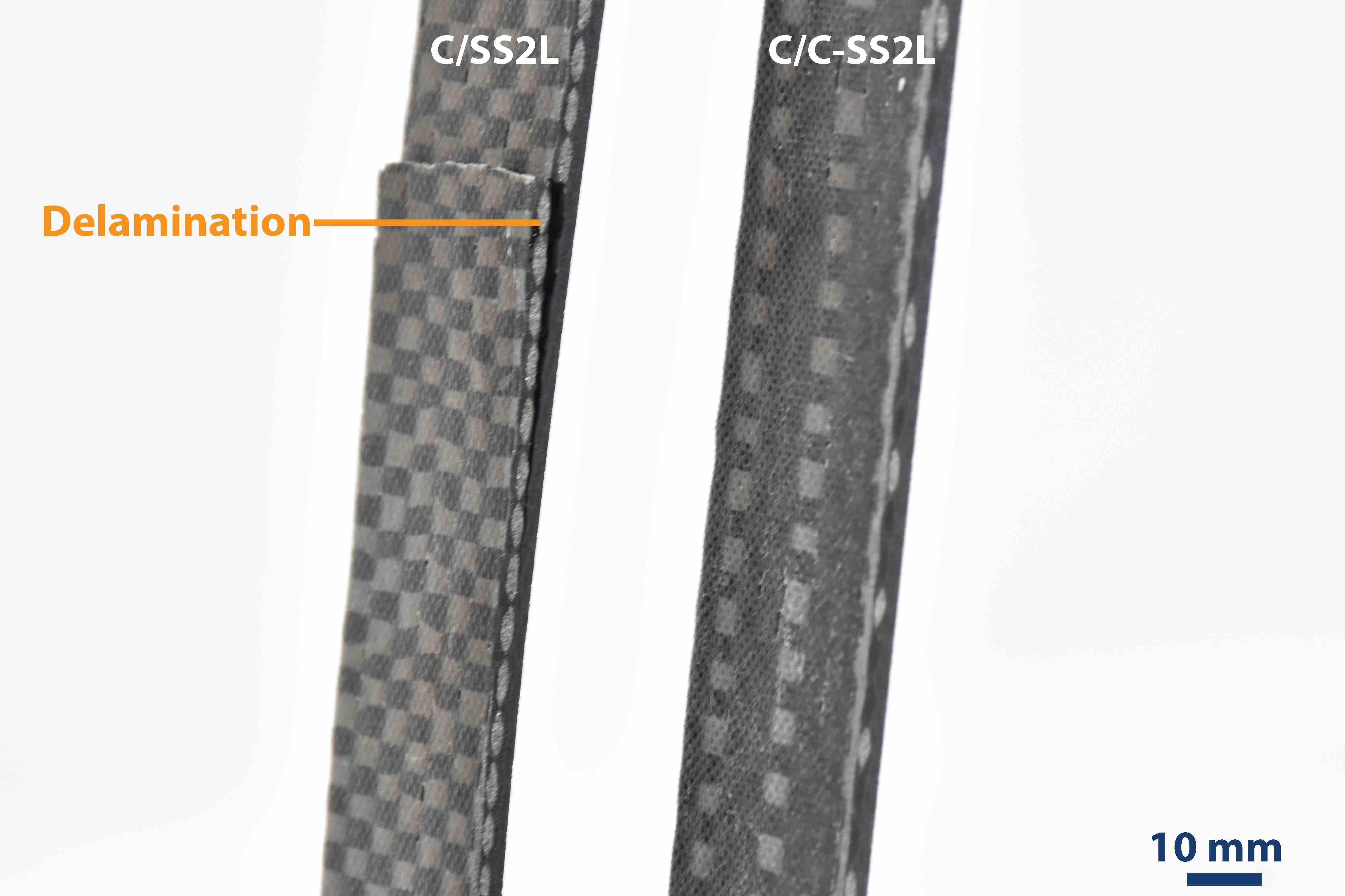}
}
\caption{Comparison of failure modes in C/SS2L and C/C-SS2L composites under 4-point bending: (a) loading surface and (b) bottom surface. C/SS2L composite exhibited delamination due to weak interlaminar bonding, due to all metallic fibers in the hybrid weave. C/C-SS2L composites resisted delamination, with carbon fibers in the hybrid weave improving the interlaminar bonding.}
\label{img:4PBFail}
\end{figure}

\textbf{Figures}~\ref{img:ResTIG} (a, b) present a summary of the residual flexural properties of C/C composites compared to composites containing one and two layers of hybrid layers. We observe that pristine samples with hybrid weaves exhibited lower flexural strength than the reference C/C composites. This reduction can be attributed to the stainless steel yarns in the bottom layers, which undergo tensile loading. Recall that these tests were conducted with the samples positioned such that the top impact face incorporating the hybrid layers was facing downward. As we discussed in \textbf{Appendix} \ref{si:ResTens}, the presence of stainless steel yarns reduced the tensile properties of the composite (\textbf{App. Figure} \ref{siimg:restensupd}) due to the lower tensile properties of the stainless steel yarn, which affected the flexural properties. The drop in flexural properties became more pronounced as we moved from single-layer to two-layer hybrid weaves due to the increased stainless steel yarn content in the bottom layers of the flexure sample. However, composites with hybrid weaves containing reduced stainless steel content exhibited a less significant decline in flexural properties, as shown in \textbf{Figure}~\ref{img:ResTIG} (a). We speculate this behavior has two main reasons: (i) the presence of 24K carbon fibers in the weft and a reduced amount of stainless steel fibers in the C/C-SS weave, which improved the stiffness and strength, and (ii) differences in interlayer bonding between hybrid layers and the underlying C/C weave layers. To elaborate, in C/SS hybrid weaves, bonding occurs between steel yarns and the carbon fiber in the underlying C/C layers. In contrast, in C/C-SS weaves, both carbon fiber and steel yarns within the hybrid weave bond to the carbon fiber in the underlying C/C layers. Since the stainless steel fibers lack sizing, bonding with the C/C layers would be weaker, whereas carbon fibers in C/C-SS weaves enhance bond strength. We can observe this behavior in \textbf{Figure}~\ref{img:4PBFail} where the C/SS2L composite displayed delamination, which was not observed in the C/C-SS2L sample, highlighting the improved bonding in the latter. Previous studies also report a reduction in flexural properties due to the presence of an LSP  with short vertical conductive fibers, as the V-fiber layer introduced defects, such as voids at the interface between the short V-fibers and in-plane carbon fiber fabric \cite{kumar2023enhanced}.

From \textbf{Figures}~\ref{img:ResTIG} (a, b), we observed that the summary plots exhibited a dip behavior for flexural properties, with the lowest values corresponding to the central samples containing the highest damage. The individual stress-strain responses for the three samples across all impacted composites are presented in \textbf{App. Figure} \ref{siimg:ResTIGSvS}. To quantify the reduction in flexural properties, we calculated the flexural modulus ratio and flexural strength ratio by normalizing the flexural properties of the damaged sample with the average flexural properties of the pristine left (L) and right (R) samples. The ratios, which indicate the loss in flexural properties, are summarized in \textbf{App. Figures}~\ref{siimg:ResTIG} (a, b) to highlight the residual performance of reference C/C composite compared to the hybrid weave composites. The drops in flexural properties for all composites subjected to the quasi-static arcs are summarized in \textbf{Table} \ref{tab:residualflex}.

For the C/C composites, we observed the highest reduction in both modulus and strength, consistent with severe through-thickness damage. In contrast, for the C/SS1L sample, the reductions in flexural strength and modulus were negligible due to the hybrid weave in the top layer, which mitigated the extent of damage caused by the electric arc. However, the C/SS2L sample exhibited a moderate decrease in performance due to damage in the two hybrid layers and weaker bonding between the stainless steel fibers in both layers, which led to early delamination in the damaged area. This brittle failure behavior can be observed in \textbf{App. Figure}~\ref{siimg:ResTIGSvS} (c). 

In the C/C-SS1L composites, we found that the flexural modulus and strength were reduced by moderate values. These drops in flexural properties may result from the lower stainless steel content, which did not mitigate the damage as effectively as in C/SS1L. For the C/C-SS2L composite samples, the flexural modulus declined further due to damage across the two hybrid layers. However, unlike the C/SS2L composite, the reduction in flexural strength was smaller, and no early delamination was observed. We speculate that this improved performance in residual flexural strength resulted from the presence of weft carbon fibers in the hybrid weaves, which enhanced the interlayer bonding.

\begin{table}[h!]
    \centering
    \renewcommand{\arraystretch}{1.4}
    \caption{Summary of maximum reduction in flexural properties of composites subjected to quasi-static arc and simulated lightning strike impacts. \\
    * For the quasi-static arc, the maximum drop was calculated by comparing the damaged center sample to the average of the pristine left (L) and right (R) samples. For the lightning strike, the damaged sample was compared to the average of the outermost pristine samples (L3 and R3). This is shown in \textbf{Figure}~\ref{img:ressampleillus}.} 
    \resizebox{\columnwidth}{!}{
    \begin{tabular}{lcccc}
    \hline
         \textbf{Material} &  \multicolumn{2}{c|}{\textbf{Quasi-static Arc}} & \multicolumn{2}{|c}{\textbf{Simulated Lightning Strike}}  \\ \hline
         
         & \textbf{Flexural Modulus ($\%$)} & \textbf{Flexural Strength ($\%$)} & \textbf{Flexural Modulus ($\%$)} & \textbf{Flexural Strength ($\%$)}    \\ 
        
        \textit{C/C} & 36.51 & 36.93 & 93.32 &  92.63    \\ 
        
        \textit{C/SS1L} & 0 & 0 & 56.45 & 61.68     \\ 
    
        \textit{C/SS2L} & 9.64 & 18.24 & 9.48 & 40.79    \\ 
        
        \textit{C/C-SS1L} & 7.40 & 3.90 & 71.43 & 68.01  \\ 
    
        \textit{C/C-SS2L} & 8.83 & 8.05 & 73.05 & 67.05  \\  \hline
    
    \end{tabular}
    }
    \label{tab:residualflex}
    \end{table}

\begin{figure}[h!]
\centering
\subfigure[]{
\includegraphics[width=0.45\textwidth]{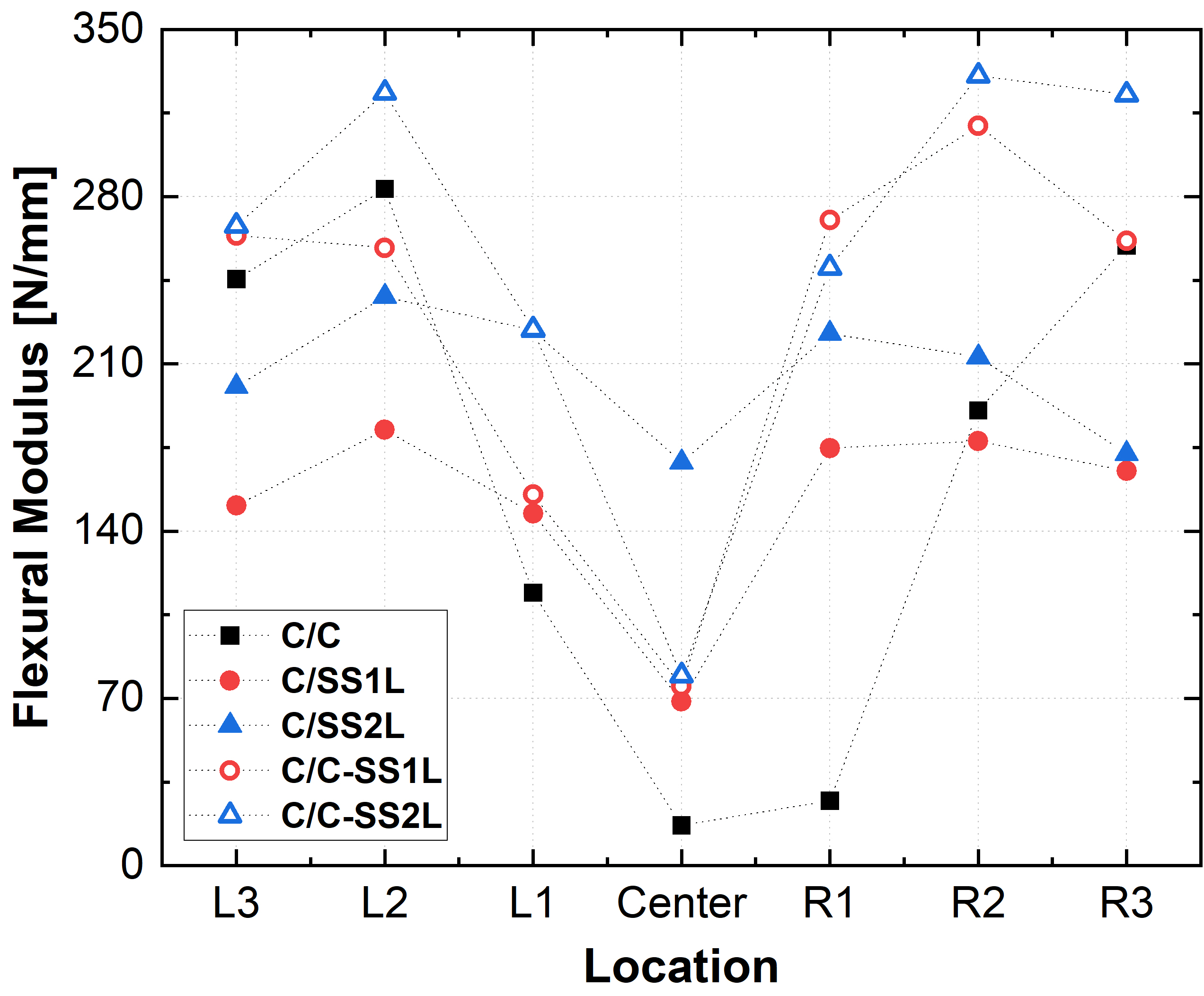}
\label{img:ResLS1}
}
\subfigure[]{
\includegraphics[width=0.45\textwidth]{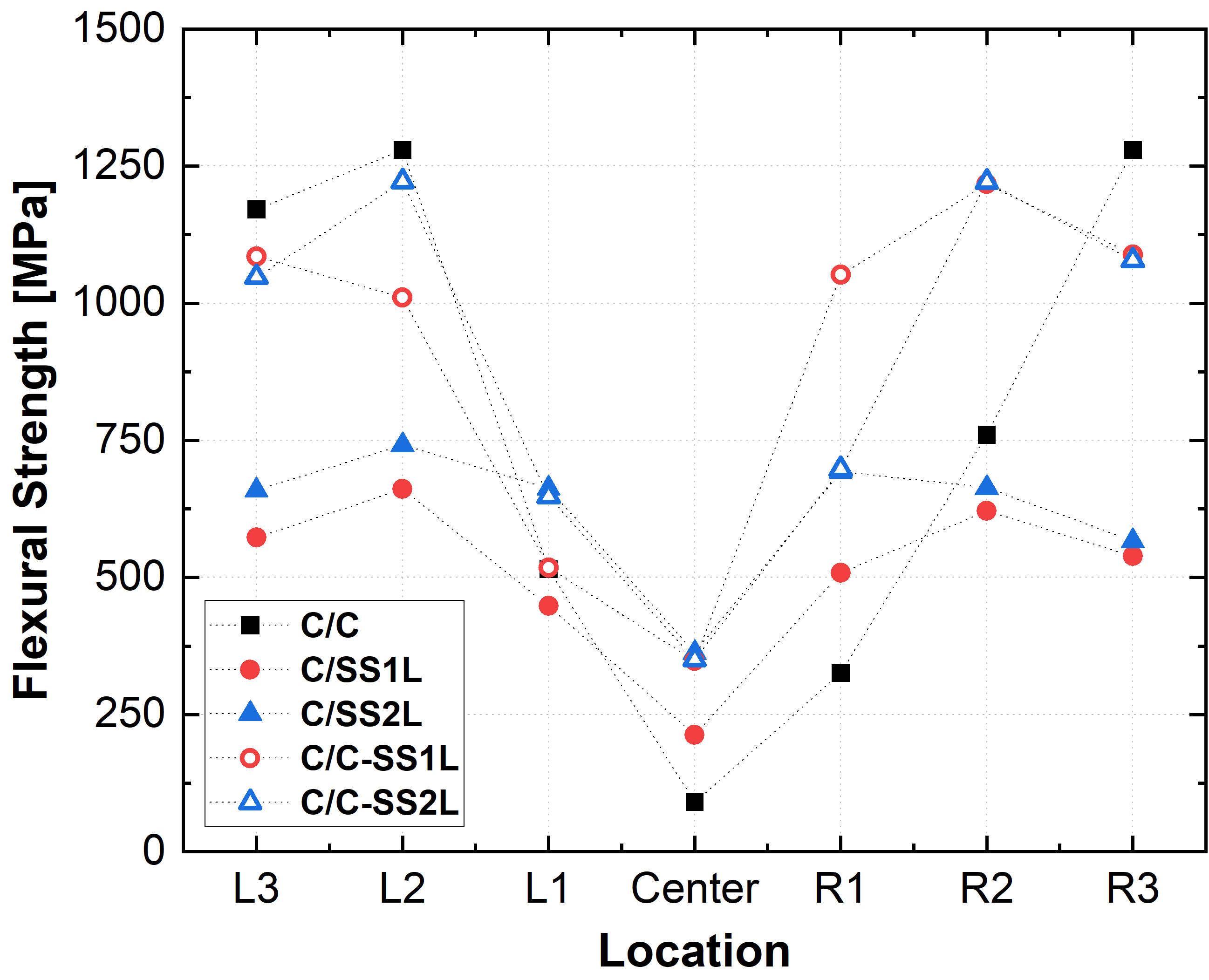}
\label{img:ResLS2}
}  
\caption{Comparison of (a) flexural modulus and (b) flexural strength of seven samples extracted from composites exposed to simulated lightning strikes.}
\label{img:ResLS}
\end{figure}

\paragraph{Simulated lightning strike:}\label{res:resLS}
\textbf{Figures}~\ref{img:ResLS} (a, b) present the flexural properties of composites after a simulated lightning strike. Each sample was sectioned into seven samples, with the outermost samples (L3 and R3) representing pristine conditions, as illustrated in \textbf{Figure}~\ref{img:ressampleillus}. As previously observed in \textbf{Section} \ref{res:resTIGimp}, hybrid weaves showed lower pristine flexural strength than the reference C/C composite due to stainless steel yarns undergoing tension on the bottom surface. However, compared to quasi-static arc impact, simulated lightning strikes caused more severe damage, resulting in higher reductions in flexural properties. To evaluate residual performance, flexural modulus and strength were normalized using the average of the pristine samples (L3 and R3). The residual property ratios are summarized in \textbf{App. Figures} \ref{siimg:ResLS} (a, b), with maximum reductions across configurations listed in \textbf{Table} \ref{tab:residualflex}.

The C/C composite exhibited the most extensive loss in flexural properties due to extensive through-thickness delamination across all layers, as shown in \textbf{Figure} \ref{img:uCTLS} (f). In C/SS1L, we observed that the composites retained significantly more strength and stiffness, indicating improved performance from the conductive hybrid weave on the top layer. The C/SS2L composite demonstrated the best retention of flexural performance. This behavior is attributed to the presence of stainless steel yarns in both the warp and weft directions, which enhanced current dissipation and reduced the extent of damage.

On the other hand, C/C-SS1L showed greater reductions in flexural properties compared to C/SS1L due to lower stainless steel content and delamination beneath the hybrid weave (\textbf{Figure} \ref{img:uCTLS} (i)). C/C-SS2L also exhibited similar reductions as C/C-SS1L, attributed to the reduced amount of stainless steel yarns and damage in the two hybrid weave layers. Notably, compared to the C/C composite, all hybrid weave composites also constrained the in-plane damage area, as only 3 of 7 samples exhibited reductions in flexural properties, while 4 of 7 C/C samples were significantly degraded.

Overall, composites with hybrid weaves as LSPs exhibited enhanced damage tolerance compared to the reference C/C composite under quasi-static arc and simulated lightning strikes. Including stainless steel yarns lowers surface electrical resistivity, reducing Joule heating and minimizing damage. Further, reducing the stainless steel yarns in C/C-SS weaves provided additional bonding benefits due to carbon–carbon fiber interfaces, but were less effective than C/SS configurations. These results emphasize the trade-off between damage tolerance and mechanical bonding in hybrid LSP designs.

\section{Conclusions}\label{conc}
In this study, we presented a novel drapable LSP solution through hybrid weaving of carbon fibers with metallic (stainless steel) yarns. Fundamentally, we examined the performance of hybrid stainless steel/carbon woven composites under quasi-static arc impacts and simulated lightning strike conditions. We also discussed how hybridization parameters influence the effectiveness of these hybrid composites in dissipating current during an electric arc impact. Specifically, we developed novel hybrid weaves with stainless steel yarns oriented in the weft direction, varying the ratios of stainless steel to carbon yarns. These ratios ranged from 100\% stainless steel yarns (C/SS) to 33\% stainless steel (C/C-SS) in the weft. We manufactured carbon-carbon (C/C) woven composites without lightning strike protection (LSP) as a reference. In addition, we fabricated composites with one and two layers of hybrid weaves to investigate the influence of hybrid configurations on damage in composites due to electric arc tests. This approach demonstrates that strategically introducing architectures by hybridizing woven fabrics with metallic fibers can enhance damage resilience under electrical impacts and simulated lightning strikes. Furthermore, due to reduced damage severity, composites with hybrid weaves showed superior residual flexural properties compared to the reference C/C composites.

The key outcomes of this work are:

\begin{itemize}
    \item Incorporating stainless steel yarns woven with carbon fiber yarns improved the resilience of composite materials to electrical impacts and lightning strikes. The interwoven metallic fibers in the top layers of composites reduced the in-plane electrical resistivity, leading to improved current dissipation and reduced Joule heating at the impact site. The novel hybrid weave is also highly drapable due to the fabric form, making it suitable for complex-shaped structures. 
    \item Under quasi-static arc conditions, adding a single layer of hybrid weave reduced the through-thickness damage and lowered surface temperature compared to the reference composite without LSP (C/C). Adding another layer of 90-degree in-plane rotated hybrid weave further reduced damage and surface temperature, although the improvement was marginal compared to composites with a single hybrid layer. 
    \item Similar trends were observed for simulated lightning strikes, with hybrid weave composites exhibiting reduced damage areas on both the impact and back faces relative to the reference C/C composites. As the samples were not grounded on the top layer, through-thickness damage was observed for all composites, with and without the hybrid weave. However, adding hybrid weave layers to the top decreased mass loss from 2.98\% in the C/C reference composite to just 0.038\% for samples with two C/SS weave layers.
    \item Following quasi-static arc impact, the flexural modulus and strength of the C/C composite dropped by 36.30\% and 35.80\%, respectively. In contrast, composites with hybrid weaves experienced a \textbf{maximum} reduction of 23.80\% in flexural modulus and 19.10\% in flexural strength. These reductions were attributed to the extent of damage in composites with fewer stainless steel weaves and weak interlayer bonding between stainless steel yarns across two hybrid weave layers. 
    \item After simulated lightning strikes, the flexural modulus and strength of the C/C composite dropped by 93.32\% and 92.63\%, respectively. In contrast, composites with hybrid weaves exhibited significantly better retention. The reductions ranged from a \textbf{maximum} of 73.10\% in flexural modulus and 68.01\% in flexural strength, to a \textbf{minimum} of 9.48\% in flexural modulus and 40.79\% in flexural strength, respectively. 
\end{itemize}

Our investigations revealed, for the first time, that a drapable hybrid weave can serve as an effective LSP for composite structures, providing efficient current dissipation while minimizing damage extent. The ratio of metallic fibers in the weave can be strategically varied across different areas to optimize weight, interlaminar bonding, and lightning strike protection. Additionally, the hybridization can be incorporated into various weave patterns, such as twill and satin, to improve the fabric's drapability. The fabric form of the LSP enables easy integration as prepregs into industrial-scale manufacturing without process modifications, making these architected textiles ideal reinforcements for creating safer, more durable woven composite structures with higher damage tolerance and enhanced mechanical properties.

\section*{Funding}
This research was partly funded by the NSF CAREER award through the {\em{Mechanics of Materials and Structures (MOMS) Program}} [\#: 2046476].

\section*{Acknowledgements}
The authors would like to acknowledge Mike Hughes at the Design Innovation Lab, University of Wisconsin-Madison, for his assistance with the quasi-static arc impacts. The authors also acknowledge Eric Kazyak for providing access to the LEXT OLS5100 laser microscope and Maia Rauh for her assistance with the loom setups, both at the University of Wisconsin-Madison.

\section*{Author Contributions}
Author H.R.T. contributed to the conceptualization of the methodology, formal analysis, investigating, visualization, verification, and writing -- preparing the original draft. Authors V.S. and M.O. contributed to the manufacturing and investigating. Authors E.C. and C.D.L. contributed to the manufacturing. Authors M.D.R.H. and D.W. contributed to lightning strike testing. Authors P.K.R., R.V.L., and D.J. contributed to micro-computed tomography scans. Author P.P. contributed to the conceptualization of the methodology, writing -- reviewing and editing, visualization, verification, supervision, project administration, and funding acquisition.

\section*{Data Availability}
The data that support the findings of this study are available from the corresponding author, PP, upon reasonable request.


{\footnotesize
\bibliographystyle{unsrt}
\bibliography{LSP_bib}

\begin{thebibliography}{10}

\bibitem{feraboli2009damage}
P.~Feraboli and M.~Miller.
\newblock Damage resistance and tolerance of carbon/epoxy composite coupons
  subjected to simulated lightning strike.
\newblock {\em Composites Part A: Applied Science and Manufacturing},
  40(6-7):954--967, 2009.

\bibitem{feraboli2010damage}
P.~Feraboli and H.~Kawakami.
\newblock Damage of carbon/epoxy composite plates subjected to mechanical
  impact and simulated lightning.
\newblock {\em Journal of aircraft}, 47(3):999--1012, 2010.

\bibitem{chakravarthi2011carbon}
D.~K. Chakravarthi, V.~N. Khabashesku, R.~Vaidyanathan, J.~Blaine,
  S.~Yarlagadda, D.~Roseman, Q.~Zeng, and E.~V. Barrera.
\newblock Carbon fiber--bismaleimide composites filled with nickel-coated
  single-walled carbon nanotubes for lightning-strike protection.
\newblock {\em Advanced Functional Materials}, 21(13):2527--2533, 2011.

\bibitem{bai2022physics}
R.~Bai, B.~Chen, J.~Colmars, and P.~Boisse.
\newblock Physics-based evaluation of the drapability of textile composite
  reinforcements.
\newblock {\em Composites Part B: Engineering}, 242:110089, 2022.

\bibitem{li2023impact}
Y.~Li, F.~Wang, C.~Huang, J.~Ren, D.~Wang, J.~Kong, T.~Liu, and L.~Long.
\newblock Impact damage reduction of woven composites subject to pulse current.
\newblock {\em Nature Communications}, 14(1):5046, 2023.

\bibitem{feng2024physics}
H.~Feng, S.~P. Subramaniyan, H.~Tewani, and P.s Prabhakar.
\newblock Physics-constrained neural network for design and feature-based
  optimization of weave architectures.
\newblock {\em Composites Part A: Applied Science and Manufacturing},
  187:108465, 2024.

\bibitem{tewani2024archii}
H.~Tewani, J.~Cyvas, K.~Perez, and P.~Prabhakar.
\newblock Ar$\chi$i-textile composites: Role of weave architecture on mode-i
  fracture energy in woven composites.
\newblock {\em Composites Part A: Applied Science and Manufacturing}, page
  108499, 2024.

\bibitem{kumar2020factors}
V.~Kumar, T.~Yokozeki, C.~Karch, A.~A. Hassen, C.~J. Hershey, S.~Kim, J.~M.
  Lindahl, A.~Barnes, Y.~K. Bandari, and V.~Kunc.
\newblock Factors affecting direct lightning strike damage to fiber reinforced
  composites: A review.
\newblock {\em Composites Part B: Engineering}, 183:107688, 2020.

\bibitem{das2021brief}
S.~Das and T.~Yokozeki.
\newblock A brief review of modified conductive carbon/glass fibre reinforced
  composites for structural applications: Lightning strike protection,
  electromagnetic shielding, and strain sensing.
\newblock {\em Composites Part C: Open Access}, 5:100162, 2021.

\bibitem{wang2024challenges}
Y.~Wang, Y.~Fan, and O.I. Zhupanska.
\newblock Challenges and future recommendations for lightning strike damage
  assessments of composites: Laboratory testing and predictive modeling.
\newblock {\em Materials}, 17(3):744, 2024.

\bibitem{gagne2014lightning}
M.~Gagn{\'e} and D.~Therriault.
\newblock Lightning strike protection of composites.
\newblock {\em Progress in Aerospace Sciences}, 64:1--16, 2014.

\bibitem{rajesh2018damage}
P.S.M. Rajesh, F.~Sirois, and D.~Therriault.
\newblock Damage response of composites coated with conducting materials
  subjected to emulated lightning strikes.
\newblock {\em Materials \& Design}, 139:45--55, 2018.

\bibitem{kawakami2011lightning}
H.~Kawakami and P.~Feraboli.
\newblock Lightning strike damage resistance and tolerance of scarf-repaired
  mesh-protected carbon fiber composites.
\newblock {\em Composites Part A: Applied Science and Manufacturing},
  42(9):1247--1262, 2011.

\bibitem{guo2019enhanced}
Y.~Guo, Y.~Xu, Q.~Wang, Q.~Dong, X.~Yi, and Y.~Jia.
\newblock Enhanced lightning strike protection of carbon fiber composites using
  expanded foils with anisotropic electrical conductivity.
\newblock {\em Composites Part A: Applied Science and Manufacturing},
  117:211--218, 2019.

\bibitem{kumar2019interleaved}
V.~Kumar, S.~Sharma, A.~Pathak, B.P. Singh, S.R. Dhakate, T.~Yokozeki,
  T.~Okada, and T.~Ogasawara.
\newblock Interleaved mwcnt buckypaper between cfrp laminates to improve
  through-thickness electrical conductivity and reducing lightning strike
  damage.
\newblock {\em Composite Structures}, 210:581--589, 2019.

\bibitem{bigand2025destructive}
A.~Bigand, C.~Espinosa, and J-M. Bauchire.
\newblock Destructive and non-destructive analysis of lightning-induced damage
  in protected and painted composite aircraft laminates.
\newblock {\em Aerospace}, 12(5):446, 2025.

\bibitem{gou2010carbon}
J.~Gou, Y.~Tang, F.~Liang, Z.~Zhao, D.~Firsich, and J.~Fielding.
\newblock Carbon nanofiber paper for lightning strike protection of composite
  materials.
\newblock {\em Composites Part B: Engineering}, 41(2):192--198, 2010.

\bibitem{zhang2019lightning}
J.~Zhang, X.~Zhang, X.~Cheng, Y.~Hei, L.~Xing, and Z.~Li.
\newblock Lightning strike damage on the composite laminates with carbon
  nanotube films: Protection effect and damage mechanism.
\newblock {\em Composites Part B: Engineering}, 168:342--352, 2019.

\bibitem{xia2020fabrication}
Q.~Xia, H.~Mei, Z.~Zhang, Y.~Liu, Y.~Liu, and J.~Leng.
\newblock Fabrication of the silver modified carbon nanotube film/carbon fiber
  reinforced polymer composite for the lightning strike protection application.
\newblock {\em Composites Part B: Engineering}, 180:107563, 2020.

\bibitem{kumar2023enhanced}
V.~Kumar, W.~Lin, Y.~Wang, R.~Spencer, S.~Saha, C.~Park, P.~Yeole, N.~S.
  Hmeidat, C.~Herring, M.~L. Rencheck, D.~K. Pokkalla, A.~A. Hassen,
  M.~Theodore, U.~Vaidya, and V.~Kunc.
\newblock Enhanced through-thickness electrical conductivity and lightning
  strike damage response of interleaved vertically aligned short carbon fiber
  composites.
\newblock {\em Composites Part B: Engineering}, 253:110535, 2023.

\bibitem{hirano2016lightning}
Y.~Hirano, T.~Yokozeki, Y.~Ishida, T.~Goto, T.~Takahashi, D.~Qian, S.~Ito,
  T.~Ogasawara, and M.~Ishibashi.
\newblock Lightning damage suppression in a carbon fiber-reinforced polymer
  with a polyaniline-based conductive thermoset matrix.
\newblock {\em Composites Science and Technology}, 127:1--7, 2016.

\bibitem{wang2018fabrication}
B.~Wang, Y.~Duan, Z.~Xin, X.~Yao, D.~Abliz, and G.~Ziegmann.
\newblock Fabrication of an enriched graphene surface protection of carbon
  fiber/epoxy composites for lightning strike via a percolating-assisted resin
  film infusion method.
\newblock {\em Composites Science and Technology}, 158:51--60, 2018.

\bibitem{raimondo2018multifunctional}
M.~Raimondo, L.~Guadagno, V.~Speranza, L.~Bonnaud, P.~Dubois, and K.~Lafdi.
\newblock Multifunctional graphene/poss epoxy resin tailored for aircraft
  lightning strike protection.
\newblock {\em Composites Part B: Engineering}, 140:44--56, 2018.

\bibitem{das2022thickness}
S.~Das, V.~Kumar, J.~Lee, T.~Yokozeki, and T.~Okada.
\newblock Thickness threshold study of polyaniline-based lightning strike
  protection coating for carbon/glass fiber reinforced polymer composites.
\newblock {\em Composite Structures}, 280:114954, 2022.

\bibitem{langot2023performance}
J.~Langot, E.~Gourcerol, A.~Serbescu, D.~Brassard, K.~Chizari, M.~Lapalme,
  A.~Desautels, F.~Sirois, and D.~Therriault.
\newblock Performance of painted and non-painted non-woven nickel-coated carbon
  fibers for lightning strike protection of composite aircraft.
\newblock {\em Composites Part A: Applied Science and Manufacturing},
  175:107772, 2023.

\bibitem{langot2022comparative}
J.~Langot, E.~Gourcerol, A.~Serbescu, D.~Brassard, K.~Chizari, A.~Lapalme,
  M.and~Desautels, Fr. Sirois, and D.~Therriault.
\newblock A comparative study on the performance of nickel-based technologies
  for lightning strike protection of composite aircraft.
\newblock In {\em 2022 36th International Conference on Lightning Protection
  (ICLP)}, pages 170--175. IEEE, 2022.

\bibitem{zhao2018development}
Z.J. Zhao, G.J. Xian, J.G. Yu, J.~Wang, J.F. Tong, J.H. Wei, C.C. Wang,
  P.~Moreira, and X.S. Yi.
\newblock Development of electrically conductive structural bmi based cfrps for
  lightning strike protection.
\newblock {\em Composites Science and Technology}, 167:555--562, 2018.

\bibitem{liu2021multifunctional}
H.~Liu, Y.~Guo, Y.~Zhou, G.~Wan, Z.~Chen, and Y.~Jia.
\newblock Multifunctional nickel-coated carbon fiber veil for improving both
  fracture toughness and electrical performance of carbon fiber/epoxy composite
  laminates.
\newblock {\em Polymer Composites}, 42(10):5335--5347, 2021.

\bibitem{de2010experimental}
E.~de~Bilbao, D.~Soulat, G.~Hivet, and A.~Gasser.
\newblock Experimental study of bending behaviour of reinforcements.
\newblock {\em Experimental Mechanics}, 50:333--351, 2010.

\bibitem{rueden2017imagej2}
C.T. Rueden, J.~Schindelin, M.C. Hiner, B.E. DeZonia, A.E. Walter, E.T. Arena,
  and K.W. Eliceiri.
\newblock Imagej2: Imagej for the next generation of scientific image data.
\newblock {\em BMC bioinformatics}, 18(1):529, 2017.

\bibitem{lampkin2019epoxy}
S.~Lampkin, W.~Lin, M.~Rostaghi-Chalaki, K.~Yousefpour, Y.~Wang, and J.~Kluss.
\newblock Epoxy resin with carbon nanotube additives for lightning strike
  damage mitigation of carbon fiber composite laminates.
\newblock In {\em Proceedings of the American Society for
  Composites-Thirty-fourth Technical Conference}, volume~10, 2019.

\bibitem{yousefpour2021design}
K.~Yousefpour, M.~Rostaghi-Chalaki, J.~Warden, D.~Wallace, and C.~Park.
\newblock Design and construction of an impulse current generator for lightning
  strike experiments.
\newblock {\em International Journal of Electrical and Computer Engineering},
  15(11):372--375, 2021.

\bibitem{kumar2018effect}
V.~Kumar, T.~Yokozeki, T.~Okada, Y.~Hirano, T.~Goto, T.~Takahashi, and
  T.~Ogasawara.
\newblock Effect of through-thickness electrical conductivity of cfrps on
  lightning strike damages.
\newblock {\em Composites Part A: Applied Science and Manufacturing},
  114:429--438, 2018.

\bibitem{jang2025protection}
W.-H. Jang, D.~Hong, S.~Mallesh, J.~Lee, C.~Park, C.-G. Kim, W.-H. Choi, and
  Y.~Nam.
\newblock Protection concept for foamed radar-absorbing sandwich composites
  with high-conductive film against lightning strike impacts.
\newblock {\em Composites Part A: Applied Science and Manufacturing},
  190:108660, 2025.

\bibitem{ASTM6272}
D6272-17 E01.
\newblock {\em ASTM Standards - Standard Test Method for Flexural Properties of
  Unreinforced and Reinforced Plastics and Electrical Insulating Materials by
  Four-Point Bending}, 2020.

\bibitem{ASTM3039}
ASTM D3039/D3039M.
\newblock {\em ASTM Standards - Standard Test Method for Tensile Properties of
  Polymer Matrix Composite Materials}, 2008.

\end{thebibliography}
}

\newpage

\section*{Supplementary information}

\appendix
\renewcommand\thefigure{\thesection.\arabic{figure}}
\renewcommand{\theHfigure}{\thesection.\arabic{figure}}
\setcounter{figure}{0} 


\section{Infrared (IR) thermography} \label{si:thermal}

\begin{figure}[h!]
\centering
\includegraphics[width=0.9\textwidth]{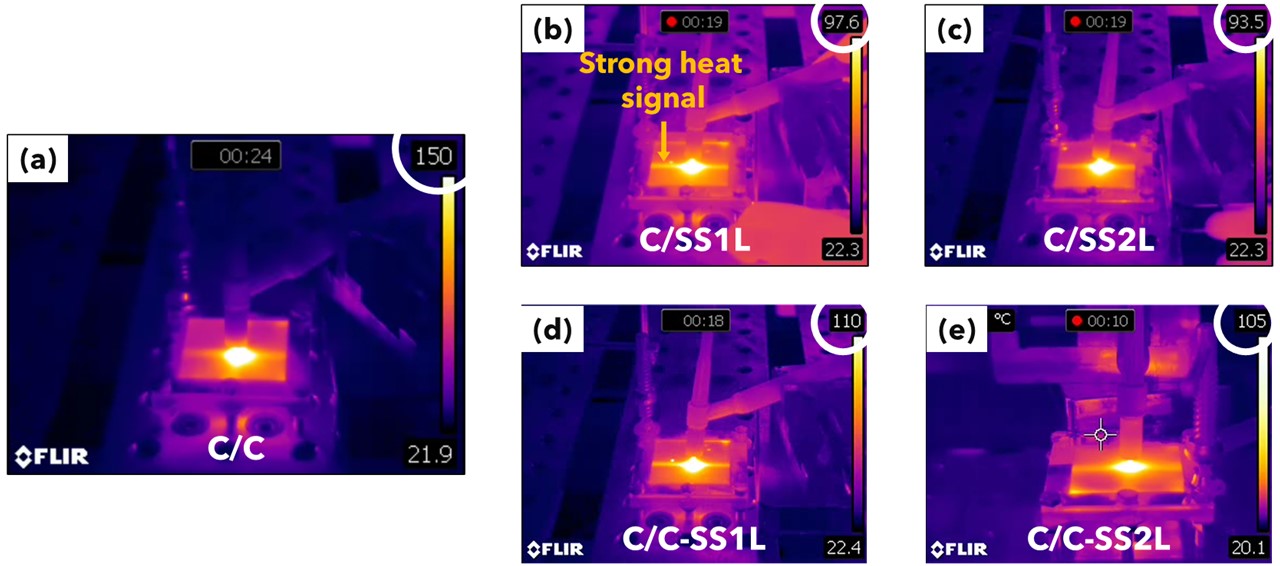}
\caption{FLIR camera images captured 3 seconds after quasi-static arc impacts for (a)C/C; (b) C/SS1L; (c) C/SS2L; (d) C/C-SS1L; and (e) C/C-SS2L. The maximum temperature recorded in each case is shown on the top right corner of each image.}
\label{siimg:thermalTIG}
\end{figure}

\begin{figure}[h!]
\centering
\includegraphics[width=0.9\textwidth]{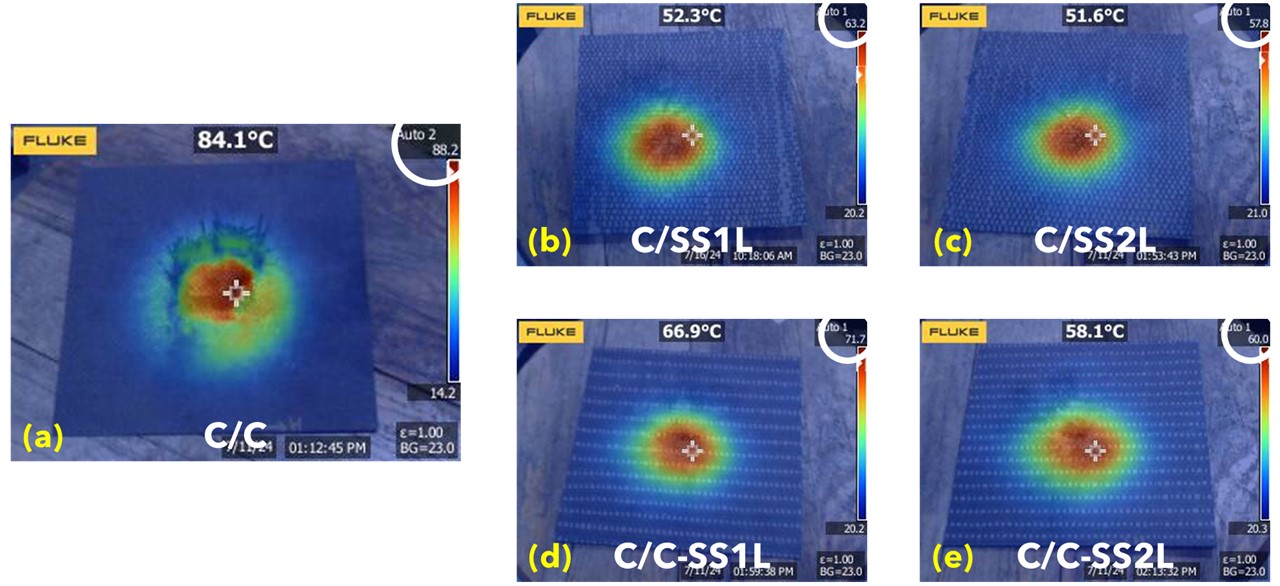}
\caption{Fluke thermal camera images captured 1 minute after simulated lightning strikes for (a)C/C; (b) C/SS1L; (c) C/SS2L; (d) C/C-SS1L; and (e) C/C-SS2L. The maximum temperature recorded in each case is shown on the top right corner of each image.}
\label{siimg:thermalLS}
\end{figure}

\setcounter{figure}{0}

\section{Tensile performance}\label{si:ResTens}
 
\begin{figure}[h!]
\centering
\subfigure[]{
\includegraphics[width=0.36\textwidth]{Images-LSP/FIbEdit.jpg}
\label{siimg:allyarns}
}
\subfigure[]{
\includegraphics[width=0.45\textwidth]{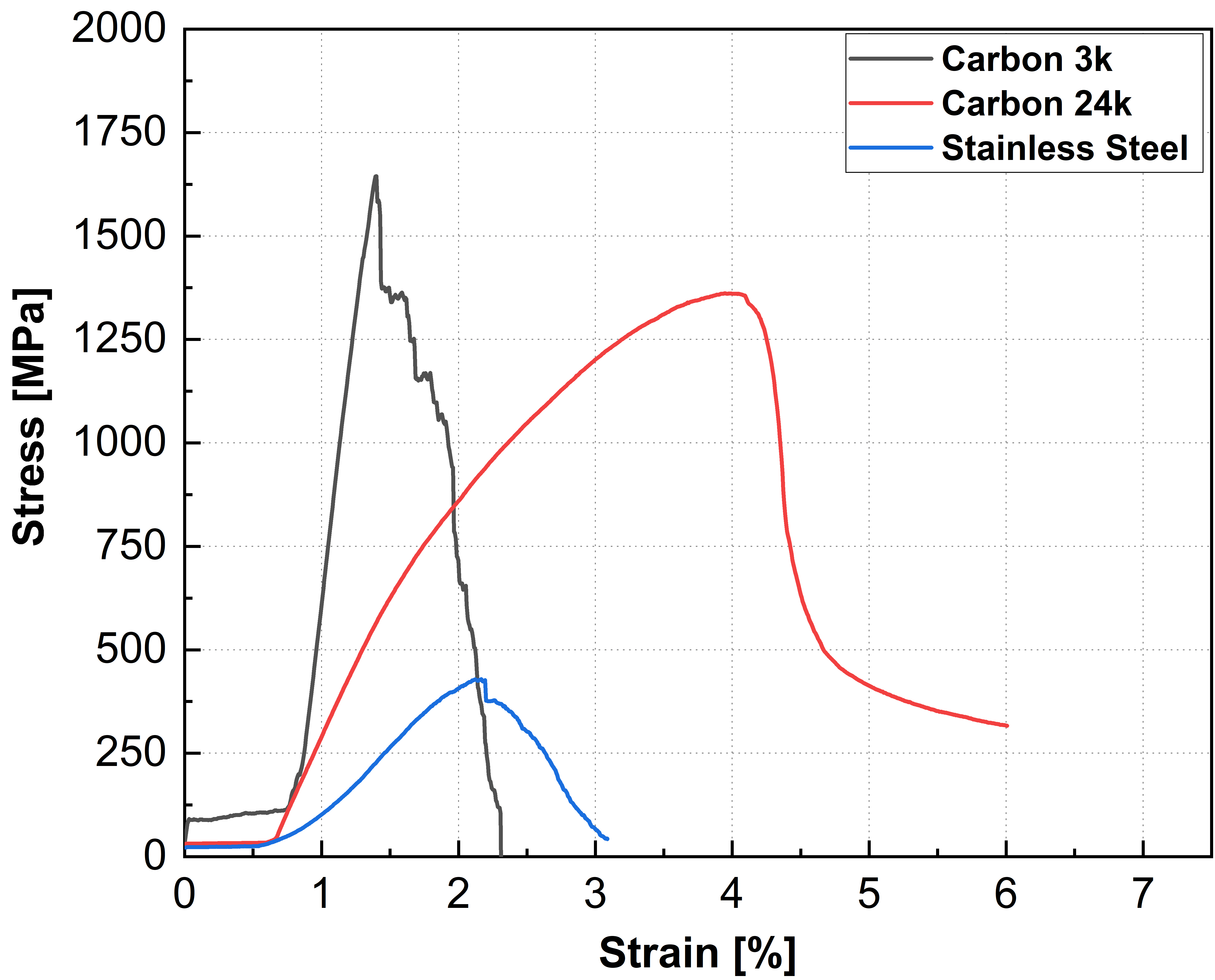}
\label{siimg:tensyarn}
}
\subfigure[]{
\includegraphics[width=0.45\textwidth]{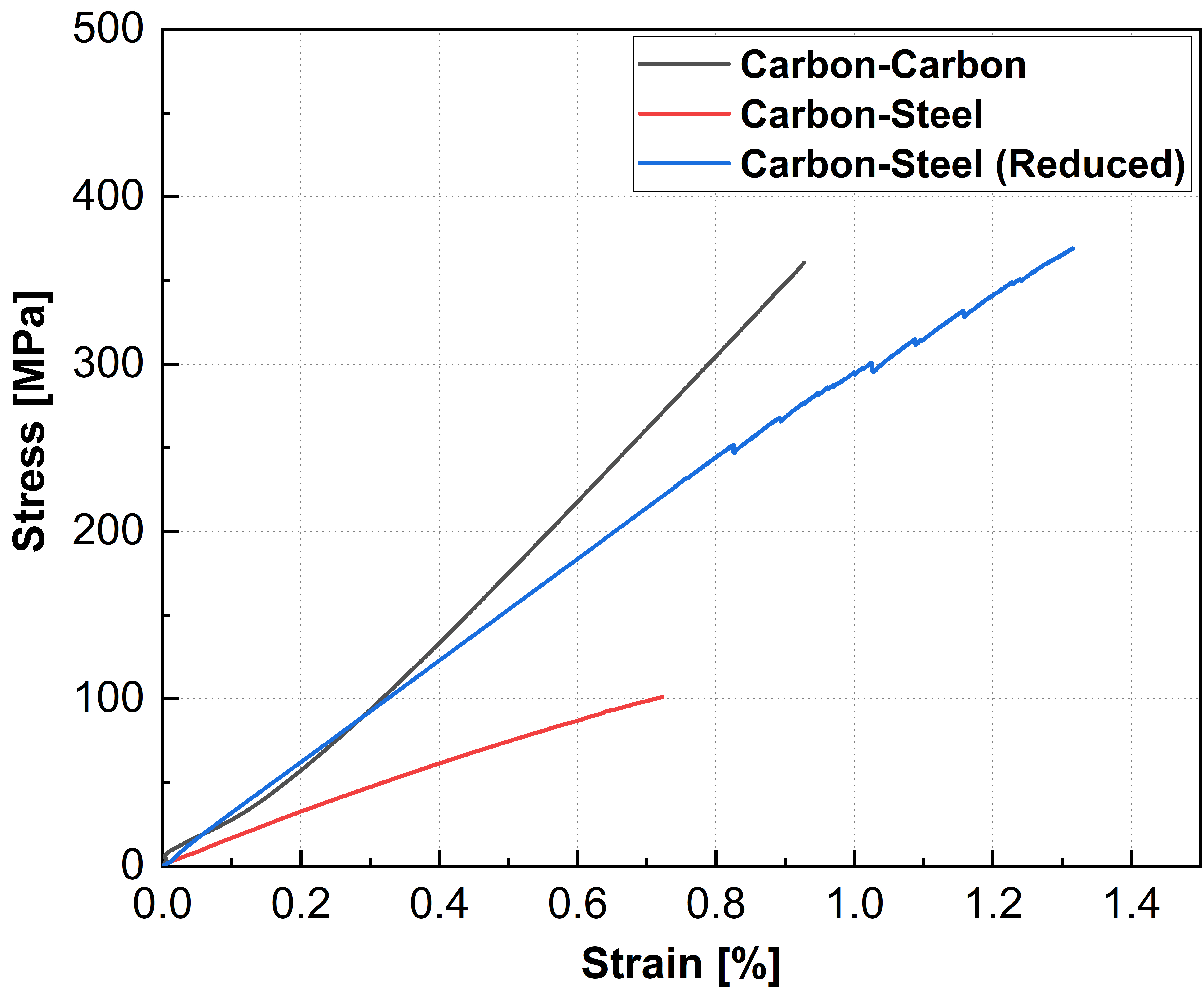}
\label{siimg:tens}
}
\subfigure[]{
\includegraphics[width=0.496\textwidth]{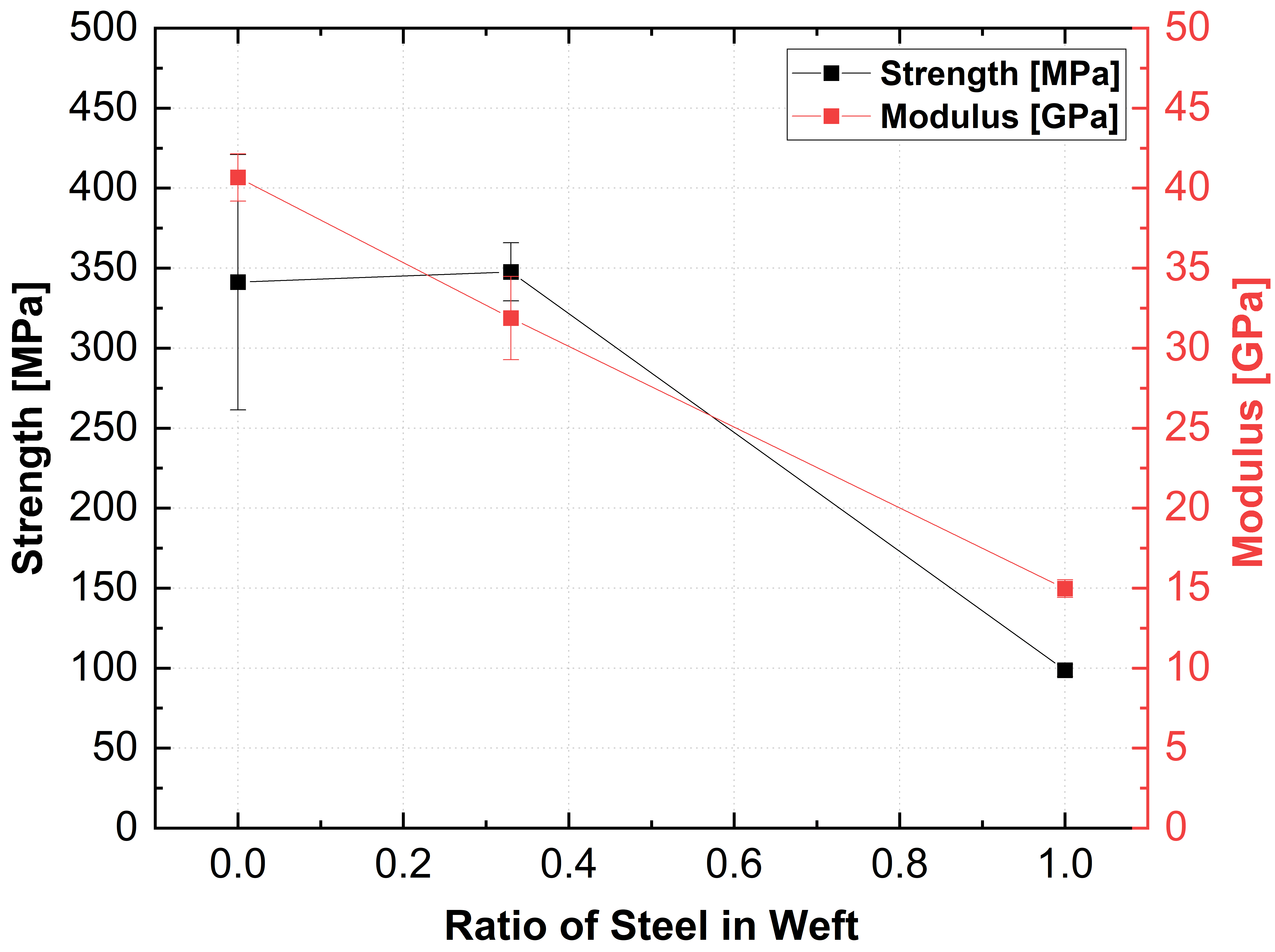}
\label{siimg:tenssumm}
}
\caption{(a) 3K carbon fiber, 24K carbon fiber, and stainless steel fiber yarns used in the current study; (b) Yarn tensile performance for all yarn types; and (c) Representative stress-strain response with (d) summarized tensile properties of composites with carbon/carbon weave and hybrid carbon/steel weaves}
\label{siimg:restensupd}
\end{figure}

We conducted uniaxial tests for each yarn type shown in \textbf{Figure} \ref{siimg:restensupd}(a), 3K carbon, 24K carbon, and stainless steel fibers, to characterize their mechanical response. Each yarn was twisted into a tight circular cross-section to determine its cross-sectional area, and the diameter was measured accordingly. A 100mm sample of each fiber yarn was randomly cut from the spool and secured in grips with a 25mm gripping length on each side. The tests were then conducted at a rate of 2mm/min using an MTS Criterion C42 load frame equipped with a 5kN load cell. For the composite samples, we performed uniaxial tensile tests on the Admet 2613 tabletop test frame with a load cell capacity of 50 kN to evaluate the influence of hybridization on mechanical response. To prevent stress concentration in the gripping area, we attached Garolite G-10/FR4 Sheets tabs with a length of 50mm at the ends of the sample. We performed these tests in compliance with ASTM D3039 \cite{ASTM3039} at a 2mm/minute crosshead displacement rate.

From \textbf{Figure} \ref{siimg:restensupd}(b), we can observe that the 3K carbon fiber yarn exhibited higher strength and modulus compared to both the 24K carbon fiber and stainless steel fiber yarns. However, the 24K carbon fiber yarns demonstrated the highest strain to failure. In contrast, the stainless steel fiber yarns had the lowest strength and modulus, failing at a lower strain value. We speculate that the extremely low diameter of the metallic fiber yarns limited their ability to stretch further. The tensile response of these fiber yarns influenced the overall tensile performance, as illustrated in \textbf{Figures} \ref{siimg:restensupd}(c) and (d). The carbon-carbon weave composite, composed of 3K carbon fiber yarns in both the warp and weft directions, exhibited superior tensile strength and modulus. On the other hand, composites containing a hybrid weave with 3K carbon and stainless steel fiber yarns in the warp and weft directions, respectively, exhibited the lowest tensile strength and modulus. Although the hybrid weave is proposed to serve as a sacrificial layer for LSP, the tensile properties of the hybrid weave can be further optimized. By incorporating 24K carbon fiber and reducing the amount of stainless steel fiber yarns in the weft direction, we observed an improvement in tensile strength and modulus, as well as a delay in failure due to the higher strain at failure of the 24K carbon fiber yarn. Furthermore, reducing the quantity of metallic fiber yarns led to a decrease in the density of hybrid weave composites from 1.91 $g/cm^3$ to 1.66 $g/cm^3$ (by 13.09\%).

\setcounter{figure}{0}

\section{Residual flexural performance}\label{si:ResFlexTIG}

\begin{figure}[h!]
\centering
\subfigure[]{
\includegraphics[width=0.31\textwidth]{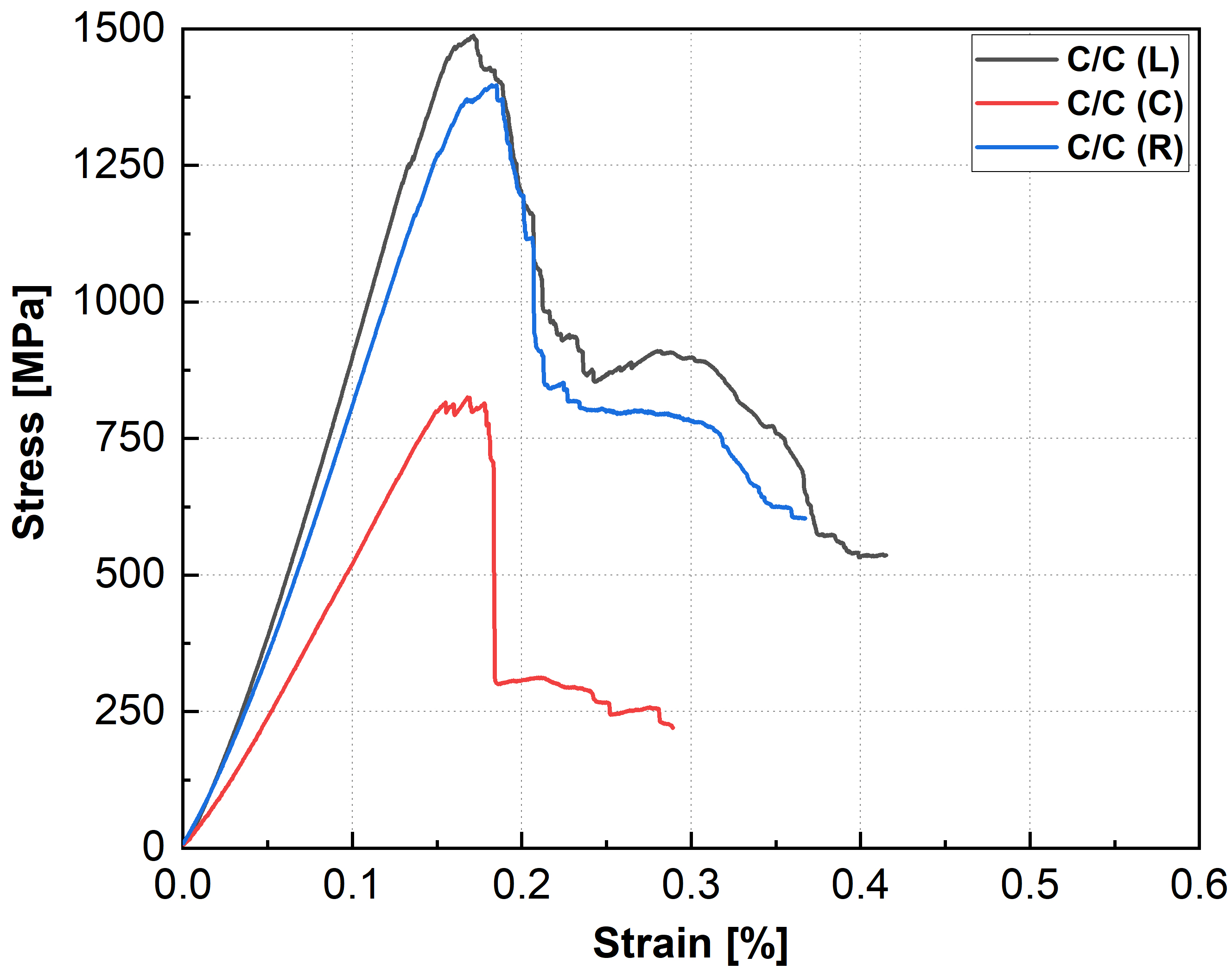}
}
\subfigure[]{
\includegraphics[width=0.31\textwidth]{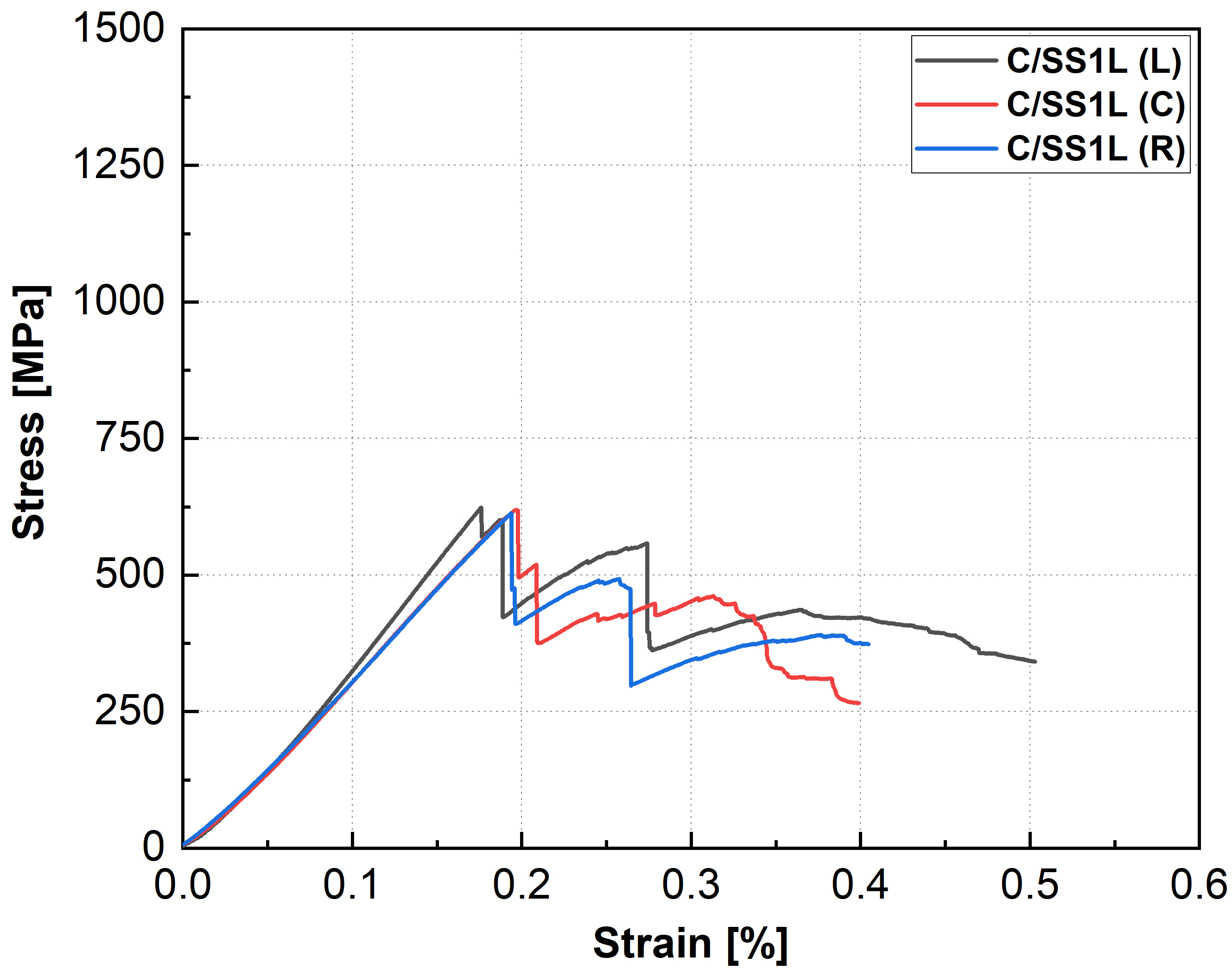}
}
\subfigure[]{
\includegraphics[width=0.31\textwidth]{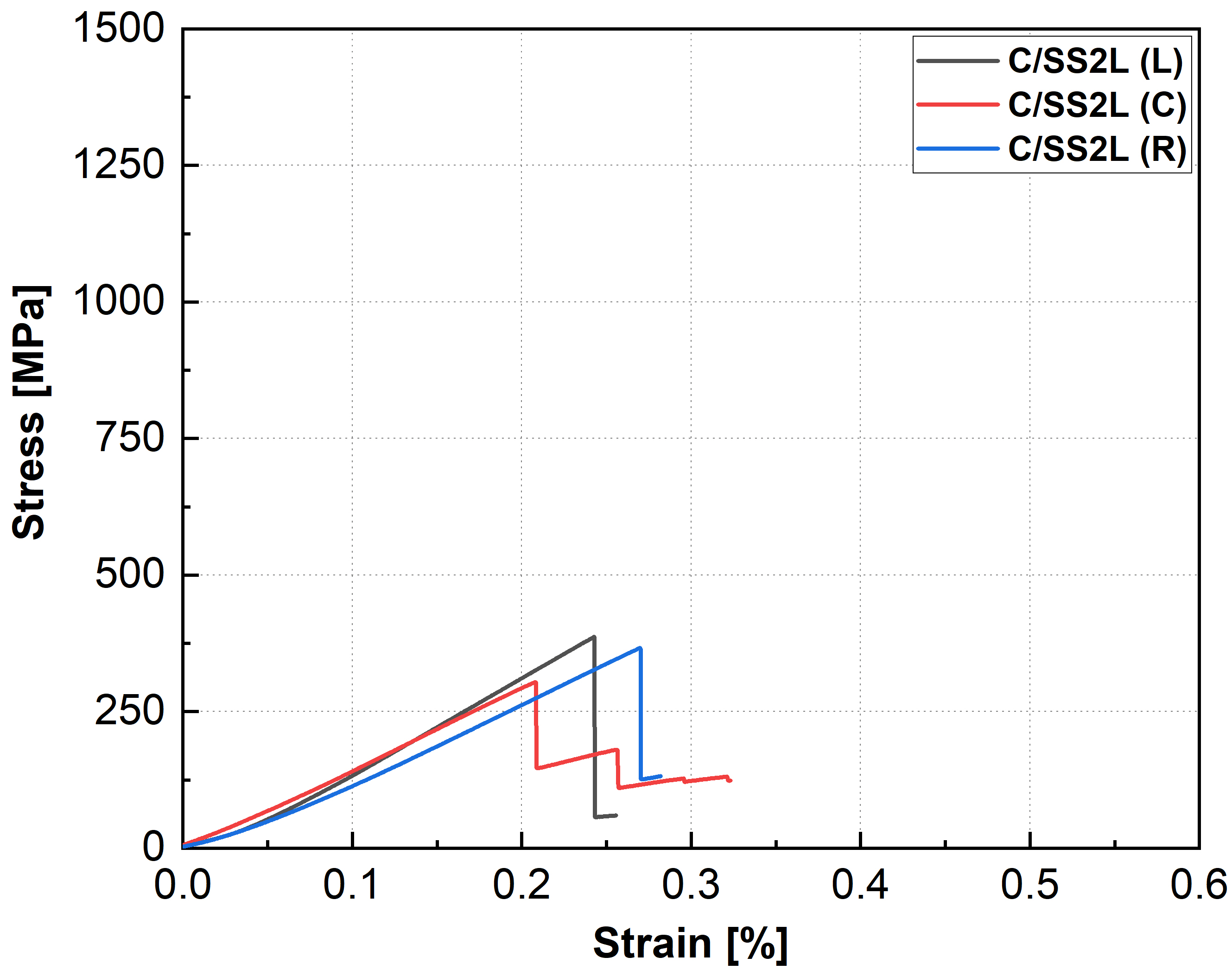}
}
\subfigure[]{
\includegraphics[width=0.31\textwidth]{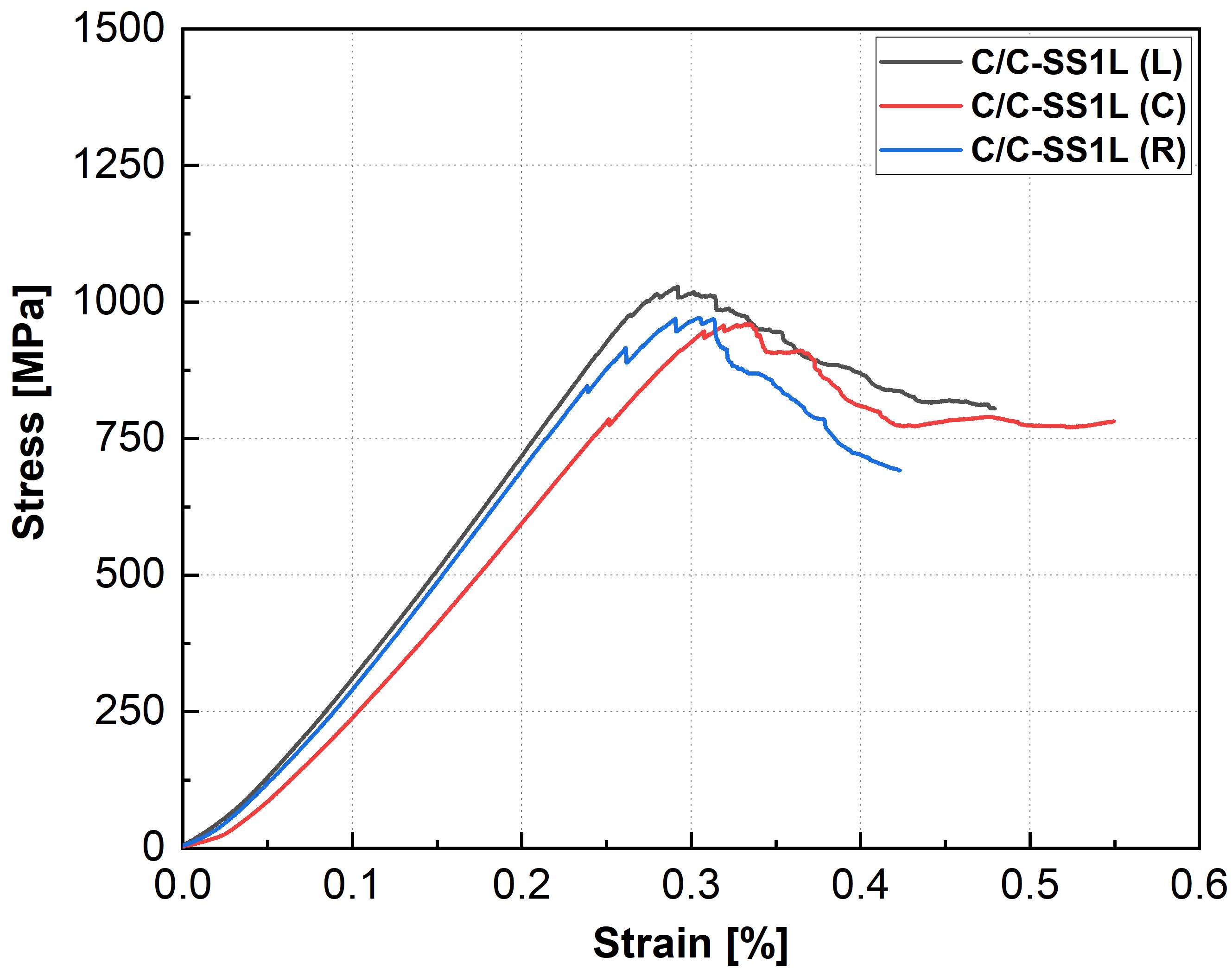}
}
\subfigure[]{
\includegraphics[width=0.31\textwidth]{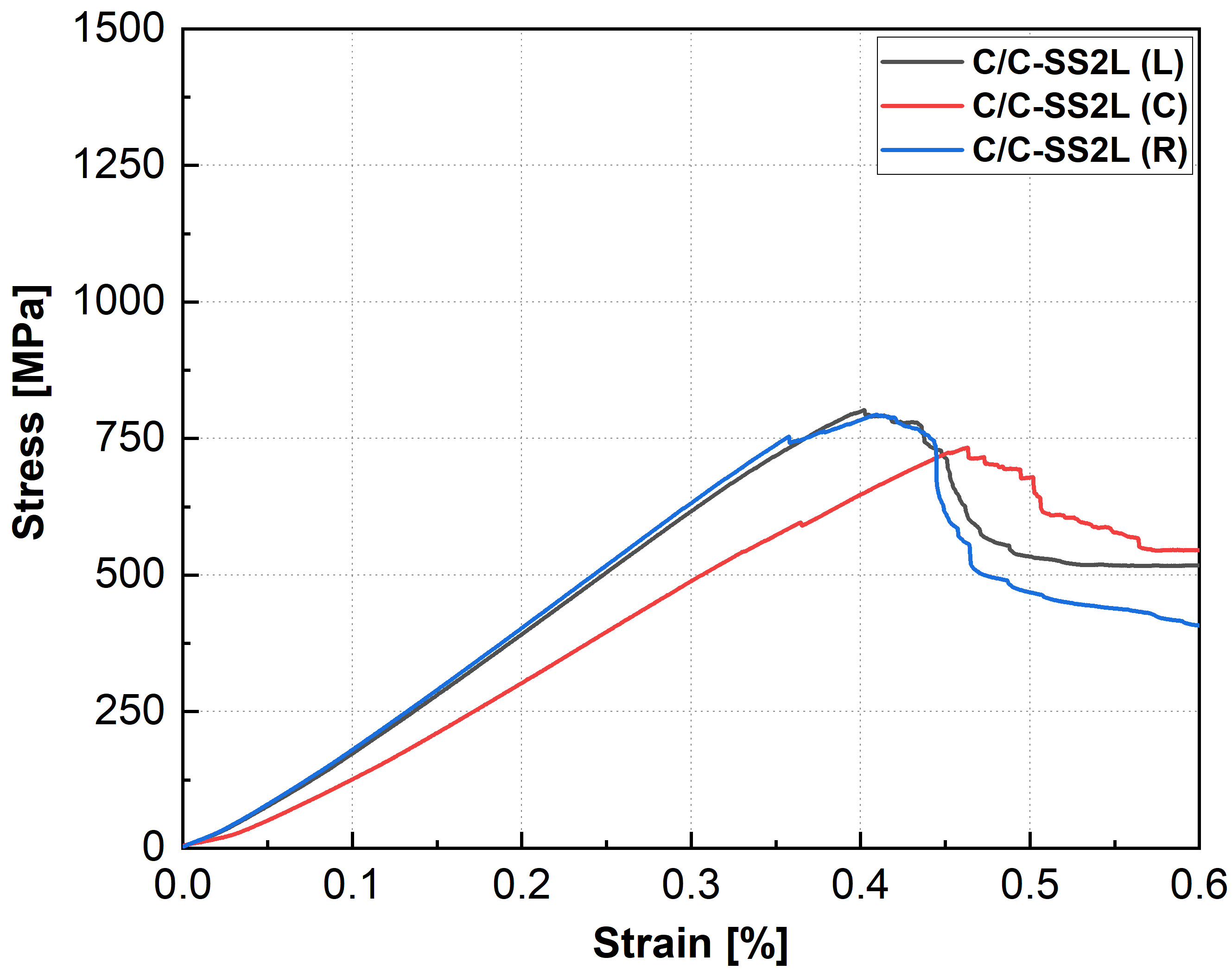}
}
\caption{Four-point bend stress-strain response for the three samples obtained from composites impacted with quasi-static electric arcs. Three samples were taken from a single impacted composite: "Left" and "Right" from either side of the impact site, and "Center" from the central region containing the damage.}
\label{siimg:ResTIGSvS}
\end{figure} 

\begin{figure}[h!]
\centering
\subfigure[]{
\includegraphics[width=0.45\textwidth]{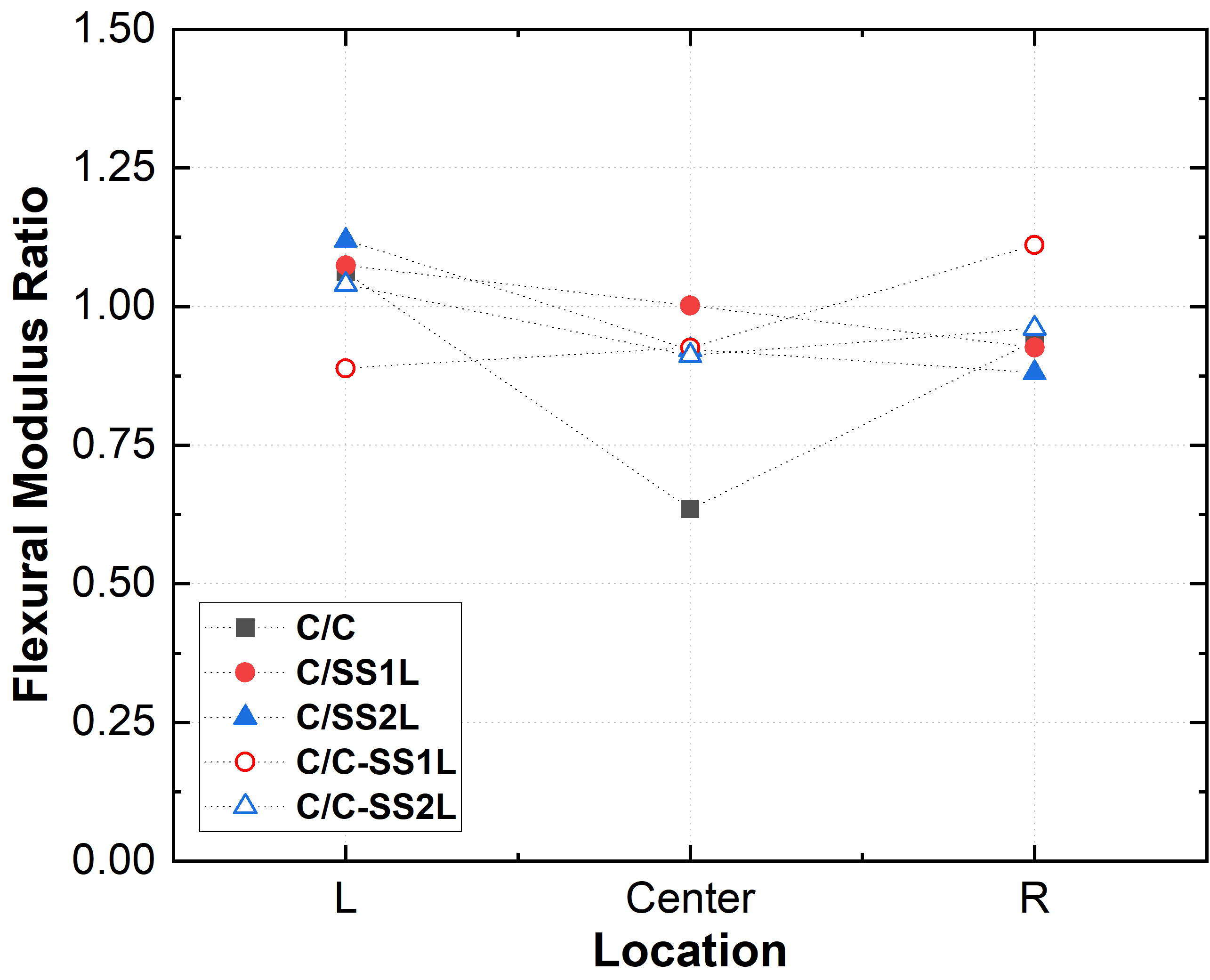}
\label{siimg:ResTIGNorm1}
}
\subfigure[]{
\includegraphics[width=0.45\textwidth]{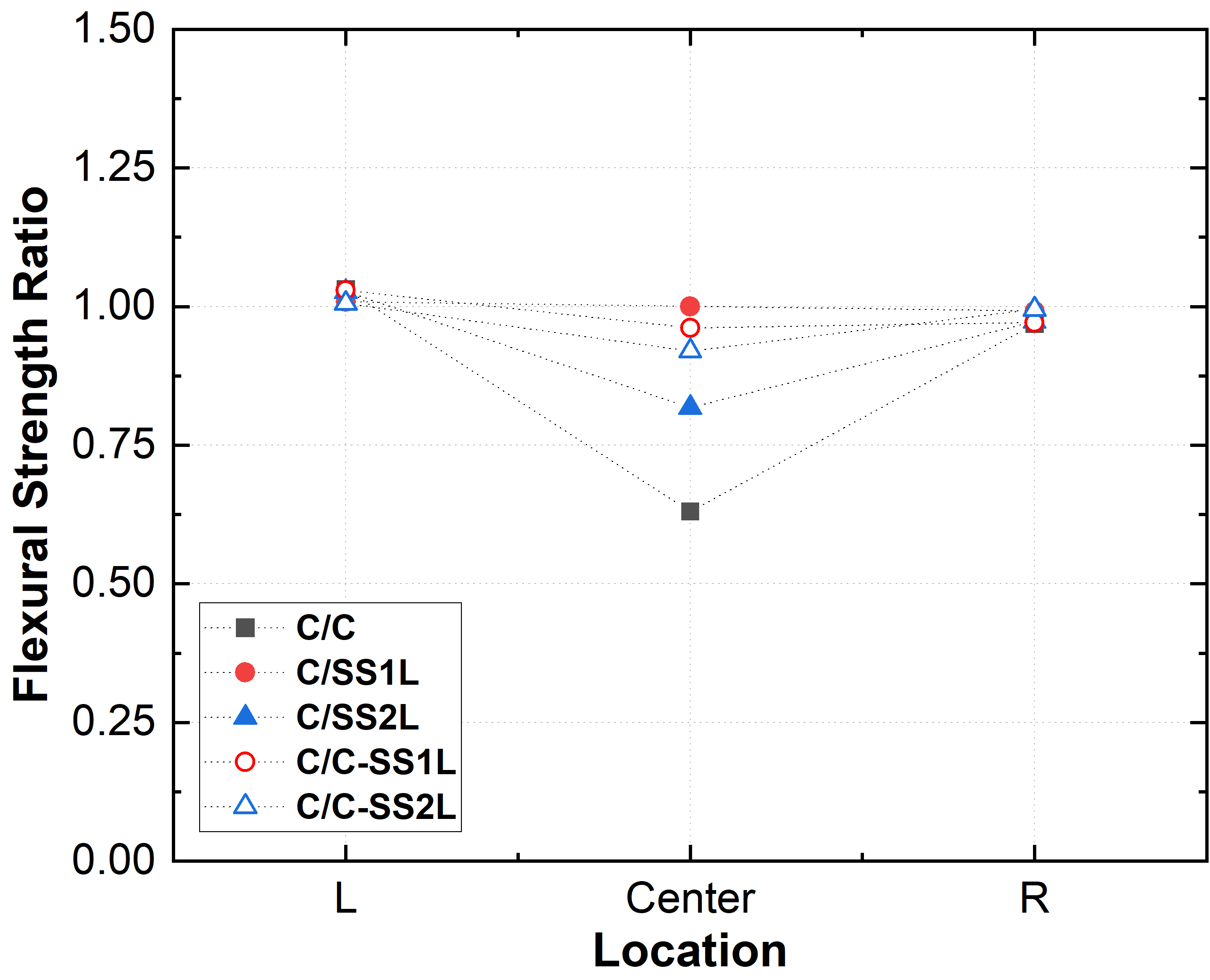}
\label{siimg:ResTIGNorm2}
}
\caption{Comparison of (a) flexural modulus ratio ($\frac{E_i}{Average(E_L,E_R)}$) and (b) flexural strength ratio ($\frac{\sigma_i}{Average(\sigma_L,\sigma_R)}$) of three samples extracted from composites exposed to quasi-static arc impacts, illustrating the extent of reduction in mechanical performance. ($E$ and $\sigma$ are flexural modulus and strength, respectively. $i$ represents the location)}
\label{siimg:ResTIG}
\end{figure}

\begin{figure}[h!]
\centering
\subfigure[]{
\includegraphics[width=0.31\textwidth]{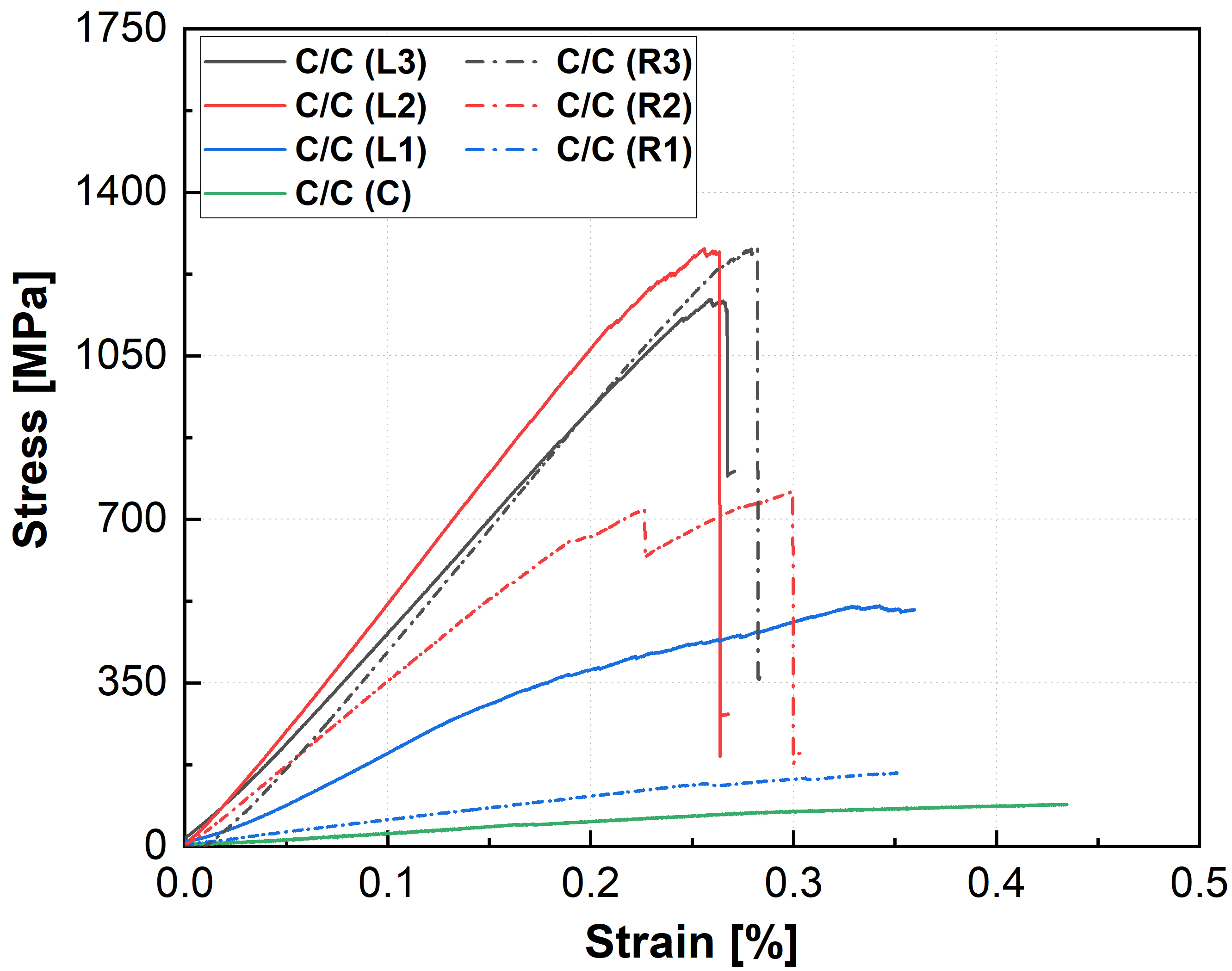}
}
\subfigure[]{
\includegraphics[width=0.31\textwidth]{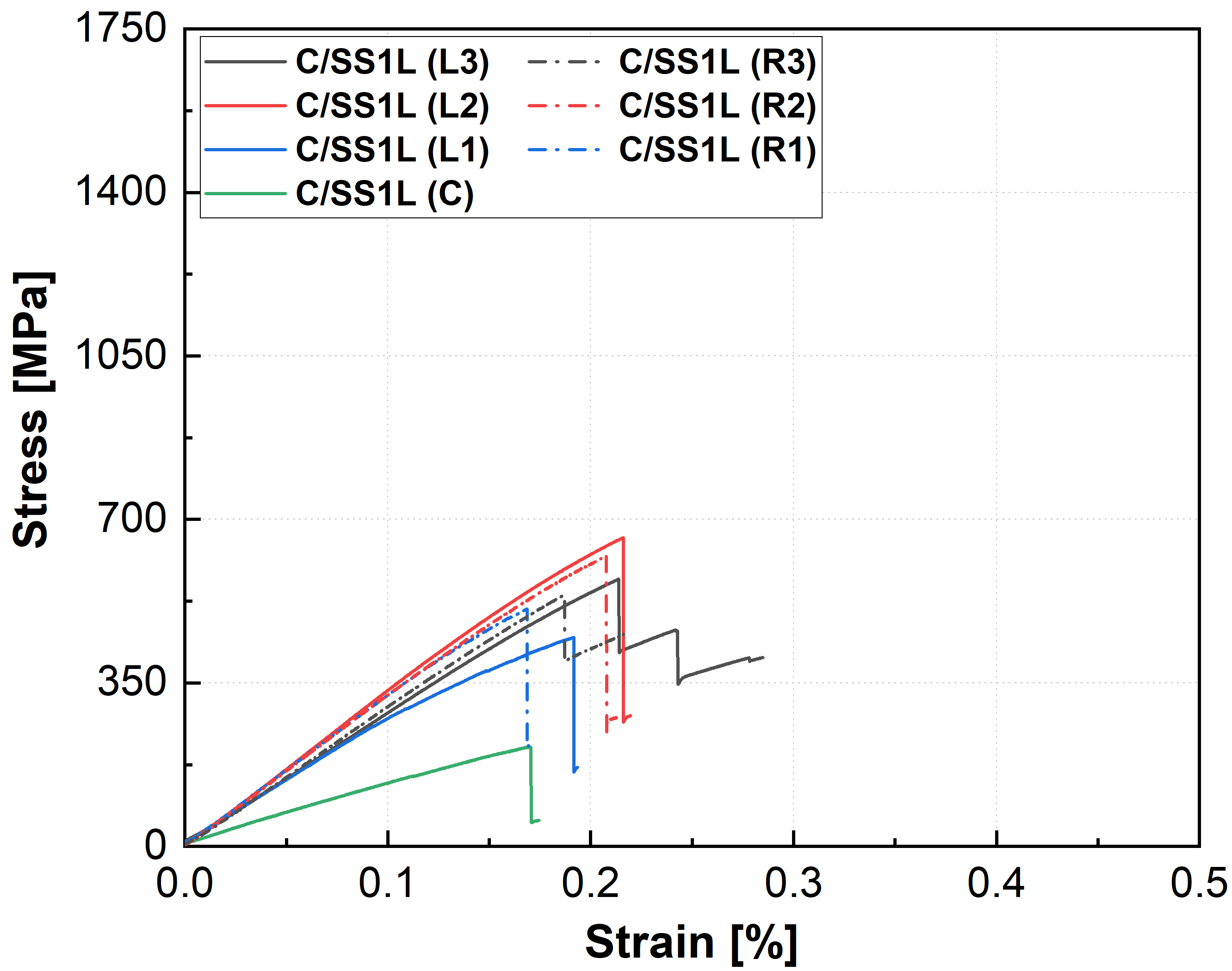}
}
\subfigure[]{
\includegraphics[width=0.31\textwidth]{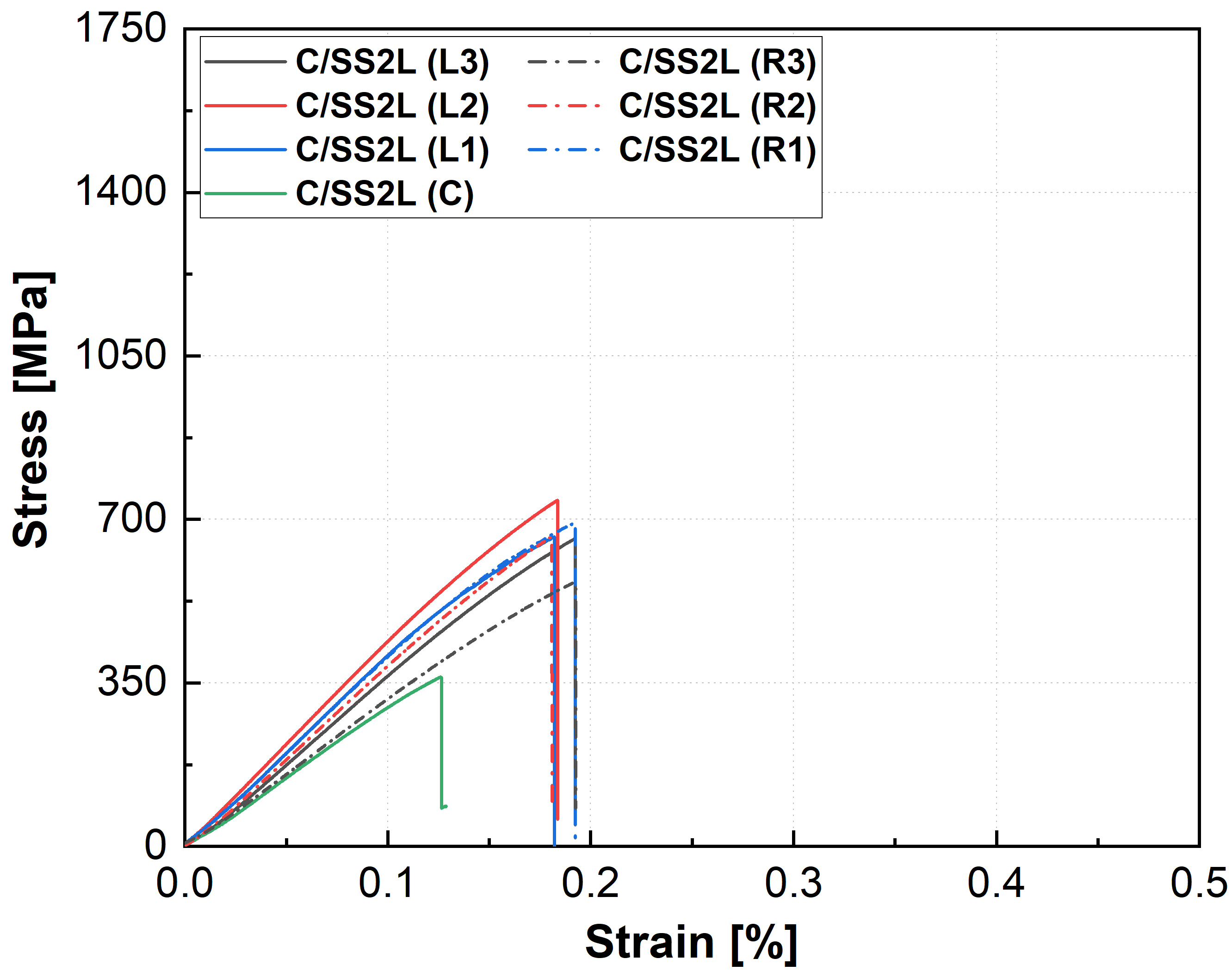}
}
\subfigure[]{
\includegraphics[width=0.31\textwidth]{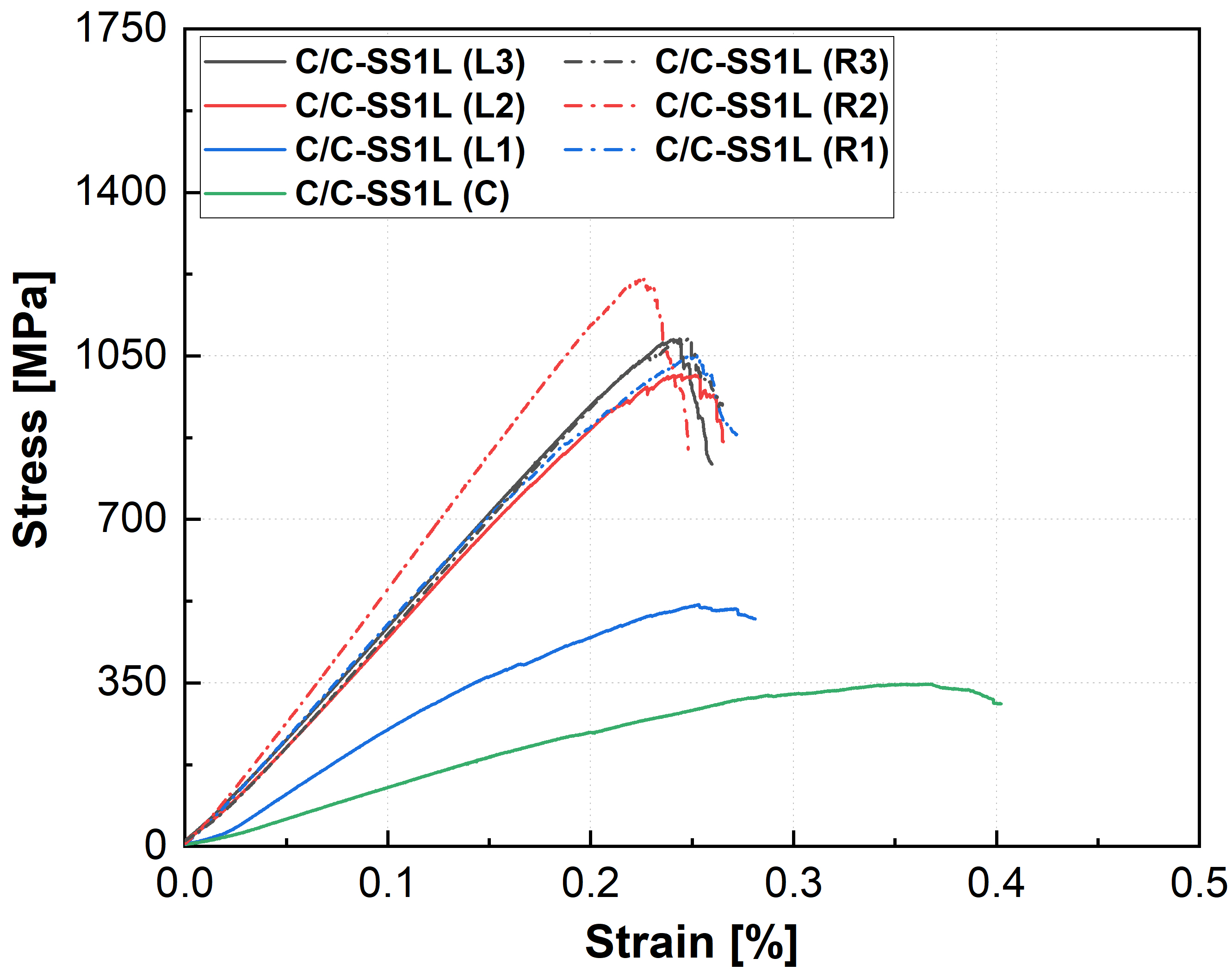}
}
\subfigure[]{
\includegraphics[width=0.31\textwidth]{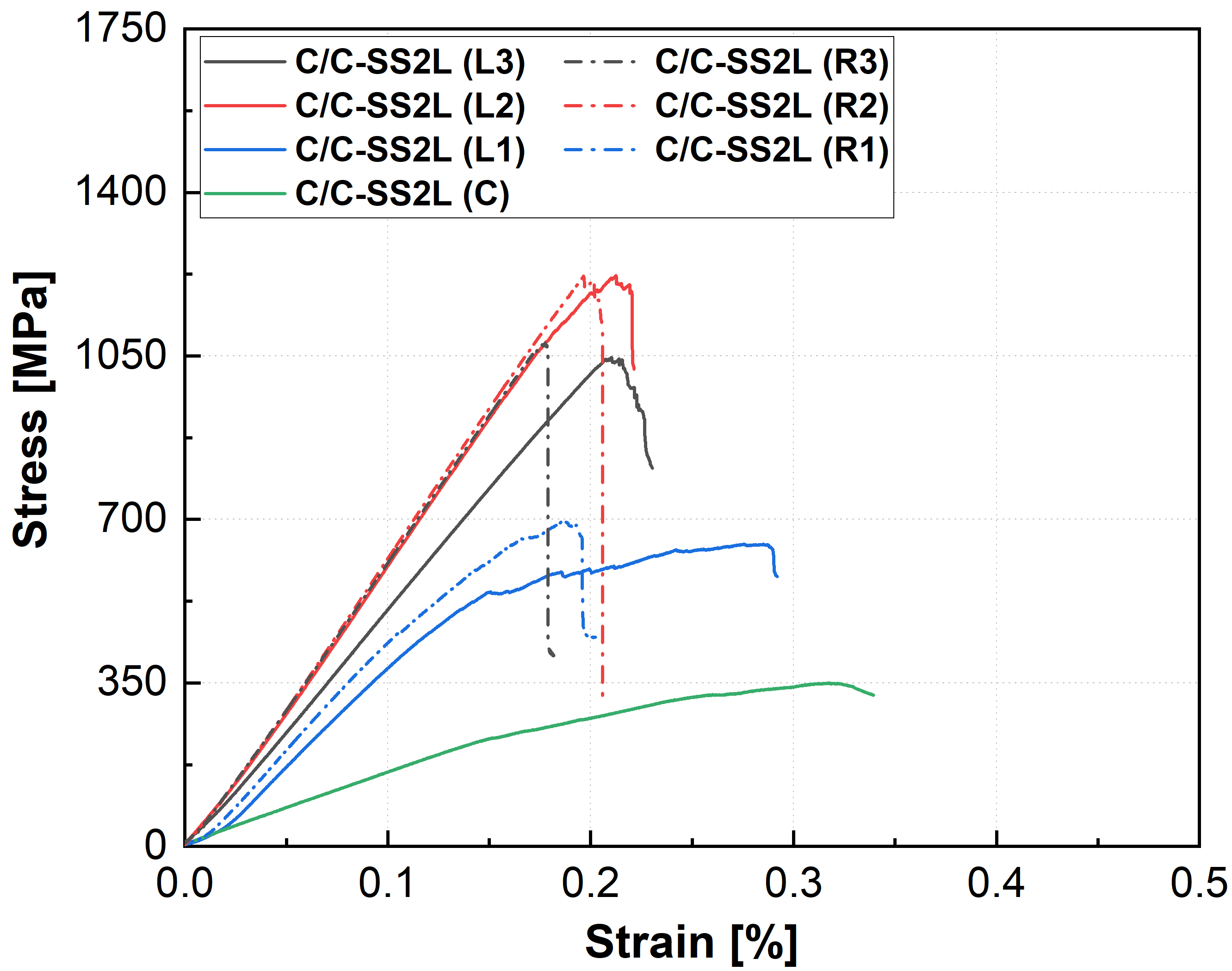}
}
\caption{Four-point bend stress-strain response for the seven samples obtained from composites impacted with simulated lightning strikes. Seven samples were taken from a single impacted composite: three from "Left" and three from "Right" from either side of the impact site, and "Center" from the central region containing the damage.}
\label{siimg:ResLSSvS}
\end{figure}   

\begin{figure}[h!]
\centering
\subfigure[]{
\includegraphics[width=0.45\textwidth]{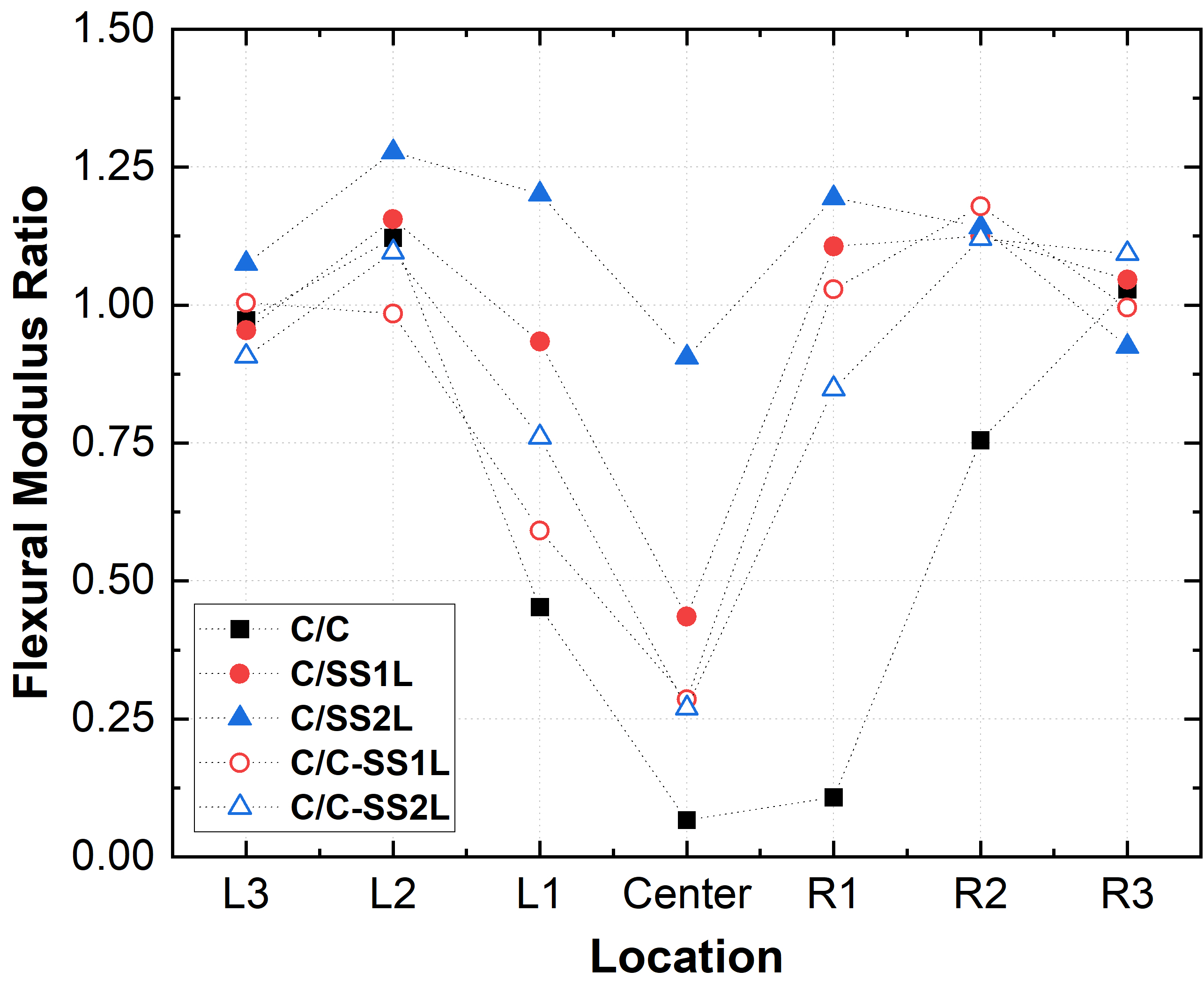}
\label{siimg:ResLSNorm1}
}
\subfigure[]{
\includegraphics[width=0.45\textwidth]{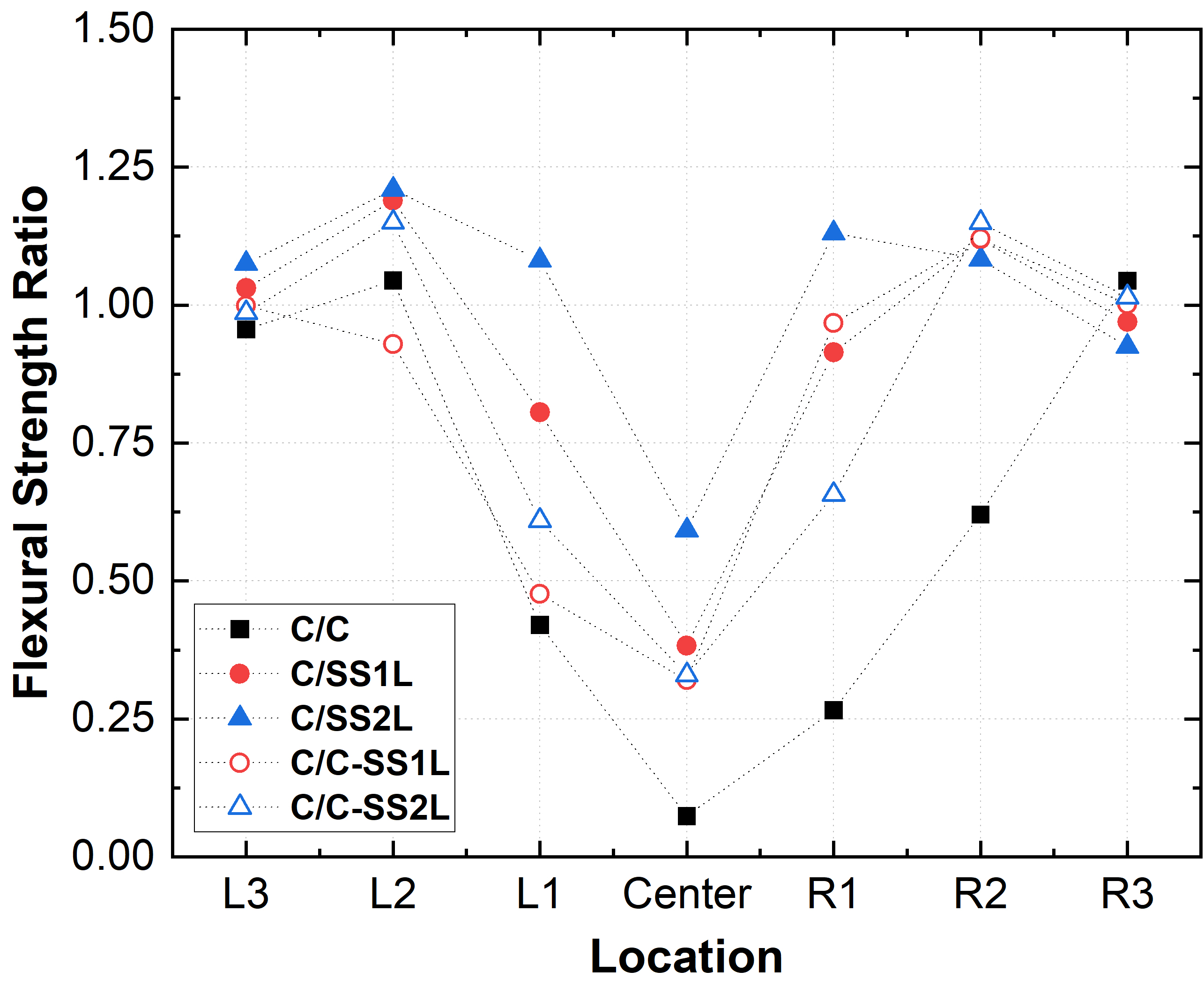}
\label{siimg:ResLSNorm2}
}
\caption{Comparison of (a) flexural modulus ratio ($\frac{E_i}{Average(E_{L3},E_{R3})}$) and (b) flexural strength ratio ($\frac{\sigma_i}{Average(\sigma_{L3},\sigma_{R3})}$) of seven samples extracted from composites exposed to simulated lightning strikes, illustrating the extent of reduction in mechanical performance. ($E$ and $\sigma$ are flexural modulus and strength, respectively. $i$ represents the location)}
\label{siimg:ResLS}
\end{figure}

\end{document}